\documentclass[journal]{IEEEtran}
\usepackage{cite}
\usepackage{epsfig}
\usepackage{pifont}
\usepackage{amssymb}
\usepackage{enumitem}
\usepackage{multirow}
\usepackage[cmex10]{amsmath, nccmath}
\usepackage{upgreek}
\usepackage{graphicx}
\usepackage{mathtools}
\usepackage{bm}
\newtheorem{theorem}{Theorem}

\newtheorem{lemma}{Lemma}
\newtheorem{definition}{Definition}

\usepackage{balance}
\newcommand{\norm}[1]{\left\lVert#1\right\rVert}
\newcommand{\cs}{{S}}
\newcommand{\ske}{{\mathrm{Sk}}}
\newcommand{\cB}[1]{\left\{#1\right\}}
\newcommand{\nB}[1]{\left(#1\right)}

\newcommand{\PX}{P_X}
\newcommand{\PY}{P_Y}
\newcommand{\QY}{Q_Y}
\newcommand{\PXo}{{P_X^{(\mr{o})}}}
\newcommand{\PXi}{{P_X^{(\mr{i})}}}
\newcommand{\QYs}{{Q_Y^*}}
\newcommand{\PXdag}{{P_X^\dagger}}
\newcommand{\PXs}{{P_X^*}}
\newcommand{\W}{P_{Y|X}}
\newcommand{\Wx}{P_{Y|X = x}}
\newcommand{\Wxn}{P_{\mb{Y}|\mb{X} = \mb{x}}}
\newcommand{\hi}{\mb{h}^{(\mr{i})}}
\newcommand{\ho}{\mb{h}^{(\mr{o})}}

\newcommand{\E}[1]{\mathbb E\left[#1\right]}
\newcommand{\Prob}[1]{\mathbb P\left[#1\right]}

\newcommand{\bs}[1]{\boldsymbol{#1}}
\newcommand{\mc}[1]{\mathcal{#1}}
\newcommand{\mb}[1]{\mathbf{#1}}
\newcommand{\ms}[1]{\mathsf{#1}}
\newcommand{\mr}[1]{\mathrm{#1}}
\newcommand{\bigo}[1]{O\left(#1\right)}

\newcommand{\Qinv}{Q^{-1}(\epsilon_n)}
\newcommand{\Phat}{\hat{P}_{\mb{x}}}
\newcommand{\Var}[1]{\mathrm{Var}\left[#1\right]}

\newcommand{\thmref}[1]{Theorem~\ref{#1}}

\newcommand{\secref}[1]{Section~\ref{#1}}
\newcommand{\lemref}[1]{Lemma~\ref{#1}}

\newcommand{\appref}[1]{Appendix~\ref{#1}}

\newcommand{\figref}[1]{Fig.~\ref{#1}}

\DeclareMathOperator*{\argmin}{arg\,min}

\usepackage{hyperref}
\usepackage{url}
\usepackage{cases}
\hypersetup{colorlinks=true, linkcolor=blue, citecolor=blue, urlcolor = blue}
\usepackage{ifthen}
\newif\ifshowtodo
\showtodotrue
\renewcommand{\IEEEproofindentspace}{1\parindent}

\newcommand{\VersionLength}{long}
\providecommand{\ver}{\ifthenelse{\equal{\VersionLength}{long}}}

\allowdisplaybreaks
\begin{document}
\title{Third-order Analysis of Channel Coding in the Small-to-Moderate Deviations Regime}

\author{Recep~Can~Yavas,~\IEEEmembership{Member,~IEEE}, Victoria~Kostina,~\IEEEmembership{Senior~Member,~IEEE}, and Michelle~Effros,~\IEEEmembership{Fellow,~IEEE}
\thanks{Manuscript received January 31, 2023; revised April 24, 2024; accepted July 1, 2024.}
\thanks{When this work was completed, R. C.~Yavas, V.~Kostina, and M.~Effros were all with the Department of Electrical Engineering, California Institute of Technology, Pasadena, CA~91125, USA. R. C. Yavas is currently with CNRS@CREATE, 138602, Singapore (e-mail:  vkostina, effros@caltech.edu, recep.yavas@cnrsatcreate.sg). This work was supported in part by the National Science Foundation (NSF) under grant CCF-1817241 and CCF-1956386. This paper was presented in part at ISIT 2022 \cite{yavas2022ThirdISIT}.}}
\date{\today}
\IEEEoverridecommandlockouts
\maketitle 
\begin{abstract}
This paper studies the third-order characteristic of nonsingular discrete memoryless channels and the Gaussian channel with a maximal-power constraint.
The third-order term in our expansions employs a new quantity here called the \emph{channel skewness}, which affects the approximation accuracy more significantly as the error probability decreases. For the Gaussian channel, evaluating Shannon's 1959 random coding bound and Vazquez-Vilar's 2021 meta-converse bound in the central limit theorem (CLT) regime enables exact computation of the channel skewness. For discrete memoryless channels, this work generalizes Moulin's 2017 bounds on the asymptotic expansion of the maximum achievable message set size for nonsingular channels from the CLT regime to include the moderate deviations (MD) regime, thereby refining Altu\u{g} and Wagner's 2014 MD result. For an example binary symmetric channel and most practically important $(n, \epsilon)$ pairs, including $n \in [100, 500]$ and $\epsilon \in [10^{-10}, 10^{-1}]$, an approximation up to the channel skewness is the most accurate among several expansions in the literature.
A derivation of the third-order term in the type-II error exponent of binary hypothesis testing in the MD regime is also included; the resulting third-order term is similar to the channel skewness. 
\end{abstract}

\section{Introduction}
The fundamental limit of channel coding is the maximum achievable message set size $M^*(n, \epsilon)$ given a channel $\W$, a blocklength $n$, and an average error probability $\epsilon$. Since determining $M^*(n, \epsilon)$ exactly is difficult for arbitrary triples $(\W, n, \epsilon)$, the literature investigating the behavior of $M^*(n, \epsilon)$ studies three asymptotic regimes: the central limit theorem (CLT) regime, where the error probability bound is kept constant and analyses bound the convergence of rate to capacity as $n$ grows; the large deviations (LD) regime, also called the error exponent regime, where the rate is kept constant and analyses bound the convergence of error probability to zero as $n$ grows; and the moderate deviations (MD) regime, where the error probability decays sub-exponentially to zero, and the rate approaches the capacity more slowly than $O(1/\sqrt{n})$. 
Provided more resources (in this case the blocklength), we would typically expect to see improvements in both the achievable rate and the error probability, an effect not captured by asymptotics that fix either rate or error probability.  
Emerging applications in ultra-reliable low-latency communication such as tele-surgery and tactile internet have delay constraints as small as 1 ms \cite{shirvan}, which corresponds to blocklengths of a few hundreds, and error probability constraints as small as $10^{-9}$. The fact that the accuracy of asymptotic expansions deteriorates at short blocklengths further motivates interest in refining the asymptotic expansions of the maximum achievable channel coding rate. 

\subsection{Literature Review}
Channel coding analyses in the CLT regime fix a target error probability $\epsilon \in (0, 1)$ and approximate $\log M^{*}(n, \epsilon)$ as the blocklength $n$ approaches infinity.
Examples of such results include Strassen's expansion~\cite{strassen} for discrete memoryless channels (DMCs) with capacity $C$, positive $\epsilon$-dispersion $V_{\epsilon}$ (defined in \cite[Sec. IV]{polyanskiy2010Channel}), and maximal error probability constraint $
\epsilon$, showing 
\begin{align}
    \log M^*(n, \epsilon) = nC - \sqrt{n V_{\epsilon}} Q^{-1}(\epsilon) + O(\log n). \label{eq:Gaussianapp}
\end{align}
Polyanskiy \emph{et al.} \cite{polyanskiy2010Channel} and Hayashi \cite{hayashi} revisit Strassen's result, showing that the same asymptotic expansion holds for the average (over the codebook and channel statistics) error probability constraint, deriving lower and upper bounds on the coefficient of the $\log n$ term, and extending the result to Gaussian channels with maximal and average power constraints. In all asymptotic expansions below, the \emph{average}  error probability criterion is employed.


For channel coding in the LD regime, one fixes a rate $R = \frac{\log M}{n}$ strictly below the channel capacity and seeks to characterize the minimum achievable error probability $\epsilon^{*}(n, R)$ as the blocklength $n$ approaches infinity. In this regime, $\epsilon^{*}(n, R)$ decays exponentially with $n$. For $R$ above the critical rate, \cite[Ch. 5]{gallager1968book} derives the error exponent $E(R)$, where
\begin{align}
    \epsilon^*(n, R) = e^{-n E(R) + o(n)}. \label{eq:Gallag}
\end{align}
Bounds on the $o(n)$ term in \eqref{eq:Gallag} appear in \cite{altug2014refinement, altug2021onexact, honda2018, segura2018}. For the Gaussian channel with a maximal-power constraint, Shannon \cite{shannon1959Probability} derives LD-regime achievability and converse bounds with an $o(n)$ term that is tight up to an $O(1)$ gap. Erseghe \cite{erseghe2016} gives an alternative proof of these LD approximations using the Laplace integration method. A recent paper by Vazquez-Vilar \cite{vazquez2021} derives refined non-asymptotic converse bounds for the Gaussian channel under maximal and average power constraints and analyzes these bounds in the LD regime.  



The CLT and LD asymptotic approximations in \eqref{eq:Gaussianapp} and \eqref{eq:Gallag}, respectively, become less accurate as the $(n, \epsilon)$ pair gets farther away from the regime that considered in their derivation. For example, the CLT approximation falls short if $\epsilon$ is small since there is a hidden $Q^{-1}(\epsilon)^2$ term inside the $O(\log n)$ term, and $Q^{-1}(\epsilon)$ approaches $\infty$ as $\epsilon$ approaches 0. 
To better understand this phenomenon, consider the class of nonsingular channels, which is the class of DMCs $\W$ for which there exist channel inputs $x_1, x_2 \in \mc{X}$ and a channel output $y \in \mc{Y}$ for which $P_{Y|X}(y|x_1) > P_{Y|X}(y|x_2) > 0$ and the focus in this work. Recall that in the CLT regime, the third-order $O(\log n)$ term  is equal to $\frac{1}{2} \log n + O(1)$ \cite{polyanskiy2010Channel, tomamichel2013converse} for all nonsingular channels. In this work, we show that when $\epsilon_n$ decays sub-exponentially to zero, the third-order term becomes roughly $\frac{1}{2} \log n + S Q^{-1}(\epsilon_n)^2 \approx \frac{1}{2} \log n + 2 S \log \frac{1}{\epsilon_n}$, where the $S$ term, which is the \emph{channel skewness} defined shortly. As this equation makes clear, the $O(1)$ term in the CLT expansion depends on $\epsilon_n$ through the function $Q^{-1}(\epsilon_n)^2$, which can dominate the $\frac{1}{2} \log n$ term for $\epsilon_n$ small enough. 
For example, for $\epsilon_n$ polynomially decaying to zero, i.e.,  $\epsilon_n = \frac{1}{n^r}$ for some $r > 0$, the coefficient $\frac{1}{2}$ of $\log n$ in \cite{polyanskiy2010Channel} becomes inaccurate. Further, if $\epsilon_n$ decays more quickly, then this inaccuracy becomes more extreme. For example, for $\epsilon_n = \exp\{-n^{1/2}\}$, the second-order term becomes $\sqrt{2V} n^{3/4}$, and the third-order term becomes $2 S \sqrt{n}$.  Similarly, the LD approximation falls short if the rate $R$ is large since the second-order term $o(n)$ in the error exponent grows arbitrarily large as the rate approaches the capacity. 

The inability of the CLT and LD regimes to provide accurate approximations for a wide range of $(n, \epsilon)$ pairs and the hope of deriving more accurate (yet computable) approximations to the finite blocklength rate motivate the study of the MD regime, which simultaneously considers low error probabilities and high achievable rates. For DMCs with positive dispersion $V$ and a sequence of sub-exponentially decaying $\epsilon_n$ values, Altu\u{g} and Wagner \cite{altug2014moderate} and Polyanskiy \emph{et al.} \cite{polyanskiy2010Moderate} show that
\begin{IEEEeqnarray}{rCl}
     \log M^*(n, \epsilon_n) = nC - \sqrt{n V_{\epsilon_n}} Q^{-1}(\epsilon_n) + o(\sqrt{n} Q^{-1}(\epsilon_n)). \IEEEeqnarraynumspace \label{eq:altug}
\end{IEEEeqnarray}
This result implies that the CLT approximation to the maximum achievable message set size $\log M^*(n, \epsilon_n) \approx nC - \sqrt{n V}Q^{-1}(\epsilon_n)$ as in \eqref{eq:Gaussianapp}, remains valid in the MD regime, leaving open the rate of convergence to that bound. 

To discuss the accuracy of the CLT approximation \eqref{eq:Gaussianapp}, for any given channel, fix an average error probability $\epsilon$ and blocklength $n$. We define the channel's \emph{non-Gaussianity} as
\begin{align}
    \zeta(n, \epsilon) \triangleq \log M^*(n, \epsilon) - (nC - \sqrt{n V_{\epsilon}}Q^{-1}(\epsilon)), \label{eq:CLT}
\end{align}
which captures the third-order term in the expansion of $\log M^*(n, \epsilon)$ around $nC$.


According to Strassen's expansion \eqref{eq:Gaussianapp}, $\zeta(n, \epsilon) = O(\log n)$. Subsequent works include several refinements to that result. The results of \cite{polyanskiy2010Channel} imply that the non-Gaussianity of a DMC with finite input alphabet $\mc{X}$ and output alphabet $\mc{Y}$ is bounded as
\begin{align}
    O(1) \leq \zeta(n, \epsilon) \leq \left(|\mc{X}|-\frac{1}{2}\right) \log n + O(1). \label{eq:lowupbound}
\end{align}
Further, restricting the class of channels considered enables further improvements to \eqref{eq:lowupbound}. We next briefly define several channel characteristics and the corresponding refinements. 

Each definition relies on the channel transition probability kernel $[P_{Y|X}(y|x)]$ from $x$ to $y$, with rows corresponding to channel inputs and columns corresponding to channel outputs. See \secref{sec:singular} for formal definitions.
Singular channels are channels for which all entries in each column of the transition matrix are 0 or $p$ for some constant $p 
\in (0, 1]$; nonsingular channels are channels that do not satisfy this property. While the binary symmetric channel (BSC) is nonsingular, the binary erasure channel (BEC) is singular. Gallager-symmetric channels are channels whose output alphabet can be partitioned into subsets so that for each subset of the transition probability kernel that uses inputs as rows and outputs of the subset as columns has the property that each row (respectively, column) is a permutation of each other row (respectively, column) \cite[p.~94]{gallager1968book}. Both the BSC and BEC are Gallager-symmetric. Cover--Thomas-symmetric channels \cite[p.~190]{cover} are the channels for which all rows (and respectively columns) of the transition probability kernel are permutations of each other; the family of Cover--Thomas symmetric channels is a subset of the class of Gallager-symmetric channels. The BSC is Cover--Thomas-symmetric; the BEC is not.
For Gallager-symmetric, singular channels, $\zeta(n, \epsilon) = O(1)$ \cite{altug2021onexact}. 
For nonsingular channels, the random coding union (RCU) bound improves the lower bound in \eqref{eq:lowupbound} to $\frac{1}{2} \log n + O(1)$ \cite[Cor.~54] {polyanskiy2010thesis}. For DMCs with positive $\epsilon$-dispersion, Tomamichel and Tan \cite{tomamichel2013converse} improve the upper bound to $\frac{1}{2} \log n + O(1)$. A random variable is called lattice if it takes values on a lattice with probability 1 and is called non-lattice otherwise. For nonsingular channels with positive $\epsilon$-dispersion and non-lattice information density, Moulin \cite{moulin2017log} shows\footnote{There is a sign error in \cite[eq. (3.1)-(3.2)]{moulin2017log}, which then propagates through the rest of the paper. The sign of the terms with $\ske(\PXs)$ should be positive rather than negative in both equations. The error in the achievability result originates in \cite[eq. (7.15) and (7.19)]{moulin2017log}, where it is missed that $\ske(-X) = -\ske(X)$ for any random variable $X$. The error in the converse result also stems from the sign error in \cite[eq. (6.8)]{moulin2017log}.}
\begin{align}
    \zeta(n, \epsilon) &\geq\frac{1}{2} \log n + \underline{\cs} \, Q^{-1}(\epsilon)^2 + \underline{B} + o(1)  \label{eq:achrho}\\
    \zeta(n, \epsilon) &\leq
    \frac{1}{2} \log n + \overline{\cs} \, Q^{-1}(\epsilon)^2 + \overline{B} + o(1), \label{eq:convrho}
\end{align}
where $\underline{\cs}$, $\overline{\cs}$. $\underline{B}$, and  $\overline{B}$ are constants that depend on the channel parameters. It is possible to extend Moulin's expansions in \eqref{eq:achrho}--\eqref{eq:convrho} to all DMCs with lattice information densities. To do that, we should use the continuity-corrected Edgeworth expansion given in \cite[Ch.~3.16]{kolassa} instead of the standard Edgeworth expansion for non-lattice random variables. We can further refine the achievability bound in \eqref{eq:achrho} by considering the tie-breaking strategy from \cite{haim2013}. Specifically, Haim \emph{et al.} \cite[Sec.~IV]{haim2013} argue that if there is a tie between $k$ messages in the largest information density, then an equiprobable random decoding rule among ties succeeds with probability $\frac{1}{k}$; analyzing \cite[Th.~1]{haim2013} instead of the RCU bound in the CLT regime would improve $\underline{B}$ in \eqref{eq:achrho}.

In \cite{polyanskiy2010Channel}, Polyanskiy \emph{et al.} consider the Gaussian channel with a maximal-power constraint $P$, under which every codeword has power less than or equal to $nP$, showing for the CLT domain that the non-Gaussianity $\zeta(n, \epsilon, P)$ is bounded as
\begin{align}
    O(1) \leq \zeta(n, \epsilon, P) \leq \frac{1}{2} \log n + O(1). \label{eq:GaussPol}
\end{align}
Tan and Tomamichel \cite{tan2015Third} improve \eqref{eq:GaussPol} to 
\begin{align}
    \zeta(n, \epsilon, P) = \frac{1}{2} \log n + O(1),
\end{align}
which means that in the CLT regime, the non-Gaussianity of the Gaussian channel is the same as that of nonsingular DMCs with positive $\epsilon$-dispersion. 

%

The MD result in \eqref{eq:altug}
can be expressed as
\begin{IEEEeqnarray}{rCl}
    \zeta(n, \epsilon_n) = o(\sqrt{n} Q^{-1}(\epsilon_n)). \IEEEeqnarraynumspace \label{eq:moderateexp}
\end{IEEEeqnarray}
Polyanskiy and Verd\'u \cite{polyanskiy2010Moderate} show \eqref{eq:moderateexp} using the MD result in \cite[Th.~3.7.1]{dembobook}. While that MD result is tight enough to prove several second-order MD asymptotics in information theory, it is not tight enough to refine the third- and higher-order terms. Polyanskiy and Verd\'u also extend \eqref{eq:moderateexp} to the Gaussian channel with a maximal power constraint. In \cite{chubb2017quantum}, Chubb \emph{et al.} extend the second-order MD expansion in \eqref{eq:moderateexp} to quantum channels. In \cite[Th.~2]{yavas2021VLSF}, the current authors derive an asymptotic expansion of the maximum achievable rate for variable-length stop-feedback codes, where $\epsilon_n = \frac{1}{\sqrt{n \log n}}$. In \cite[Lemma~3]{sakai2021Third}, Sakai \emph{et al.} derive a third-order asymptotic expansion  for the minimum achievable rate of lossless source coding, where $\epsilon_n$ decays polynomially with $n$; this third-order expansion can be extended to all MD sequences using the tools presented here. A second-order MD analysis of lossy source coding appears in \cite{tan2012moderatelossy}.

Since binary hypothesis testing (BHT) is closely related to several information-theoretic problems and admits a CLT approximation similar to that of channel coding \cite{polyanskiy2010Channel}, BHT is a topic of some interest in this work. Refined asymptotics for BHT receive significant attention from the information theory community. If the type-I error probability is a constant $1-\alpha \in (0, 1)$ independent of the number of samples $n$ (i.e., in the Stein regime), the minimum achievable type-II error probability $\beta$ is a function of $\alpha$ and $n$, and a CLT approximation to the type-II error exponent, $-\log \beta_{\alpha}$, appears in \cite[Sec.~2]{strassen} and \cite[Lemma~58]{polyanskiy2010Channel}. In \cite{strassen}, Strassen considers testing $P^{\otimes n}$ against $Q^{\otimes n}$ and identifies the skewness term in the type-II error exponent. To do this, he relies on the Edgeworth expansion given in \eqref{eq:Edge}, below. In \cite[Lemma~58]{polyanskiy2010Channel}, Polyanskiy \emph{et al.} extend Strassen's result to the case of independent but not necessarily identical distributions but do not derive a bound on the skewness term. In \cite[Th.~18]{moulin2017log}, Moulin refines \cite[Lemma~58]{polyanskiy2010Channel} by deriving the skewness term in the semistrong non-lattice and lattice cases.\footnote{There is a typo in \cite[eq. (6.8)]{moulin2017log}. The sign of the third term in \cite[eq. (6.8)]{moulin2017log} should be plus rather than minus.} In the LD (or Chernoff) regime, where both error probabilities decay exponentially, the type-I and type-II error exponents appear in, e.g.,\cite[eq.~(11.196)-(11.197)]{cover}. A second-order MD analysis of BHT appears in~\cite{sason2012moderate}. In \cite[Th.~11]{chen2020lossless}, Chen \emph{et al.} derive the third-order asymptotic expansion of the type-II error probability region in the CLT regime for composite hypothesis testing that considers a single null hypothesis and $k$ alternative hypotheses. The second-order term in their result includes an extension of the $Q^{-1}(\cdot)$ function to $k$-dimensional Gaussian random vectors.

Casting optimal coding problems in terms of hypothesis testing elucidates the fundamental limits of coding. In \cite[Th.~5]{sgb}, Shannon \emph{et al.} use a BHT result to derive lower bounds on the error probability in channel coding in the LD regime.
Polyanskiy \emph{et al.} derive a converse result \cite[Th.~27]{polyanskiy2010Channel} in channel coding using the minimax of the type-II error probability of BHT, the $\beta_{\alpha}$ function; they call this converse the meta-converse bound. Kostina and Verd\'u prove a converse result \cite[Th.~8]{kostina2012lossyJ} for fixed-length lossy compression of stationary memoryless sources using the $\beta_{\alpha}$ function. This result is extended to lossless joint source-channel coding in \cite{kostina2013joint}. For lossless data compression, Kostina and Verd\'u give lower and upper bounds \cite[eq.~(64)]{kostina2012lossyJ} on the minimum achievable codebook size in terms of  $\beta_{\alpha}$. For lossless multiple access source coding, also known as Slepian--Wolf coding, Chen \emph{et al.} derive a converse result \cite[Th.~19]{chen2020lossless} in terms of the composite hypothesis testing version of the $\beta_{\alpha}$ function. Composite hypothesis testing is also used in a random access channel coding scenario to decide whether any transmitter is active \cite{yavas2020Random}. The works in \cite{polyanskiy2010Channel, kostina2012lossyJ, kostina2013joint, chen2020lossless, yavas2020Random} derive second- or third-order asymptotic expansions for their respective problems by using the asymptotics of the $\beta_{\alpha}$ function.


\subsection{Contributions of This Work}
The accuracy of Strassen's CLT approximation \eqref{eq:Gaussianapp}, giving $\zeta(n, \epsilon) = O(\log n)$, decreases significantly when the blocklength $n$ is small and the error probability $\epsilon$ is small. As discussed earlier, this problem arises because of the hidden $ Q^{-1}(\epsilon)^2$ term inside the non-Gaussianity \eqref{eq:CLT} \cite{moulin2017log}. Recall that $Q^{-1}(
\epsilon)^2$ approaches $2 \log \frac{1}{\epsilon}$, which in turn grows without bound as $\epsilon \to 0$. 
To capture this phenomenon, we define the \emph{channel skewness} operationally~as 
\begin{align}
    \cs \triangleq \lim \limits_{\epsilon \to 0} \liminf\limits_{n \to \infty} \frac{\zeta(n, \epsilon) - \frac{1}{2} \log n}{Q^{-1}(\epsilon)^2}. \label{eq:skewnessdef}
\end{align}
The channel skewness serves as the third-order fundamental channel characteristic after channel capacity and dispersion \cite[Sec. IV]{polyanskiy2010Channel}. The skewness of the information density (see \eqref{eq:skewness}, below) plays a critical role in characterizing the channel skewness. Throughout the paper, we use $\overline{S}$ and $\underline{S}$ to represent upper and lower bounds on the channel skewness~$S$. 
Our contributions in this paper are summarized as follows.
\begin{itemize}
    \item We study nonsingular DMCs with positive dispersion, showing that the MD-regime lower and upper bounds on the non-Gaussianity in \eqref{eq:achrho}--\eqref{eq:convrho} hold up to the skewness term; this result justifies why the skewness approximations remain accurate even for error probabilities as small as $10^{-10}$ and blocklengths as short as $n \leq 500$.
    \item For Cover--Thomas-symmetric channels \cite[p.~190]{cover} (e.g., the BSC), the lower and upper bounds in \eqref{eq:achrho}--\eqref{eq:convrho} match, and we derive the term that is one order higher than the channel skewness.
    \item We compute the channel skewness of the Gaussian channel with a maximal-power constraint by deriving refined bounds in the CLT regime; the resulting approximations have an accuracy similar to that of Shannon's LD approximations from \cite{shannon1959Probability}.
    \item We derive tight bounds on the minimum achievable type-II error probability for BHT in the MD regime; our bounds yield a fourth-order asymptotic expansion that includes the third and fourth central moments of the log-likelihood ratio. The converse in our refined result for Cover--Thomas-symmetric channels (described in the previous bullet) is a direct application of this expansion. Our expansion is also potentially useful in other applications, such as extending the results from \cite{blahut1974, kostina2012lossyJ, kostina2013joint, chen2020lossless, yavas2020Random}, which rely on the BHT asymptotics, to the MD regime.  
\end{itemize}
We next detail each of these contributions.

A sequence of error probabilities $\{\epsilon_n\}_{n = 1}^{\infty}$ is said to be a \emph{small-to-moderate deviations (SMD) sequence} if
\begin{align}
\lim_{n \to \infty} \frac{1}{n} \log \frac{1}{\epsilon_n} = \lim_{n \to \infty} \frac{1}{n} \log \frac{1}{1-\epsilon_n} = 0.  \label{eq:range}
\end{align}
Since $Q^{-1}(1 - \epsilon) = - Q^{-1}(\epsilon)$ and $\lim\limits_{\epsilon \to 0} \frac{Q^{-1}(\epsilon)}{\sqrt{2 \log \frac{1}{\epsilon}}} = 1$ \cite[Lemma~5.2]{csiszarbook}, the condition in \eqref{eq:range} is equivalent to $Q^{-1}(\epsilon_n) = o(\sqrt{n})$.
The family of SMD sequences includes all error probability sequences except for the LD sequences, which approach 0 or 1 exponentially quickly. It therefore extends the family of MD error probability sequences to include, for example, sequences that sub-exponentially approach 1, e.g., $\frac{1}{n^r}, \exp\{-n^s\}, 1 - \frac{1}{n^r}, 1 - \exp\{-n^s\}$ with $r > 0$ and $s \in (0, 1)$ and sequences in the CLT regime (where $\epsilon_n =  \epsilon \in (0, 1)$ is a constant independent of $n$).
We show in Theorems \ref{thm:mainAch}--\ref{thm:mainConv} in \secref{sec:nons}, below, that if the channel is nonsingular with positive dispersion and the error probability sequence $\{\epsilon_n\}$ is SMD \eqref{eq:range}, then $\zeta(n, \epsilon_n)$ in \eqref{eq:moderateexp} is bounded~as
\begin{align}
    \zeta(n, \epsilon_n) &\geq \frac{1}{2} \log n + \underline{\cs} \, Q^{-1}(\epsilon_n)^2 \notag \\
    &\quad + \bigo{\frac{Q^{-1}(\epsilon_n)^3}{\sqrt{n}}} + O(1)  \label{eq:achA}\\
     \zeta(n, \epsilon_n) &\leq \frac{1}{2} \log n + \overline{\cs}\, Q^{-1}(\epsilon_n)^2  \notag \\
     &\quad + \bigo{\frac{Q^{-1}(\epsilon_n)^3}{\sqrt{n}}} + O(1), \label{eq:convA}
\end{align}
where the constants $\underline{\cs}$ and $\overline{\cs}$ are the same ones as in \eqref{eq:achrho}--\eqref{eq:convrho}. The bounds \eqref{eq:achA}--\eqref{eq:convA} generalize \eqref{eq:achrho}--\eqref{eq:convrho} to non-constant error probabilities $\epsilon_n$ at the expense of not bounding the constant term; 
\eqref{eq:achA}--\eqref{eq:convA} do not require the assumption that the information density is non-lattice as in \cite{moulin2017log}. The non-Gaussianity $\zeta(n, 
\epsilon)$ gets arbitrarily close to $O(\sqrt{n})$ as the decay of $\epsilon_n$ approaches exponential decay, rivaling the dispersion term in \eqref{eq:Gaussianapp}. Thus, refining the third-order term as we do in \eqref{eq:achA}--\eqref{eq:convA} is especially significant in the MD regime. 
The achievability bound \eqref{eq:achA} arises from analyzing the RCU bound in \cite[Th.~16]{polyanskiy2010Channel}; the converse bound \eqref{eq:convA} uses the non-asymptotic converse bound in \cite[Prop.~6]{tomamichel2013converse} and the saddlepoint result in \cite[Lemma~14]{moulin2017log}. For $\epsilon_n$ in the MD regime (i.e., \eqref{eq:range} holds with either $\epsilon_n \to 0$ or $\epsilon_n \to 1$), neither the Berry-Esseen theorem used in \cite{polyanskiy2010Channel} nor the refined Edgeworth expansion used in \cite{moulin2017log} to treat the constant $\epsilon$ case is sharp enough for the precision in \eqref{eq:achA}--\eqref{eq:convA}. We replace these tools with the MD bounds found in \cite[Ch.~8]{petrov1975}.  In our analysis of the RCU bound, we also refine \cite[Lemma~47]{polyanskiy2010Channel}, which is commonly used in CLT regime approximations, giving Theorems~\ref{thm:chaganty}--\ref{thm:lattice}.

Since both the Edgeworth expansion and the LD result used in \cite{moulin2017log} take distinct forms for lattice and non-lattice random variables, the constant terms $\underline{B}$ and $\overline{B}$ in \eqref{eq:achrho}--\eqref{eq:convrho} depend on whether the information density $\imath(X; Y)$ is a lattice or  non-lattice random variable. In \cite{moulin2017log}, Moulin focuses primarily on channels with non-lattice information densities; the only example channel with a lattice information density that he studies  is the BSC, which he analyzes separately in \cite[Th.~7]{moulin2017log}. Our analysis shows that a single proof holds for lattice and non-lattice cases if we do not attempt to bound the $O(1)$ term. 


For Cover--Thomas-symmetric channels, $\underline{S} = \overline{S} = S$, and we refine \eqref{eq:achA}--\eqref{eq:convA} in \thmref{thm:refinedAchConv} in \secref{sec:refined} below by deriving the coefficient of the $\bigo{\frac{Q^{-1}(\epsilon_n)^3}{\sqrt{n}}}$ term. For the BSC and a wide range of $(n, \epsilon)$ pairs, our asymptotic approximation for the maximum achievable rate using terms up to the channel skewness, i.e., $\zeta(n, \epsilon) \approx \frac{1}{2} \log n + \cs \, Q^{-1}(\epsilon)^2$, is more accurate than both of Moulin's bounds from \eqref{eq:achrho} and \eqref{eq:convrho}; the accuracy of our approximation is similar to that of the saddlepoint approximations in \cite{segura2018, honda2018}, which are achievability bounds. Moreover, for the BSC with an $(n, \epsilon)$ pair satisfying $\epsilon \in [10^{-10}, 10^{-1}]$ and $n \in [100, 500]$, including the $\bigo{\frac{Q^{-1}(\epsilon_n)^3}{\sqrt{n}}}$ term from \thmref{thm:refinedAchConv} in our approximation yields a less accurate approximation than is obtained by stopping at the channel skewness (see \figref{fig:rates}). This highlights the importance of channel skewness relative to the higher-order terms in characterizing the channel.

\thmref{thm:Gaussian}, in \secref{sec:Gaussian}, below, derives lower and upper bounds on the non-Gaussianity of the Gaussian channel with a maximal-power constraint in the CLT regime. Our bounds yield the channel skewness term exactly. We derive these bounds by analyzing Shannon's random coding bound in \cite[eq.~19]{shannon1959Probability} and Vazquez-Vilar's meta-converse bound in \cite[Th.~3]{vazquez2021} in the CLT regime. The achievability bound uses a tight approximation to a quantile of the noncentral $t$-distribution, and the converse bound uses the asymptotic expansion of the minimum type-II error probability for a test between two Gaussian distributions. The prior techniques from \cite[Th.~54]{polyanskiy2010Channel} and \cite{tan2015Third} are not sharp enough to derive the channel skewness.

Using the MD results in \cite[Ch. 8]{petrov1975} and the strong LD results in \cite{chaganty}, in \thmref{thm:NP} in \secref{sec:NP}, below, we derive the asymptotics of BHT in the MD regime, characterizing the minimum achievable type-II error of a hypothesis test that chooses between two product distributions given that the type-I error is an SMD sequence \eqref{eq:range}. Our result refines \cite{sason2012moderate} to the third-order term. 


A summary of the literature on the asymptotic expansions in channel coding for both DMCs and the Gaussian channel for different error probability regimes appears in Table~\ref{tab}. The LD regime seeks expressions of the form
    $\epsilon^*(n, R) \approx \frac{K}{n^r} \exp\{- n E(R)\}$.
In Table~\ref{tab}, $E(R), r$, and $K$ are called the second-, third-, and fourth-order terms. 

\begin{table*}[htbp!] 
\caption{The summary of the literature and our contributions for the asymptotic expansions in channel coding}
\centering
\label{tab}
\begin{tabular}{ccccccc}
\textbf{Paper} & \textbf{Channel} & \textbf{Bound} & \textbf{Regime} & \textbf{Order of expansion} & \textbf{Requires latticeness} & \textbf{Skewness term} \\
\hline 
\cite{strassen, polyanskiy2010Channel, hayashi}    & DMC    &  Ach + Conv   & CLT     & 2      & \ding{55}     & \ding{55}    \\
\cite[Th.~53]{polyanskiy2010thesis} & Nonsingular DMC    & Ach  & CLT   & 3  & \ding{55}  & \ding{55}    \\
\cite{tomamichel2013converse} & Nonsingular DMC    &Conv  & CLT   & 3  & \ding{55}  & \ding{55}    \\
\cite{moulin2017log}    & Nonsingular DMC    & Ach + Conv  & CLT  &  4   & \ding{51}    & \ding{51}  \\
\cite{altug2014moderate, polyanskiy2010Moderate}  & DMC  & Ach + Conv & MD  & 2       & \ding{55}    & \ding{55}    \\
\cite{altug2014refinement}  & Singular + Nonsingular DMC & Ach & LD    & 3    & \ding{55}   & \ding{55}   \\
\cite{honda2018}    & Nonsingular DMC & Ach & LD   & 4   & \ding{55}   & \ding{55}                     \\
Theorems~\ref{thm:mainAch}--\ref{thm:mainConv} & Nonsingular DMC & Ach + Conv  & CLT + MD   & 3   & \ding{55}  & \ding{51}  \\
\cite{tan2015Third}  & Gaussian & Ach   & CLT  & 3    & -   & \ding{55}  \\
\cite[Th.~54]{polyanskiy2010Channel}  & Gaussian & Conv   & CLT  & 3    & -   & \ding{55}  \\
\cite{shannon1959Probability, erseghe2016}  & Gaussian &Ach + Conv & LD  & 4    & -    & \ding{55}   \\
\cite{vazquez2021}  & Gaussian &Conv & LD  & 4    & -    & \ding{55}   \\
Theorem~\ref{thm:Gaussian}   & Gaussian &Ach + Conv  & CLT  & 4  & -    & \ding{51}    
\end{tabular}
\end{table*}

The remainder of the paper is organized as follows. We define notation and give preliminaries needed to formally present our results in \secref{sec:notation}.  \secref{sec:main} presents and discusses the main results. Proofs of the main results appear in Sections~\ref{sec:proofs}--\ref{app:Gaussian}.

\section{Notation and Preliminaries} \label{sec:notation}
\subsection{Notation}
For any $k \in \mathbb{N}$, we let $[k] \triangleq \{1, \dots, k\}$. 
We denote random variables by capital letters (e.g., $X$) and realizations of random variables by lowercase letters (e.g., $x$). We use boldface letters (e.g., $\mathbf{x}$) to denote vectors,  calligraphic letters (e.g., $\mathcal{X}$) to denote alphabets and sets, and sans serif font (e.g., $\mathsf{A})$ to denote matrices. The $i$-th entry of a vector $\mb{x}$ is denoted by $x_i$, and the $(i, j)$-th entry of a matrix $\ms{A}$ is denoted by $\ms{A}_{i,j}$. The $n \times n$ identity matrix is denoted by $\ms{I}_n$. For a symmetric matrix $\ms{A}$ supported on $\mc{X} \times \mc{X}$ and a subset $\mc{B} \subseteq \mc{X}$, $\ms{A}_{\mc{B}}$ denotes the $|\mc{X}| \times |\mc{X}|$ matrix where $(\ms{A}_{\mc{B}})_{x, x'} = \ms{A}_{x, x'}$ for $x, x' \in \mc{B}$ and $(\ms{A}_{\mc{B}})_{x, x'} = 0$, otherwise. For a vector $\mb{x}$ supported on $\mc{X}$, $\mb{x}_{\mc{B}}$ is defined similarly. The row space of a matrix $\ms{A}$ is denoted by $\mathrm{row}(\ms{A})$.

The sets of real numbers and complex numbers are denoted by $\mathbb{R}$ and $\mathbb{C}$, respectively. All-zero and all-one vectors are denoted by $\bs{0}$ and $\bs{1}$, respectively. A vector inequality $\mb{x} \leq \mb{y}$ for $\mb{x}, \mb{y} \in \mathbb{R}^d$ is understood element-wise, i.e., $x_i \leq y_i$ for all $i \in [d]$. We denote the inner product $\sum_{i = 1}^d x_i y_i$ by $\langle \mb{x}, \mb{y} \rangle$. We use $\| \cdot \|_{\infty}$ and $\norm{\cdot}_2$ to denote the $\ell_{\infty}$ and $\ell_2$ norms, i.e., $\| \mb{x} \|_{\infty} \triangleq \max\limits_{i \in [d]} |x_i|$ and $\norm{\mb{x}}_2 \triangleq \sqrt{\langle \mb{x}, \mb{x} \rangle}$. The multivariate normal distribution with mean $\boldsymbol{\mu}$ and covariance matrix $\ms{\Sigma}$ is denoted by $\mc{N}(\bs{\mu}, \ms{\Sigma})$.

The set of all distributions on the channel input alphabet $\mc{X}$ (respectively the channel output alphabet $\mc{Y}$) is denoted by $\mc{P}$ (respectively $\mc{Q})$. The support of a vector $\mb{h} \in \mathbb{R}^{|\mc{X}|}$ is denoted by $\mr{supp}(\mb{h}) \triangleq \{x \in \mc{X} \colon h_x \neq 0\}$. We write $X \sim \PX$ to indicate that $X$ is distributed according to $\PX \in \mc{P}$. Given a distribution $\PX \in \mc{P}$ and a transition probability kernel $\W$ from $\mc{X}$ to $\mc{Y}$, we write $\PX \times \W$ to denote the joint distribution of $(X, Y)$, and $\PY$ to denote the marginal distribution of $Y$, i.e., $\PY(y) = \sum_{x \in \mc{X}} \PX(x) \W(y|x)$ for all $y \in \mc{Y}$. Given a transition probability kernel $\W$, the distribution of $Y$ given $X = x$ is denoted by $\Wx$. For an arbitrary vector $\mb{h}$ supported on $\mc{X}$, $\mb{h} \to \W \to \tilde{\mb{h}}$ denotes the relationship $\tilde{h}_y = \sum_{x \in \mc{X}} h_x P_{Y|X}(y|x)$ for all $y \in \mc{Y}$.

For a sequence $\mathbf{x} = (x_1, \dots, x_n)$, the empirical distribution (or type) of $\mathbf{x}$ is denoted by
\begin{align}
    \hat{P}_{\mathbf{x}}(x) = \frac{1}{n} \sum_{i = 1}^n 1\{ x_i = x \}, \quad \forall\, x \in \mc{X}.
\end{align}
The set of length-$n$ types is denoted by $\mc{P}_n = \{P_X \in \mc{P} \colon n P_X(x) \in \mathbb{Z} \,\,\, \forall \, x \in \mc{X}\}$.
A lattice random variable is a random variable taking values in $\{a + kd \colon k \in \mathbb{Z}\}$, where $d \in \mathbb{R}_+$ is the \emph{span} of the lattice. We say that a random vector $\mb{X} = (X_1, \dots, X_n)$ is non-lattice if each of $X_i$, $i \in [n]$ is non-lattice, and $X_i$ is lattice if each of $X_i$, $i \in [n]$ is lattice. The case where some of the coordinates of $\mb{X}$ are lattice and the rest of the coordinates are non-lattice is excluded in this paper. We measure information in nats, and logarithms and exponents have base~$e$.

As is standard, $f(n) = O(g(n))$ means $\limsup_{n \to \infty} \left \lvert \frac{f(n)}{g(n)} \right \rvert < \infty$, and $f(n) = o(g(n))$ 
means $\lim_{n \to \infty} \left \lvert \frac{f(n)}{g(n)}\right \rvert = 0$. We use $Q(\cdot)$ to represent the complementary Gaussian cumulative distribution function (CDF) $Q(x) \triangleq \frac{1}{\sqrt{2 \pi}} \int_{x}^{\infty} \exp\cB{-\frac{t^2}{2}} \mathrm{d}t$ and $Q^{-1}(\cdot)$ to represent its functional inverse.

The skewness of a random variable $X$ is denoted by
\begin{align}
    \ske(X) \triangleq \frac{\E{(X-\E{X})^3}}{\Var{X}^{3/2}}. \label{eq:skewness}
\end{align}
Recall that the residual in the Berry--Esseen theorem has the form $\frac{c}{\sqrt{n}} \frac{\E{|X-\E{X}|^3}}{\Var{X}^{3/2}}$ for a global constant $c > 0$ (e.g., \cite[Ch. XVI.5, Th. 1]{feller1971introduction}), which is quite similar to $\ske(X)$. The skewness $\ske(X)$ appears in the asymptotically tight residual known as the Edgeworth expansion \cite[p.~7]{esseen} (see also \cite[Ch.~XVI.4, Th. 1]{feller1971introduction})
\begin{align}
    &\Prob{\frac{1}{\sqrt{n \Var{X_1}}} \sum_{i = 1}^n (X_i - \E{X_1}) \leq x} \notag \\
    &\quad = Q(-x) + \frac{\ske(X_1)}{6 \sqrt{n}}(1-x^2) \phi(x) + o\left(\frac{1}{\sqrt{n}}\right), \label{eq:Edge}
\end{align}
where the random variables $X_i$, $i \in [n]$, are independent and identically distributed (i.i.d.) and non-lattice, and $\phi(\cdot)$ is the standard Gaussian density. 
The skewness of a random variable plays a critical role in our expansions of the maximum achievable channel coding rate. 

\subsection{Definitions Related to Information Density}
The relative entropy between distributions $P$ and $Q$ on a common alphabet, the second and third central moments of the log-likelihood ratio, and the skewness of the log-likelihood ratio are denoted by 
\begin{align}
    D(P \|Q) &\triangleq \E{\log \frac{P(X)}{Q(X)}} \\
    V(P \|Q) &\triangleq  \Var{\log \frac{P(X)}{Q(X)}} \\
    T(P \|Q) &\triangleq  \E{\left(\log \frac{P(X)}{Q(X)} - D(P\|Q)\right)^3} \\
    S(P \|Q) &\triangleq  \frac{T(P \|Q)}{V(P \| Q)^{3/2}},
\end{align}
where $X \sim P$. Let $\PX \in \mc{P}$ and $\QY \in \mc{Q}$, and let $\W$ be a transition probability kernel from $\mc{X}$ to $\mc{Y}$.
The conditional versions of the above quantities are denoted by
\begin{align}
    D(\W \| \QY|\PX) &\triangleq \sum_{x \in \mathcal{X}} \PX(x) D(\Wx \| \QY)\\
    V(\W \| \QY|\PX) &\triangleq \sum_{x \in \mathcal{X}} \PX(x) V(\Wx \| \QY) \\
    T(\W \| \QY|\PX) &\triangleq \sum_{x \in \mathcal{X}} \PX(x) T(\Wx \| \QY) \\
    \ske(\W \| \QY|\PX) &\triangleq \frac{T(\W \| \QY|\PX)}{V(\W \| \QY|\PX)^{3/2}}.
\end{align}
Let $(X, Y) \sim \PX \times \W$. The information density is defined as
\begin{align}
    \imath(x; y) &\triangleq \log \frac{\W(y|x)}{\PY(y)}, \quad \forall \, x \in \mc{X}, y \in \mc{Y}.
\end{align}
In the remainder of the paper, we assume that the channel $\W$ is clear from the context and is fixed, and we eliminate it from the input arguments of quantities such as mutual information and dispersion. We define the following moments of the random variable $\imath(X; Y)$. 
\begin{itemize}[leftmargin=*]
\item The mutual information
\begin{align}
    I(\PX) \triangleq \E{\imath(X; Y)} = D(\W \| \PY | \PX), 
\end{align}
\item the unconditional information variance
\begin{IEEEeqnarray}{rCl}
    V_{\mathrm{u}}(\PX) &\triangleq& V(\PX \times \W \| \PX \times \PY) \notag \\
    &=& \Var{\imath(X; Y)}, \IEEEeqnarraynumspace
\end{IEEEeqnarray}
\item the unconditional information third central moment
\begin{align}
    T_{\mathrm{u}}(\PX) &\triangleq T(\PX \times \W \| \PX \times \PY) \\
     &= \E{(\imath(X; Y) - I(\PX))^3},
\end{align}
\item the unconditional information skewness
\begin{align}
    \ske_{\mathrm{u}}(\PX) &\triangleq \ske(\imath(X; Y)) = \frac{T_{\mathrm{u}}(\PX)}{V_{\mathrm{u}}(\PX)^{3/2}},
\end{align}
\item the conditional information variance
\begin{IEEEeqnarray}{rCl}
    V(\PX) &\triangleq& V(\W \| \PY | \PX) = \E{\Var{\imath(X; Y) | X}}, \IEEEeqnarraynumspace
\end{IEEEeqnarray}
\item the conditional information third central moment
\begin{IEEEeqnarray}{rCl}
    T(\PX) \triangleq T(\W \| \PY | \PX), \IEEEeqnarraynumspace
\end{IEEEeqnarray}
\item the conditional information skewness
\begin{IEEEeqnarray}{rCl}
    \ske(\PX) \triangleq \frac{T(\W \| \PY | \PX )}{V(\W \| \PY | \PX)^{3/2}}, \IEEEeqnarraynumspace
\end{IEEEeqnarray}
\item the reverse dispersion \cite[Sec. 3.4.5]{polyanskiy2010thesis}
\begin{align}
    V_{\mr{r}}(\PX) &\triangleq \E{\Var{\imath(X; Y) | Y}}.
\end{align}
\end{itemize}


\subsection{Discrete Memoryless Channel}
A DMC is characterized by a finite input alphabet $\mathcal{X}$, a finite output alphabet $\mc{Y}$, and a transition probability kernel $\W$, where $\W(y|x)$ is the probability that the output of the channel is $y \in \mathcal{Y}$ given that the input to the channel is $x \in \mc{X}$. The $n$-letter input-output relation of a DMC is 
\begin{align}
    \W^n(\mathbf{y} | \mathbf{x}) = \prod_{i = 1}^n \W(y_i | x_i).
\end{align}

We proceed to define the channel code. 
\begin{definition}
An $(n, M, \epsilon)$-code for a DMC $\W$ comprises an encoding function
\begin{align}
    \ms{f} \colon [M] \to \mathcal{X}^n,
\end{align}
and a decoding function 
\begin{align}
    \ms{g} \colon \mc{Y}^n \to [M],
\end{align}
that satisfy an average error probability constraint
\begin{align}
    1 - \frac{1}{M} \sum_{m = 1}^M \W^n(\ms{g}^{-1}(m) | \ms{f}(m)) \leq \epsilon. \label{eq:averageerror}
\end{align}
\end{definition}
The maximum achievable message set size $M^*(n, \epsilon)$ under the average error probability criterion is defined as
\begin{align}
    M^*(n, \epsilon) \triangleq \max\{M \colon \exists \, \text{an } (n, M, \epsilon)\text{-code}\}. \label{eq:Mstardef}
\end{align}

\subsection{Definitions Related to the Optimal Input Distribution}
The capacity of a DMC $\W$ is 
\begin{align}
    C \triangleq \max_{\PX \in \mc{P}} I(\PX).
\end{align}
We denote the set of capacity-achieving input distributions by
\begin{align}
    \mc{P}^\dagger \triangleq \{\PX \in \mc{P} \colon I(\PX) = C\}.
\end{align}
Even if there are multiple capacity-achieving input distributions ($|\mc{P}^\dagger| > 1$), the capacity-achieving output distribution is unique ($\PX, \PX' \in \mc{P}^\dagger$ implies $\sum_{x \in \mc{X}} \PX(x) \W(y|x) = \sum_{x \in \mc{X}} \PX'(x) \W(y|x)$ for all $y \in \mc{Y}$) \cite[Cor.~2 to Th.~4.5.2]{gallager1968book}. We denote the unique capacity-achieving output distribution by $\QYs \in \mc{Q}$; $\QYs$ satisfies $\QYs(y) > 0$ for all $y \in \mathcal{Y}$ for which there exists an $x \in \mc{X}$ with $\W(y|x) > 0$  \cite[Cor.~1 to Th.~4.5.2]{gallager1968book}.
For any $\PXdag \in \mc{P}^\dagger$, it holds that $V(\PXdag) = V_{\mathrm{u}}(\PXdag)$ \cite[Lemma~62]{polyanskiy2010Channel}. 

Define 
\begin{align}
V_{\min} &\triangleq \min_{\PXdag \in \mc{P}^\dagger} V(\PXdag) \\
V_{\max} &\triangleq \max_{\PXdag \in \mc{P}^\dagger} V(\PXdag).
\end{align}
The $\epsilon$-dispersion \cite{polyanskiy2010Channel} of a channel is defined as
\begin{align}
    V_{\epsilon} \triangleq \begin{cases} V_{\min} &\text{if } \epsilon < \frac 1 2 \\ V_{\max} &\text{if } \epsilon \geq \frac 1 2.\end{cases} \label{eq:Vminmax}
\end{align}
The set of dispersion-achieving input distributions is defined~as
\begin{align}
    \mc{P}^* \triangleq  \left\{\PXdag \in \mc{P}^\dagger \colon V(\PXdag) = V_{\epsilon} \right\}. 
\end{align}
Any $\PXdag \in \mc{P}^\dagger$ satisfies $D(\Wx \| \QYs) = C$ for each $x \in \mathcal{X}$ with $\PXdag(x) > 0$, and $D(\Wx \| \QYs) \leq C$ for all $x \in \mathcal{X}$ \cite[Th.~4.5.1]{gallager1968book}. Hence, the support of any capacity-achieving input distribution is a~subset~of
\begin{align}
    \mc{X}^\dagger = \{x \in \mc{X} \colon D(\Wx \| \QYs) = C\}. 
\end{align}
The support of any dispersion-achieving input distribution is a subset of
\begin{align}
    \mc{X}^* \triangleq \bigcup_{\PXs \in \mc{P}^*} \mathrm{supp}(\PXs) \subseteq \mc{X}^\dagger.
\end{align}

While analyzing the set $\mc{P}^*$ is sufficient to derive the second-order term in \eqref{eq:Gaussianapp} for $\log M^*(n, \epsilon_n)$ with an SMD sequence $\epsilon_n$, further quantities are needed to describe the optimal third-order term. 
The quantities below are used to describe the input distribution that achieves our lower bound $\underline{S}$ on the channel skewness $S$ in \eqref{eq:skewnessdef}; they also appear in \cite{moulin2017log}. Given a fixed DMC $P_{Y|X}$, the gradient and the Hessian of the mutual information $I(\PX)$ evaluated at $\PX$ are given by~\cite[eq.~(2.28)-(2.29)]{moulin2017log}
\begin{IEEEeqnarray}{rCl}
    \nabla I(\PX)_x &=& D(\Wx \| \PY) - 1 \label{eq:derivI} \\
    \nabla^2 I(\PX)_{x, x'} &=& - \sum_{y \in \mc{Y}} \frac{\W(y|x) \W(y|x')}{\PY(y)} \label{eq:hessI}\IEEEeqnarraynumspace
\end{IEEEeqnarray}
for $(x, x') \in \mathcal{X}^2$. The matrix $-\nabla^2 I(\PXdag)$ is the same for all $\PXdag \in \mc{P}^\dagger$ and is positive semidefinite. See \cite[Sec. II-D and II-E]{moulin2017log} for other properties of $-\nabla^2 I(\PXdag)$. 
We define
\begin{align}
    \ms{J} \triangleq -\nabla^2 I(\PXdag)_{x, x'}.\label{eq:J}
\end{align}
The matrices $\ms{J}_{\mc{X}^\dagger}$ and $\ms{J}_{\mc{X}^*}$ play an important role in our results and their proofs.



The following notation is used in our results in \secref{sec:nons}.
\begin{align}
    \mb{v}{(\PX)} &\triangleq \nabla V(\PX) \label{eq:vP}\\ 
    \overline{\mb{v}}{(\PX)}_x &\triangleq \sum_{x' \in \mc{X}} \PX(x') \frac{\partial V(P_{Y|X = x'} \| \PY)}{\partial \PX(x)}  \label{eq:barvP}
    \end{align}
    for $x \in \mc{X}$, and
    \begin{align}
    A_0(\PX) &\triangleq \frac{1}{8V_\epsilon} {\mb{v}{(\PX)}}^\top \tilde{\ms{J}} \mb{v}{(\PX)}, \label{eq:A0} \\
    A_1(\PX) &\triangleq  \frac{1}{8V_\epsilon} {\mb{\overline{v}}{(\PX)}}^{\top} \tilde{\ms{J}} \overline{\mb{v}}{(\PX)}, \label{eq:A1}
\end{align}
where
\begin{align}
    \tilde{\ms{J}} \triangleq \ms{J}_{\mc{X}^*}^{+} - \frac{1}{\mb{1}^\top \ms{J}_{\mc{X}^*}^+ \mb{1}} (\ms{J}_{\mc{X}^*}^{+} \mb{1})(\ms{J}_{\mc{X}^*}^{+} \mb{1})^\top, \label{eq:tildeJ}
\end{align}
and $\ms{J}_{\mc{X}^*}^+$ denotes the Moore-Penrose pseudo-inverse\footnote{Given that $\ms A = \ms U \ms{\Sigma} \ms{ V}^\top$ is the singular value decomposition of $\ms A$, $\ms A^+ \triangleq  \ms V \ms{\Sigma}^{-1} \ms{U}^\top$. The expression in \eqref{eq:tildeJ} is the compact version of \cite[Lemma~1 (iv)-(v)]{moulin2017log}.} of $\ms{J}_{\mc{X}^*}$.
One important property of $A_0(P_X)$ and $A_1(P_X)$ is that for Cover--Thomas-symmetric channels, $A_0(P_X) = A_1(P_X) = 0$ under the equiprobable input distribution, which remains optimal in terms of skewness.
See \cite[Lemma~2]{moulin2017log} for more properties of these quantities.

\subsection{Singularity of a DMC}\label{sec:singular}
The following definition divides  DMCs into two groups for which the non-Gaussianity behaves differently.
    An input distribution-channel pair $(\PX, \W)$ is \emph{singular} \cite[Def. 1]{altug2014refinement} if for all $(x, \overline{x}, y) \in \mc{X} \times \mc{X} \times \mc{Y}$ such that $\PX \times \W(x, y) > 0$ and $\PX \times \W(\overline{x}, y) > 0$, it holds that
    \begin{align}
        \W(y|x) = \W(y|\overline{x}).
    \end{align}
    We define the singularity parameter \cite[eq. (2.25)]{moulin2017log} 
    \begin{align}
        \eta(\PX) \triangleq 1 - \frac{V_{\mr{r}}(\PX)}{V_{\mathrm{u}}(\PX)}, \label{eq:eta}
    \end{align}
    which is a constant in $[0, 1]$. The pair $(\PX, \W)$ is singular if and only if $\eta(\PX) = 1$ \cite[Remark 1]{altug2014singular}. A channel $\W$ is singular if and only if $\eta(\PXs) = 1$ for all $\PXs \in \mc{P}^*$; it is nonsingular otherwise. An example of a singular channel is the BEC. Our focus in this paper is on nonsingular channels.

\section{Main Results} \label{sec:main}
Our first result describes the lower and upper bounds on the non-Gaussianity of nonsingular channels in the SMD regime, refining the expansion in \eqref{eq:moderateexp}. For symmetric channels, we further refine these bounds up to the $\bigo{\frac{Q^{-1}(\epsilon)^3}{\sqrt{n}}}$ term. We then derive tight lower and upper bounds for the non-Gaussianity of the Gaussian channel with a maximal-power constraint in the CLT regime, giving the exact expression for the channel skewness for that channel.
Our last result is a fourth-order asymptotic expansion (i.e., an expansion up to the $\bigo{\frac{Q^{-1}(\epsilon)^3}{\sqrt{n}}}$ term) for the logarithm of the minimum achievable type-II error probability for binary hypothesis tests between two product distributions in the SMD regime.
\subsection{Nonsingular Channels} \label{sec:nons}
Theorem~\ref{thm:mainAch} is our achievability result.
\begin{theorem}\label{thm:mainAch}
Suppose that $\epsilon_n$ is an SMD sequence \eqref{eq:range} and that
$\W$ is a nonsingular DMC with $V_{\min} > 0$. It holds that
\begin{align}
    \zeta(n, \epsilon_n) &\geq \frac{1}{2} \log n + \underline{S} Q^{-1}(\epsilon_n)^2 
     + \bigo{\frac{Q^{-1}(\epsilon_n)^3}{\sqrt{n}}} + O(1) \label{eq:achievability},
\end{align}
where
\begin{align}
    \underline{S} &\triangleq \max_{\PXs \in \mc{P}^*} \bigg( \frac{\ske_{\mathrm{u}}(\PXs) \sqrt{V_{\epsilon_n}}}{6}  + A_0(\PXs) + \frac{1 - \eta(\PXs)}{2(1+\eta(\PXs))}  \bigg). \label{eq:Slow}
\end{align}
\end{theorem}
\begin{IEEEproof}
The proof consists of two parts and extends the argument in \cite{moulin2017log} to allow sequences $\{\epsilon_n\}$ that decrease 0 or increase to 1 as permitted by \eqref{eq:range}. The first part analyzes a particular relaxation \cite[Th.~53]{polyanskiy2010thesis} of the RCU bound \cite[Th.~16]{polyanskiy2010Channel} for an arbitrary distribution $\PX \in \mc{P}$. This approach is used in the CLT regime for a third-order analysis in \cite[Th.~53]{polyanskiy2010thesis} and a fourth-order analysis in \cite{moulin2017log}; a slightly different relaxation of the RCU bound comes up in the LD regime \cite{altug2014refinement}. To bound the probability $\Prob{\imath(\mb{X}; \mb{Y}) \leq \tau}$, we replace the Edgeworth expansion in \cite[eq. (5.30)]{moulin2017log}, which gives the refined asymptotics of the Berry-Esseen theorem, with its MD version from \cite[Ch. 8, Th. 2]{petrov1975}. Note that the Edgeworth expansion yields an additive remainder term $\bigo{\frac{1}{\sqrt{n}}}$ to the Gaussian term. This remainder is too large for $\epsilon_n \leq \frac{1}{\sqrt{n}}$ in \eqref{eq:range} since it would dominate the Gaussian term in the Edgeworth expansion. Therefore, an MD result that yields a multiplicative remainder term $(1 + o(1))$ is desired. We apply the LD result from \cite[Th. 3.4]{chaganty} to bound the probability $\Prob{\imath(\mb{\overline{X}}; \mb{Y})  \geq \imath(\mb{X}; \mb{Y}) \geq \tau}$ that appears in the relaxed RCU bound, where $\mb{X}$ denotes the transmitted random codeword and  $\overline{\mb{X}}$ denotes an independent codeword drawn from the same distribution. This bound replaces the bounds in \cite[eq. (7.25)-(7.27)]{moulin2017log} and refines the LD bound \cite[Lemma~47]{polyanskiy2010Channel} used in \cite[Th.~53]{polyanskiy2010thesis}. We show an achievability result as a function of $I(\PX)$, $V_{\mathrm{u}}(\PX)$, and $\ske_{\mathrm{u}}(\PX)$. If $\PX = \PXs \in \mc{P}^*$, the resulting bound is \eqref{eq:achievability} with $A_0(\PXs)$ replaced by zero. We then optimize the bound over $\PX$ using the second-, first- and zeroth-order Taylor series expansions respectively of $I(\PX), V_{\mathrm{u}}(\PX)$, and $\ske_{\mathrm{u}}(\PX)$ around $\PXs \in \mc{P}^*$. Interestingly, the right-hand side of \eqref{eq:achievability} is achieved using
\begin{align}
    \PX &= \PXs + \mb{h}^* \in \mc{P}, \label{eq:thirdorderP}
\end{align}
instead of a dispersion-achieving input distribution $\PXs \in \mc{P}^*$ to generate i.i.d. random codewords; here
\begin{align}
     \mb{g} &= - \frac{Q^{-1}(\epsilon_n)}{2 \sqrt{n V_{\epsilon_n}}} \mb{v}(\PXs) \\
    \mb{h}^* &= \tilde{\ms{J}} \mb{g}. \label{eq:hstar}
\end{align}
 
Note that despite being in the neighborhood of a dispersion-achieving $\PXs$, our choice of $\PX$ in \eqref{eq:thirdorderP} might not belong to $\mc{P}^*$. This behavior is not seen between the first- and second-order optimal input distributions since every dispersion-achieving distribution is also capacity-achieving.

See \secref{sec:proofach} for the details of the proof.
\end{IEEEproof}

The input distribution in \eqref{eq:thirdorderP} is chosen by setting $P_X = P_X^* + \mb{h}$ for a value of $\mb{h}$ for which $\PXs + \mb{h} \in \mc{P}$ and $\mb{h} \to \bs{0}$ as $n \to \infty$; we then optimize the direction and the scaling of $\mb{h}$ with respect to the RCU bound. Intuitively, the above strategy is useful since it acknowledges both that the input distribution may vary with $n$ and that it cannot stray too far from the choice that optimizes the first- and second-order achievable rate. The optimal deviation $\mb{h}^*$ from the dispersion-achieving distribution is solved by the optimization problem
\begin{align}
    \sup_{\substack{\mb{h}\colon \mr{supp}(\mb{h}) \subseteq \mc{X}^{\dagger} \\ \mb{h}^\top \bs{1} = 0, \,\, \mb{h}_{\mc{X}^\dagger 
    \setminus \mc{X}^*} \geq \bs{0}}} \left(\mb{g}^\top \mb{h} - \frac{1}{2} \mb{h}^\top \ms{J}_{\mc{X}^\dagger} \mb{h}\right). \label{eq:opt2}
\end{align}
The optimization in \eqref{eq:opt2} is convex but does not have a closed-form solution in general. Following \cite[Appendix~B]{moulin2017log}, we get an optimization problem with a closed-form solution by 
further restricting the support of $\mb{h}$ as $\mr{supp}(\mb{h}) \subseteq \mc{X}^*$. This reduces \eqref{eq:opt2} to 
\begin{align}
      \sup_{\substack{\mb{h}\colon \mr{supp}(\mb{h}) \subseteq \mc{X}^{*} \\ \mb{h}^\top \bs{1} = 0, \,\,\, \mb{h} \in \mr{row}(\ms{J}_{\mc{X}^*})}} \left(\mb{g}^\top \mb{h} - \frac{1}{2} \mb{h}^\top \ms{J}_{\mc{X}^*} \mb{h}\right).\label{eq:opt}
\end{align}
In~\cite[Lemma~1]{moulin2017log}, Moulin shows that \eqref{eq:hstar} is the unique
$\mb{h}^*$ that achieves \eqref{eq:opt},
and $\frac{1}{2} \mb{g}^\top \tilde{\ms{J}} \mb{g} = A_0(\PXs) Q^{-1}(\epsilon_n)^2$ is the optimal value of the quadratic form in \eqref{eq:opt}.  
In \cite[Appendix~B (ii)]{moulin2017log}, Moulin shows that if $\mc{X}^\dagger = \mc{X}^*$, the values of the objectives in \eqref{eq:opt2} and \eqref{eq:opt} are equal, implying that restricting $\mr{supp}(\mb{h})$ to $\mc{X}^*$ does not yield a penalty in the achievable skewness term. 


In the second-order MD result in \cite{altug2014moderate}, Altu\u{g} and Wagner apply the non-asymptotic bound in \cite[Cor.~2 on p. 140]{gallager1968book}, which turns out to be insufficiently sharp for the derivation of the third-order term. 

Recall from \eqref{eq:Vminmax} that $V_{\epsilon_n}$ can be either $V_{\min}$ or $V_{\max}$. We require the condition $V_{\min} > 0$ in \thmref{thm:mainAch}, which implies that $V_{\epsilon_n} > 0$ for all $\epsilon_n$ sequences, since the MD (\thmref{thm:moderatePetrov} in \secref{sec:moderate}) and LD (Theorems \ref{thm:chaganty} and \ref{thm:lattice} in \secref{sec:large}) results apply only to random variables with positive variances. In the CLT regime, \cite[Th.~45 and 48]{polyanskiy2010Channel} and \cite[Prop.~9-10]{tomamichel2013converse} derive bounds on the non-Gaussianity of DMCs with $V_{\epsilon_n} = 0$. If $V_{\epsilon_n} = 0$, the scaling of the non-Gaussianity changes depending on whether or not the DMC is exotic \cite[p. 2331]{polyanskiy2010Channel} (most DMCs do not satisfy the definition of an exotic DMC), and whether $\epsilon_n$ is less than, equal to, or greater than $\frac{1}{2}$. A summary of the non-Gaussianity terms under these cases appears in \cite[Fig.~1]{tomamichel2013converse}.


Theorem~\ref{thm:mainConv} is our converse result.

\begin{theorem}\label{thm:mainConv}
Under the conditions of \thmref{thm:mainAch}, 
\begin{align}
    &\zeta(n, \epsilon_n) \leq \frac{1}{2} \log n +  \overline{S} Q^{-1}(\epsilon_n)^2 
     +  \bigo{\frac{Q^{-1}(\epsilon_n)^3}{\sqrt{n}}} +  O(1), \label{eq:converse}
\end{align}
where
\begin{align}
     &\overline{S} \triangleq \max_{\PXs \in \mc{P}^*} \bigg( \frac{\ske_{\mathrm{u}}(\PXs) \sqrt{V_{\epsilon_n}}}{6} + \frac{1}{2} + A_0(\PXs) - A_1(\PXs) \bigg). \label{eq:overS}
\end{align}
\end{theorem}
\begin{IEEEproof}
The proof of \thmref{thm:mainConv} combines the converse bound from \cite[Prop.~6]{tomamichel2013converse}, which relaxes the meta-converse bound \cite[Th.~27]{polyanskiy2010Channel}, with a saddlepoint result from \cite[Lemma~14]{moulin2017log}, which gives the saddlepoint solution to a quadratic form that arises after taking the Taylor series expansion of the main quantity in \cite[Prop.~6]{tomamichel2013converse}. Combining these results and not deriving the $O(1)$ term in \eqref{eq:converse} yields a much simpler proof than that in \cite{moulin2017log}. While \cite[proof of Th. 4]{moulin2017log} relies on the asymptotic expansion of the $\beta_{1-\epsilon}$ function,
the use of \cite[Prop.~6]{tomamichel2013converse} allows us to bypass this part. In the application of \cite[Prop.~6]{tomamichel2013converse}, similar to \cite[eq.~(6)]{tomamichel2013converse}, we use an auxiliary $n$-letter output distribution that is a convex combination of product distributions; see equation \eqref{eq:convexQY}, in \secref{sec:proofconv}, below, for details. 

The main difference between our proof technique in \secref{sec:proofconv} below and that in \cite{tomamichel2013converse} is that we set the first term in \eqref{eq:convexQY} as $(\QYs + \tilde{\mb{h}})^n \in \mc{Q}^n$, where $\QYs$ is the unique capacity-achieving output distribution, and $\tilde{\mb{h}}$ satisfies $\norm{\tilde{\mb{h}}}_{\infty} = \bigo{\frac{\Qinv}{\sqrt{n}}}$. The direction of $\tilde{\mb{h}}$ is found by solving a single-letter minimax problem involving the quantities $D(\W \| \QY | \PX)$ and $V(\W \| \QY | \PX)$, where the maximization is over $\PX \in \mc{P}$ and the minimization is over $\QY \in \mc{Q}$; here $\PX$ is assumed to be close to $\PXs$, and $\QY$ is assumed to be close to $\QYs$. See \secref{sec:proofconv} for details.
\end{IEEEproof}

      \subsection{Refined Results for Symmetric Channels} \label{sec:refined}

     If the channel satisfies $|\mc{P}^*| = 1$, $A_0(\PXs) = A_1(\PXs) = 0$, and $\eta(\PXs) = 0$, then our  achievability \eqref{eq:achievability} and converse \eqref{eq:converse} bounds yield the channel skewness \eqref{eq:skewnessdef}
    \begin{align}
        S = \frac{\ske_{\mathrm{u}}(\PXs) \sqrt{V_{\min}}}{6} + \frac{1}{2}. \label{eq:skewct}
    \end{align}
    Cover--Thomas-symmetric channels \cite[p. 190]{cover}
    satisfy all three conditions;\footnote{Channels that (i) are Cover--Thomas weakly symmetric, (ii) have  $|\mc{X}| = |\mc{Y}|$ and (iii) have a positive definite $\ms{J}$ satisfy the same three conditions \cite[Prop.~6]{moulin2017log}.}
    the BSC is an example of a Cover--Thomas symmetric channel. 
    
    \thmref{thm:refinedAchConv} below, refines the achievability and converse results in Theorems \ref{thm:mainAch}--\ref{thm:mainConv} for Cover--Thomas-symmetric channels. 
    \begin{theorem}\label{thm:refinedAchConv}
    Let $\W$ be a Cover--Thomas-symmetric channel with $\epsilon$-dispersion \eqref{eq:Vminmax} $V > 0$. If $\{\epsilon_n\}_{n = 1}^{\infty}$ is an SMD sequence \eqref{eq:range}, then
    \begin{IEEEeqnarray}{rCl}
       \IEEEeqnarraymulticol{3}{l}{\zeta(n, \epsilon_n)} \notag \\
       &=& \frac{1}{2} \log n +  S Q^{-1}(\epsilon_n)^2  -  \frac{3 (\mu_4 - 3 V^2) V - 4 \mu_3^2}{72 V^{5/2}} \frac{Q^{-1}(\epsilon_n)^3}{\sqrt{n}} \notag \\
       &&+ \bigo{\frac{Q^{-1}(\epsilon_n)^4}{n}} + \bigo{1}, \label{eq:refinedach}
    \end{IEEEeqnarray}
    where $S$ is the skewness \eqref{eq:skewct} under the uniform input distribution $\PXs$, and $\mu_k = \E{(\imath(X; Y) - C)^k}$ is the $k$-th central moment of the information density under $X \sim \PXs$.

    Further, if $\epsilon_n$ satisfies $Q^{-1}(\epsilon_n) = O(n^{1/6})$, which is equivalent to $\lim \limits_{n \to \infty} -\frac{1}{n^{1/3}} \log \frac{1}{\min\{\epsilon_n, 1- \epsilon_n\}} > 0$ (e.g., \cite[Lemma~5.2]{csiszarbook}),
    then the $\bigo{\frac{Q^{-1}(\epsilon_n)^3}{\sqrt{n}}}$ term is dominated by the $O(1)$ term, giving
    \begin{align}
        \zeta(n, \epsilon_n) = \frac{1}{2} \log n + \cs \, Q^{-1}(\epsilon_n)^2 + O(1).
    \end{align}
    \end{theorem}
    \begin{IEEEproof}
    See  \appref{app:refined}. 
    \end{IEEEproof}
    
\begin{figure*}
        \center
        \includegraphics[width=1\linewidth]{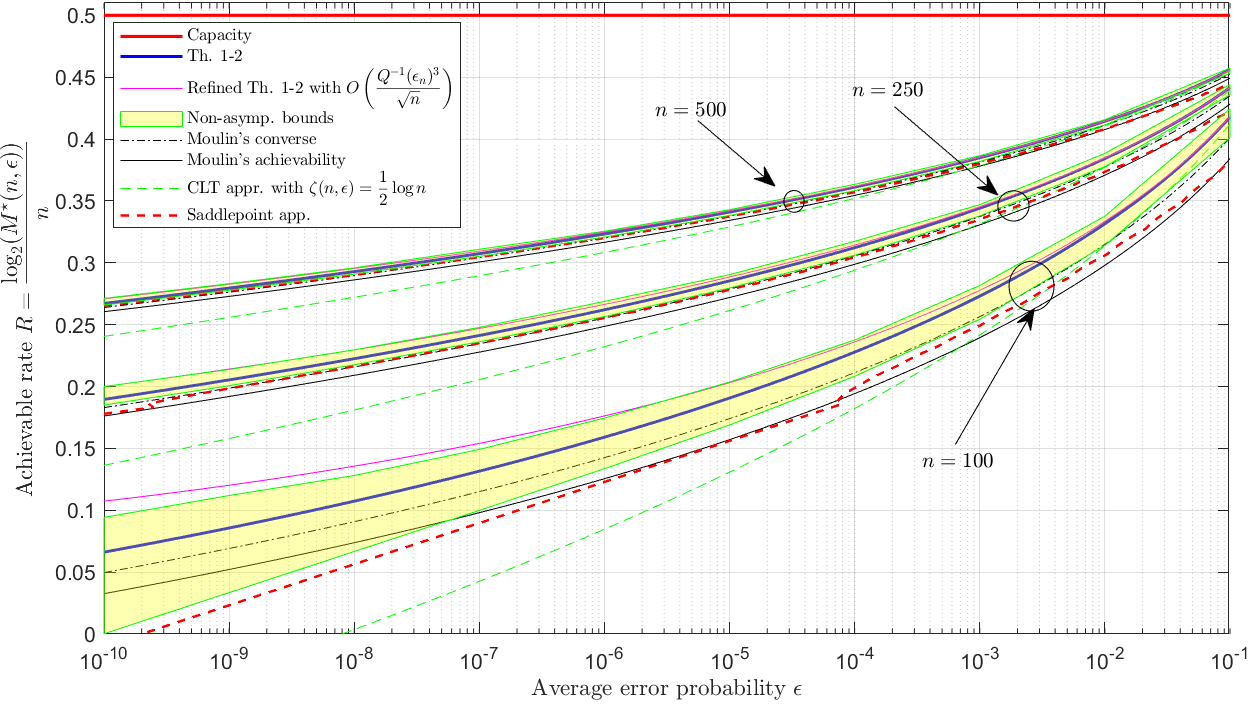}
        \caption{Achievable rate vs. average error probability for BSC(0.11): The expansions from Theorems~\ref{thm:mainAch}--\ref{thm:refinedAchConv}, excluding the $O(\cdot)$ terms, are shown for the BSC(0.11) with $\epsilon \in [10^{-10}, 10^{-1}]$ and $n = \{100, 250, 500\}$. The upper and lower boundaries of the shaded region correspond to the non-asymptotic bounds in \cite[Th. 33 and 35]{polyanskiy2010Channel}; the CLT approximation that takes $\zeta(n, \epsilon) = \frac{1}{2} \log n$ is from \cite[Th. 53]{polyanskiy2010thesis}; Moulin's results are \eqref{eq:achrho}--\eqref{eq:convrho}; the saddlepoint approximation is an achievability bound and is from \cite[Th. 1]{honda2018} and \cite[Sec. III-D]{segura2018}.}
        \label{fig:rates}
        \end{figure*}

  For the BSC with crossover probability 0.11, 
    \figref{fig:rates} 
    compares asymptotic expansions for the maximum achievable rate, $\frac{\log_2 M^*(n, \epsilon_n)}{n}$, dropping $o(\cdot)$ and $O(\cdot)$ terms except where noted otherwise. The curves plotted in \figref{fig:rates} include  Theorems~\ref{thm:mainAch} and \ref{thm:mainConv} both with and without the leading term of $\bigo{\frac{Q^{-1}(\epsilon_n)^3}{\sqrt{n}}}$ computed, various other asymptotic expansions in the CLT and LD regimes, and the non-asymptotic bounds from \cite[Th. 33 and 35]{polyanskiy2010Channel}.
    The leading term of $\bigo{\frac{Q^{-1}(\epsilon_n)^3}{\sqrt{n}}}$ in Theorems \ref{thm:mainAch} and \ref{thm:mainConv} is given in \thmref{thm:refinedAchConv}, below. Among all of these asymptotic expansions, Theorems~\ref{thm:mainAch} and \ref{thm:mainConv} ignoring the $O(\cdot)$ terms are the closest to the non-asymptotic bounds for most $(n, \epsilon)$ pairs shown, which highlights the significance of the channel skewness in obtaining accurate approximations to the finite blocklength coding rate in the medium $n$, small $\epsilon$ regime. Since Moulin's fourth-order CLT approximation requires the information density to be non-lattice, and the BSC has a lattice information density, Moulin uses a different approach to bound the $O(1)$ term for the BSC. Note that including the $\bigo{\frac{Q^{-1}(\epsilon_n)^3}{\sqrt{n}}}$ term from \thmref{thm:refinedAchConv} does not improve the accuracy because the blocklength $n \in [100, 500]$ chosen in the example is too small, which makes the $\bigo{\frac{Q^{-1}(\epsilon_n)^3}{\sqrt{n}}}$ term comparable to the skewness term.

  \addtolength{\textfloatsep}{-0in}

            In  \cite{altug2021onexact}, Altu\u{g} and Wagner show that in the LD regime, the prefactors in the lower and upper bounds on the exponentially decaying error probability for  Gallager-symmetric channels have the same order; that order depends on whether the channel is singular or nonsingular. Extending the analysis in \cite[Sec. III-C-2)]{moulin2017log} to any Gallager-symmetric channel shows that Gallager-symmetric channels satisfy $A_0(\PXs) = A_1(\PXs) = 0$. Note that $\eta(\PXs)$ is not necessarily zero (see  \cite[Sec. III-C-2)]{moulin2017log} for a counterexample), which means that for some Gallager-symmetric channels, \eqref{eq:achievability} and \eqref{eq:converse} are not tight up to the $O(1)$ term. The findings in \cite{altug2021onexact} suggest that \thmref{thm:mainAch} or \thmref{thm:mainConv} or both could be improved for some channels. The achievability bounds in \cite{altug2014refinement, altug2021onexact} bound the error probability from above as
    \begin{align}
    \epsilon \leq \Prob{\mc{D}} + (M-1) \Prob{\mc{D}^{\mathrm{c}} \cap \{\imath(\overline{\mb{X}}; \mb{Y}) \geq \imath(\mb{X}; \mb{Y})\}}, \label{eq:DD}
    \end{align}
    where 
    \begin{IEEEeqnarray}{rCl}
        \mc{D} &\triangleq & \left\{ \log \frac{\W^n(\mb{Y}|\mb{X})}{\QY^n(\mb{Y})} < \tau \right \} \label{eq:Ddef}\\
        \QY(y) &\triangleq& c \left(\sum_{x \in \mc{X}} \PX(x) \W(y|x)^{1/1+\rho}\right)^{1 + \rho}, \,\, y \in \mc{Y}. \label{eq:Q} \IEEEeqnarraynumspace
    \end{IEEEeqnarray}
    Here $\QY$ is the tilted output distribution, 
    $\rho \in [0, 1]$, $\tau \in \mathbb{R}$, and $c > 0$ is a normalization constant. Our achievability bound uses a special case of \eqref{eq:Q} with $\rho = 0$, giving $\QY = \PY$. Whether the more general bound in \eqref{eq:Q} yields an improved bound in the MD regime is a question for future work.

            \subsection{Refined Asymptotics of BHT} \label{sec:NP}
       Before describing \thmref{thm:NP}, below, we introduce binary hypothesis tests, which play a fundamental role in many coding theorems in the literature. 
	
	Let $P$ and $Q$ be two distributions on a common alphabet $\mc{X}$. Consider the binary hypothesis test
    \begin{align}
        H_0 &\colon X \sim P \\
        H_1 &\colon X \sim Q.
    \end{align}
    A randomized test between those two distributions is defined by a probability transition kernel $P_{W |X} \colon \mc{X} \to \{0, 1\}$, where $0$ indicates that the test chooses $H_0$, i.e., $X \sim P$, and $1$ indicates that the test chooses $H_1$, i.e., $X \sim Q$. We define the minimum achievable type-II error compatible with the type-I error bounded by $1-\alpha$ as \cite[eq.~(100)]{polyanskiy2010Channel} 
    \begin{align}
        \beta_{\alpha}(P, Q) \triangleq\min \limits_{P_{W|X} \colon \Prob{W = 0 | H_0} \geq \alpha} \Prob{W = 0 | H_1}. \label{eq:betaalpha}
    \end{align}
       
    
    The minimum in \eqref{eq:betaalpha} is achieved by the Neyman-Pearson test (e.g., \cite[Lemma 57]{polyanskiy2010Channel}), 
    \begin{align}
        P_{W|X}(0|x) = \begin{cases} 1 \quad &\text{if } \log \frac{dP}{dQ}(x) > \gamma \\ \tau \quad &\text{if } \log \frac{dP}{dQ}(x) = \gamma \\ 0 \quad &\text{if } \log \frac{dP}{dQ}(x) < \gamma \end{cases},
    \end{align}
    where $\log \frac{dP}{dQ}(x)$ is the log-likelihood ratio evaluated at $x \in \mathcal{X}$, $\frac{dP}{dQ}$ denotes the Radon-Nikodym derivative, and $\tau$ and $\gamma$ are chosen so that 
        $\alpha = \Prob{W = 0 | H_0}$.
        
    Let $P^{(n)} = \prod_{i = 1}^n P_i$ and $Q^{(n)} = \prod_{i = 1}^n Q_i$, where $P_i$ and $Q_i$ are distributions on a common alphabet $\mc{X}$. 
    
    Define $Z_i \triangleq \log \frac{dP_i}{dQ_i}(X_i)$, where $X_i \sim P_i$ for $i \in [n]$, and
    \begin{align}
        D_i &\triangleq \E{Z_i} = D(P_i \| Q_i) \\
        V_i &\triangleq \Var{Z_i} = V(P_i \| Q_i) \\
        \mu_{k, i} &\triangleq \E{(Z_i - D_i)^k}, \quad k \geq 3 \\
        \ske_i &\triangleq \ske(P_i \| Q_i) = \frac{\mu_{3, i}}{V_i^{3/2}}
    \end{align}
    for $i \in [n]$. Define $\overline{Z}_i \triangleq \log \frac{dP_i}{dQ_i}(\overline{X}_i)$, where $\overline{X}_i \sim Q_i$ for $i \in [n]$, and the cumulant generating function of $\overline{Z}_i$
    \begin{align}
        \kappa_i(s) &\triangleq \log \E{\exp\{s \overline{Z}_i\}}, \quad i \in [n].
    \end{align} 
    Let
    \begin{align}
        D &\triangleq \frac{1}{n} \sum_{i = 1}^n D_i \quad \quad 
        V \triangleq \frac{1}{n} \sum_{i = 1}^n V_i \\
        \ske &\triangleq \frac{1}{n} \sum_{i = 1}^n \ske_i \quad \quad \mu_{k} \triangleq \frac{1}{n} \sum_{i = 1}^n  \mu_{k, i}, \, k \geq 3,  \\
        \kappa(s) &\triangleq \frac{1}{n} \sum_{i = 1}^n \kappa_i(s).
    \end{align}
    
    \thmref{thm:NP}, below, refines \cite[Th.~18]{moulin2017log} by considering SMD sequences and also by deriving the coefficient of $\frac{Q^{-1}(\epsilon_n)^3}{\sqrt{n}}$ in the type-II error exponent.
    
    \begin{theorem}\label{thm:NP}
    Let $P_i$ and $Q_i$ be distributions on a common alphabet $\mc{X}$, and let $P_i$ be absolutely continuous with respect to $Q_i$ for $i \in [n]$. Let $\{\epsilon_n\}_{n = 1}^{\infty}$ be an SMD sequence \eqref{eq:range}. 
    Assume that
    \begin{enumerate}[label=(\Alph*), leftmargin=*]
    \item $Z_i$ satisfies Cram\'er's condition for $i \in [n]$, i.e., $\E{\exp\{s Z_i\}} < \infty$ for $s \in \mathbb{R}$ in the neighborhood of~0;
    \item $V > 0$;
    \item there exist positive constants $\beta_0$, $\beta_1$, and $c > 1$ such that $\beta_0 < |\kappa(s)| < \beta_1$ for all $s \in \mc{D} \triangleq \{s' \in \mathbb{R} \colon |s'| < c\}$, and that $\kappa(s)$ is analytic in $\mc{D}$;
    \item if the sum $\sum_{i = 1}^n \overline{Z}_i$ is non-lattice, then there exist a finite integer $\ell$, a sequence $\{w_n\}_{n = 1}^{\infty}$ satisfying $\frac{w_n}{\log n} \to \infty$, and non-overlapping index sets $\mc{I}_1, \mc{I}_2, \dots, \mc{I}_{w_n} \subset [n]$, each having size $\ell$, such that
    \begin{align}
        \sum_{i \in \mc{I}_{j}} \overline{Z}_i \text{ is non-lattice for } j \in [w_n].
    \end{align}
    \end{enumerate} 
    Then, it holds that
    \begin{align}
        &- \log \beta_{1-\epsilon_n}(P^{(n)}, Q^{(n)}) \notag \\
        &= nD - \sqrt{n V} Q^{-1}(\epsilon_n) + \frac{1}{2} \log n + \left(\frac{\ske \sqrt{V}}{6} + \frac{1}{2}\right) Q^{-1}(\epsilon_n)^2 \notag \\
        &\quad  \notag \\
        &\quad -  \frac{3 (\mu_4 - 3 V^2) V - 4 \mu_3^2}{72 V^{5/2}} \frac{Q^{-1}(\epsilon_n)^3}{\sqrt{n}} \notag \\
        &\quad + \bigo{\frac{Q^{-1}(\epsilon_n)^4}{n}} + \bigo{1}. \label{eq:betaalphaexp}
    \end{align}
    \end{theorem}
    \begin{IEEEproof}
    See \secref{sec:NPproof}.
    \end{IEEEproof}

    Example distributions $\{(P_i, Q_i)\}_{i = 1}^{n}$ that satisfy conditions (A)--(C) in \thmref{thm:NP} include the set of pairs $\{(P_i, Q_i)\}_{i = 1}^n$ where $\Var{Z_i} > 0$ for all $i \in [n]$ and $\mc{X}$ is finite. For example, if $P_i = \mathrm{Bernoulli}(p_i)$ and $Q_i = \mathrm{Bernoulli}(q_i)$ with $p_i, q_i \in (0, 1)$ and $p_i \neq q_i$ for all $i \in [n]$, then conditions (A)--(C) are satisfied. One needs to check condition (D) separately in the case where $\sum_{i = 1}^n \overline{Z}_i$ is non-lattice. If $P_1, \dots, P_n, Q_1, \dots, Q_n$ are such that $\overline{Z}_i$ is a continuous random variable for all $i \in [n]$, then condition (D) is always satisfied, and one needs to check condition (A) separately for each $Z_i$. The conditions (A)--(D) are satisfied for the Gaussian $(P_i, Q_i)$ pairs with positive variances and $P_i \neq Q_i$ for all $i \in [n]$. To highlight the purpose of condition (D), consider the sum of $n-1$ Bernoulli random variables and one Gaussian random variable, where condition (D) is violated. The resulting sum is non-lattice, but its behavior is still very close to a lattice random variable, in this case a binomial.
    
    In \figref{fig:BHT} below, we compare the asymptotic expansion in \thmref{thm:NP} with the true values from the Neyman-Pearson lemma, the CLT approximation from \cite{polyanskiy2010Channel}, and the LD approximation from \cite{cover} for BHT between two i.i.d. Bernoulli distributions.
    The first three terms on the right-hand side of \eqref{eq:betaalphaexp} constitute the CLT approximation of BHT, and are shown in \cite[Lemma 58]{polyanskiy2010Channel} in the CLT regime. The coefficient of $Q^{-1}(\epsilon_n)^2$ in the fourth term of \eqref{eq:betaalphaexp} is the skewness for BHT. The fifth term in \eqref{eq:betaalphaexp} gives the fourth-order characteristic of BHT. A direct application of \thmref{thm:NP} to the meta-converse bound \cite[Th.~27]{polyanskiy2010Channel} shows the converse part of \thmref{thm:refinedAchConv}. Together with the achievability bound of \thmref{thm:refinedAchConv}, this implies that the fourth-order characteristic of Cover--Thomas-symmetric channels and BHT are the same in the sense that $C, V, S$, and $\mu_4$ in \thmref{thm:refinedAchConv} are the same as $D, V, \frac{\ske \sqrt{V}}{6} + \frac{1}{2}$, and $\mu_4$ in \eqref{eq:betaalphaexp} evaluated at $P^{(n)} = P_{Y|X = x}^n$ and $Q^{(n)} = (Q_Y^*)^n$; here $x \in \mc{X}$ is arbitrary, and $Q_Y^*$ is the capacity-achieving output distribution. 
    
    In \thmref{thm:NP}, conditions (A) and (B) are used to apply the MD result \lemref{lem:cornish} (see \secref{sec:moderate}, below) to the sum $\sum_{i = 1}^n Z_i$; conditions (C) and (D) are used to satisfy the conditions of the LD results (Theorems~\ref{thm:chaganty} and \ref{thm:lattice} in \secref{sec:large} below) for the random variable $\sum_{i = 1}^n \overline{Z}_i$. Note that if $\sum_{i = 1}^n \overline{Z}_i$ is lattice, then each of the random variables $\overline{Z}_i$, $i \in [n]$, is lattice. 
    In the non-lattice case, the sum $\sum_{i = 1}^n \overline{Z}_i$ can be non-lattice even if one of more of the $\overline{Z}_i$ is lattice. Condition (D) of \thmref{thm:NP} requires that there are $w_n \gg \log n$ non-overlapping, non-lattice partial sums of $\overline{Z}^n$, where each partial sum is a sum of $\ell$ random variables. 
    A condition similar to condition (D) with $\ell \leq 2$ is introduced in \cite[Def. 15]{moulin2017log} for the same purpose.
    
     \begin{figure*}
        \center
        \includegraphics[width=1\linewidth]{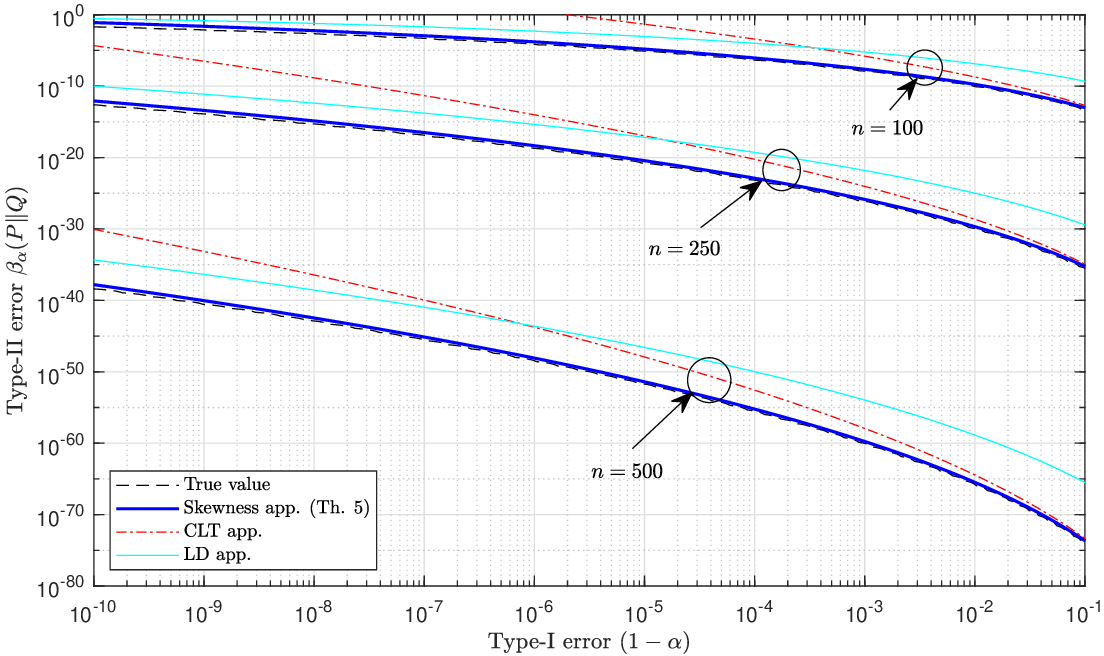}
        \caption{Type-I vs. Type-II error probability for BHT: The expansion from Theorem~\ref{thm:NP}, excluding the $O(\cdot)$ terms, is shown for $P_i = \mr{Bern}(0.6)$, $Q_i = \mr{Bern}(0.2)$, $i = 1, \dots, n$, $n \in \{100, 250, 500\}$. Our skewness approximation is compared with the true values obtained by the Neyman-Pearson lemma, the CLT approximation from \cite[Lemma 58]{polyanskiy2010Channel}, which consists of the terms up to $\frac{1}{2} \log n$, and the first-order LD approximation from \cite[Th.~11.7.1]{cover}.} 
        \label{fig:BHT}
    \end{figure*}

        \subsection{Gaussian Channel} \label{sec:Gaussian}
    The output of the memoryless Gaussian channel in response to the input $\mb{X} \in \mathbb{R}^n$ is
        \begin{align}
            \mb{Y} = \mb{X} + \mb{Z}, \label{eq:pointchannel}
        \end{align}
    where the entries of $\mb{Z}$ are drawn i.i.d. from $\mathcal{N}(0, 1)$, independent of $\mb{X}$. 
    The capacity and dispersion of the Gaussian channel are given by
    \begin{align}
        C(P) &\triangleq \frac{1}{2} \log (1 + P) \\
        V(P) &\triangleq \frac{P(P+2)}{2 (1+P)^2}.
    \end{align}
    In addition to the average error probability constraint \eqref{eq:averageerror}, an $(n, M, \epsilon, P)$ code for the Gaussian channel with a maximal-power constraint requires that each codeword has power not exceeding $nP$, i.e.,
    \begin{align}
        \norm{\mathsf{f}(m)}_2^2 \leq nP, \quad \forall \, m \in [M].
    \end{align}
    The maximum achievable message set size  $M^*(n, \epsilon, P)$ is defined similarly to \eqref{eq:Mstardef}; the corresponding non-Gaussianity is defined as
    \begin{align}
         \zeta(n, \epsilon, P) &\triangleq \log M^*(n, \epsilon, P) - (n C(P) - \sqrt{n V(P)} Q^{-1}(\epsilon)).
    \end{align}
    \thmref{thm:Gaussian}, below, gives lower and upper bounds on the non-Gaussianity $\zeta(n, \epsilon, P)$ in the CLT regime.
    
    \begin{theorem} \label{thm:Gaussian}
    Fix $\epsilon \in (0, 1)$ and $P > 0$. Then,
    \begin{align}
        \zeta(n, \epsilon, P) &\geq \frac{1}{2} \log n + S(P) Q^{-1}(\epsilon)^2 + \underline{B}(P) + \bigo{\frac{1}{\sqrt{n}}} \label{eq:Gausslower} \\
        \zeta(n, \epsilon, P) &\leq \frac{1}{2} \log n + S(P) Q^{-1}(\epsilon)^2 + \overline{B}(P) + \bigo{\frac{1}{\sqrt{n}}}, \label{eq:Gaussupper}
    \end{align}
    where 
    
    \begin{align}
        S(P) &= \frac{6 + 6 P + 4 P^2 + P^3}{6 (1 + P)^2 (2 + P)} \\
        \overline{B}(P) &= \frac{P(3 + P)}{3(1 + P)(2 + P)} + \frac{1}{2} \log(2 \pi V(P)) \label{eq:Bover}\\
        \underline{B}(P) &= \frac{P(5P + 9)}{6 (1 + P) (2 + P)} - 1 + \frac{1}{2} \log \nB{\frac{2 \pi P}{(1 + P)^2}}. \label{eq:Boverline}
    \end{align}

    \end{theorem}
    \begin{IEEEproof}
 The achievability bounds in \eqref{eq:Gausslower}, \cite[eq.~58]{shannon1959Probability}, and \cite[Th.~17]{erseghe2016} are fourth-order asymptotic expansions. They analyze the same non-asymptotic achievability bound in \cite[eq.~(19)]{shannon1959Probability}. The difference is that we analyze the tail probability of the noncentral $t$-distribution in the CLT regime while \cite{shannon1959Probability} and \cite{erseghe2016} analyze it in the LD regime. 
    Our derivation uses the Cornish-Fisher expansion of the noncentral $t$-distribution in the CLT regime \cite{fisher1960}.

    The converse bound \eqref{eq:Gaussupper} analyzes the novel meta-converse bound from \cite[Th.~3]{vazquez2021}
        \begin{align}
            &\log M^*(n, \epsilon, P) \notag \\
            &\leq \inf_{\sigma^2 > 1} - \log \beta_{1 - \epsilon}(\mathcal{N}(\sqrt{P} \bs{1}, \ms{I}_n), \mathcal{N}(\bs{0}, \sigma^2 \ms{I}_n)). \label{eq:metaG}
        \end{align}
    Since $\sigma^2 = 1 + P$ is optimal for codes whose rate approaches the capacity (see, e.g., \cite{polyanskiy2010Channel}), we first let $\sigma^2 = 1 + P + \delta_n$, where $\delta_n$ is a sequence that approaches $0$. Then we optimize the value of $\delta_n$ with respect to \eqref{eq:metaG} by analyzing the $\beta_{\alpha}$ function in the CLT regime using \cite[Th.~18]{moulin2017log}. Not setting $\delta_n = 0$ is crucial to prove the tightness of the skewness $S(P)$. The $\delta_n^*$ that achieves the minimum in \eqref{eq:metaG} is  
        \begin{align}
            \delta_n^* = - \frac{Q^{-1}(\epsilon)}{\sqrt{n}} \sqrt{\frac{2 P}{P + 2}}.
    \end{align}
    Recall that \cite[Th.~18]{moulin2017log} is the CLT version of \thmref{thm:NP}. Therefore, replacing \cite[Th.~18]{moulin2017log} with \thmref{thm:NP} shows that the expansion in \eqref{eq:Gaussupper} holds in the MD regime as well up to the skewness term $S(P) Q^{-1}(\epsilon_n)^2$.

      See \secref{app:Gaussian} for the proof details.
    \end{IEEEproof}
    
        \thmref{thm:Gaussian} yields the channel skewness of the Gaussian channel as $S(P)$ since the lower and upper bounds on the $Q^{-1}(\epsilon)^2$ term in \eqref{eq:Gausslower}--\eqref{eq:Gaussupper} match. 
    
\begin{figure*}
        \center
        \includegraphics[width=1\linewidth]{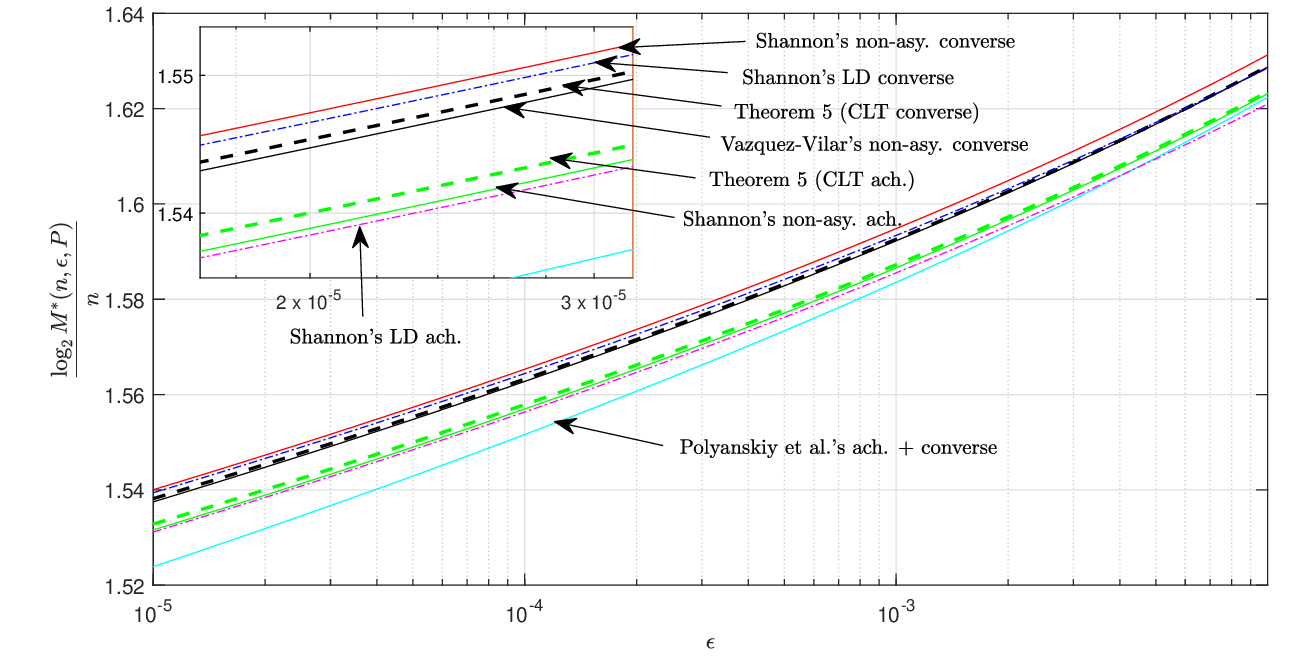}
        \caption{Achievable rate vs. average error probability for a Gaussian channel: The expansions from Theorem~\ref{thm:Gaussian}, excluding the $O(\cdot)$ term, are shown for the Gaussian channel with $P = 10$, $n = 400$, and $\epsilon \in [10^{-5}, 10^{-3}]$. Shannon's non-asymptotic bounds are from \cite[eq.~(20)]{shannon1959Probability}; Vazquez-Vilar's non-asymptotic bound is from \cite[Th.~3]{vazquez2021} where the variance of the auxiliary output distribution is optimized numerically; Shannon's LD approximations are from \cite[eq.~(51), (58)]{shannon1959Probability}; Polyanskiy \emph{et al.}'s CLT approximation that takes $\zeta(n, \epsilon, P) = \frac{1}{2} \log n$ is from  \cite{polyanskiy2010Channel, tan2015Third}.} 
        \label{fig:gauss}
        \end{figure*}

    In \figref{fig:gauss}, the skewness approximations in \thmref{thm:Gaussian} are compared with Shannon's non-asymptotic bounds and the LD approximations from \cite{shannon1959Probability}, the CLT approximation from \cite{ polyanskiy2010Channel} using the achievability bound proved in \cite{tan2015Third}, and Vazquez-Vilar's novel non-asymptotic converse bound \cite[Th.~3]{vazquez2021}. For the shown $(n, \epsilon, P)$ triples, our skewness approximation \eqref{eq:Gaussupper} is the closest to the novel non-asymptotic converse bound in \cite[Th.~3]{vazquez2021}; our skewness approximation \eqref{eq:Gausslower} is the closest to Shannon's non-asymptotic achievability bound for $\epsilon  \gtrapprox 10^{-4}$; for $\epsilon  \lessapprox 10^{-4}$, Shannon's LD approximation becomes the closest.

    Since the noncentral $t$-distribution is not a sum of independent random variables, Petrov's MD expansion in \thmref{thm:moderatePetrov} below, does not apply. The proof of \thmref{thm:moderatePetrov} relies on all moments of the random variable being finite; however, the $n$-th and higher order moments of the noncentral $t$-distribution with $n$ degrees of freedom are undefined. Therefore, one needs to find another method to derive the asymptotic expansion of the CDF of the noncentral $t$-distribution in the MD regime. However, since the Cornish-Fisher expansions in general have the same skewness term in the CLT and MD regimes (see \lemref{lem:cornish}, below), we conjecture that the achievability bound in \eqref{eq:Gaussupper} holds up to the $S(P) Q^{-1}(\epsilon_n)^2$ term in the MD regime. 

    In \cite[Th.~41]{polyanskiy2010Channel}, Polyanskiy \emph{et al.} show the converse in \eqref{eq:metaG} for codes with an equal-power constraint, i.e., each codeword has power $nP$ exactly; where $\sigma^2$ is set to the capacity-achieving output variance, $\sigma^2 = 1 + P$. Then, Polyanskiy \emph{et al.} invoke the inequality (see \cite[eq.~(83)]{shannon1959Probability}) 
        \begin{align}
           M^*(n, \epsilon, P) \leq M^*\left(n + 1, \epsilon, P\right)_{\mr{eq}}\label{eq:shannonmaxeq}
        \end{align}
        to get a converse bound for the maximal-power constraint,
        where $M(n, \epsilon, P)_{\mathrm{eq}}$ is the maximum achievable message set size for the equal-power constraint.\footnote{In \cite[eq.~(23)]{vazquez2021}, 
        a slightly tighter version, which states that $M^*(n, \epsilon, P) \leq M^*\left(n + 1, \epsilon, \frac{nP}{n+1}\right)_{\mathrm{eq}}$, is proved.} Since Vazquez-Vilar's converse \eqref{eq:metaG} does not need to apply \eqref{eq:shannonmaxeq}, it is a refinement to that of Polyanskiy \emph{et al.} for the maximal-power constraint. 

        Shannon's non-asymptotic cone-packing converse in  \cite[eq.~15]{shannon1959Probability} is the tightest known converse bound under the equal-power constraint (see, e.g., \cite{vazquez2021}).   
    It coincides with Polyanskiy \emph{et al.}'s meta-converse \cite[Th.~28]{polyanskiy2010Channel} applied with the optimal auxiliary output distribution \cite[Sec. VI-F]{polyanskiy2013saddlepoint}. The converse bounds in \cite{erseghe2016, shannon1959Probability} both analyze Shannon's cone-packing converse in \cite[eq.~(20)]{shannon1959Probability}. Analyzing Shannon's cone-packing converse in combination with the inequality in \cite[eq.~(23)]{vazquez2021}
    using the CLT approximation for the noncentral $t$-distribution tails, we derive a converse bound with $S(P)$ in \eqref{eq:Gaussupper} unchanged and $\overline{B}(P)$ replaced with $\overline{B}(P)_{\mr{Sh}} = \underline{B}(P) + 1 + C(P) - \frac{P}{2 (1 + P)}$; note that $\overline{B}(P)_{\mr{Sh}} > \overline{B}(P)$ for all $P > 0$. This result implies that in the CLT regime, Vazquez-Villar's converse for the maximal-power constraint is sharper than Shannon's converse combined with the inequality in \cite[eq.~(23)]{vazquez2021}.

    \section{Proofs of Theorems \ref{thm:mainAch} and \ref{thm:mainConv}}{\label{sec:proofs}}

    We begin by giving the preliminary definitions for quantities related to the moments of a random variable $X$.
    
    \subsection{Moment and Cumulant Generating Functions}
    Below, we dedicate the letters $s$ and $t$ to real scalars and $z$ to complex scalars.
    The moment generating function (MGF) of $X$ is defined as
    \begin{align}
        \phi(z) \triangleq \E{\exp\{z X\}}, \quad z \in \mathbb{C}. 
    \end{align}
    The $j$-th central moment is denoted by
    \begin{align}
        \mu_j \triangleq \E{(X - \E{X})^j}.
    \end{align}
    The cumulant generating function (CGF) of $X$ is defined as
    \begin{align}
        \kappa(z) \triangleq \log \phi(z) = \sum_{j = 1}^\infty \kappa_j \frac{ z^j}{j!},
    \end{align}
    where $\kappa_j$ is called the $j$-th cumulant of $X$, and there exists a one-to-one relationship between $\kappa_j$ and the central moments up to the order $j$. For example,
    \begin{align}
        \kappa_1 &= \E{X} \\
        \kappa_2 &= \mu_2 \\
        \kappa_3 &= \mu_3 \\
        \kappa_4 &= \mu_4 - 3  \mu_2^2. 
    \end{align}
    We use $\phi^{(X)}(\cdot)$ and $\kappa^{(X)}(\cdot)$ to denote the MGF and CGF of $X$ when the random variable is not clear from context. 
    The $j$-th cumulant of $c X$ is given by $\kappa_j^{(cX)} = c^j \kappa_j^{(X)}$, and the CGF of $X + Y$, where $X$ and $Y$ are independent, is 
    \begin{align}
        \kappa^{(X + Y)}(z) = \kappa^{(X)}(z) + \kappa^{(Y)}(z).
    \end{align}
    
	The MGF and CGF are naturally extended to $d$-dimensional random vectors. Let $\mb{S}$ be a $d$-dimensional random vector. The MGF and CGF of $\mb{S}$ are denoted by
	\begin{align}
	    \phi(\mb{z}) &\triangleq \E{\exp\{\langle \mb{z}, \mb{S} \rangle\}}, \quad \mb{z} \in \mathbb{C}^d, \\
	    \kappa(\mb{z}) &\triangleq \log \phi(\mb{z}). 
	\end{align}

    Next, we present the supporting results used to bound the probability terms that appear in the proofs of Theorems~\ref{thm:mainAch} and~\ref{thm:mainConv}.

    \subsection{MD Asymptotics} \label{sec:moderate}
    \thmref{thm:moderatePetrov}, stated next, is an MD result that bounds the probability that the sum of $n$ independent but not necessarily identical random variables normalized by a factor $\frac{1}{\sqrt{n}}$ deviates from the mean by $o(\sqrt{n})$. The resulting probability is an SMD sequence \eqref{eq:range}.

	\begin{theorem}[Petrov {\cite[Ch.~8, Th.~2]{petrov1975}}]\label{thm:moderatePetrov}
	Let $X_1, \dots, X_n$ be independent random variables.  Let $\E{X_i} = 0$ for $i = 1, \dots, n$, $\kappa_j = \frac{1}{n} \sum_{i = 1}^n \kappa_j^{(X_i)}$ for $j \geq 2$, and $\ske = \frac{\kappa_3}{\kappa_2^{3/2}}$. Define
	\begin{align}
	    S_n &\triangleq  \frac{1}{\sqrt{n \kappa_2}} \sum_{i = 1}^n X_i \\
	    F_n(x) &\triangleq \Prob{S_n \leq x}.
	\end{align}
	Suppose that there exist positive constants $t_0$ and $H_1, \dots, H_n$ such that the MGF satisfies 
	\begin{align}
	    \phi^{(X_i)}(t) \leq H_i \label{eq:Cramercond}
	\end{align}
	for all $t \in \mathbb{R}$ such that $-t_0 < t < t_0$ and $i = 1, \dots, n$,
        \begin{align}
            \limsup_{n \to \infty} \frac{1}{n} \sum_{i = 1}^n H_i^{3/2} &< \infty \\
            \liminf_{n \to \infty} \kappa_2 &> 0. \label{eq:kappacond}
        \end{align}
	Let $x > 0$ and $x = o(\sqrt{n})$. Then, it holds that
	\begin{align}
	    &1 - F_n(x) = Q(x) \exp\left\{\frac{x^3 }{ \sqrt{n}} \lambda_n\left( \frac{x}{\sqrt{n}} \right) \right\} \nB{1 + O\left(\frac{1 + x}{\sqrt{n}}\right)} \IEEEeqnarraynumspace \label{eq:petrov1}\\
	    &F_n(-x) = Q(x) \exp\left\{\frac{-x^3 }{ \sqrt{n}} \lambda_n\left( \frac{-x}{\sqrt{n}} \right) \right\} \nB{1 + O\left(\frac{1 + x}{\sqrt{n}}\right)}, \label{eq:petrov2}\IEEEeqnarraynumspace
	\end{align}
	where 
	\begin{align}
	    \lambda_n(x) \triangleq \sum_{i = 0}^\infty a_i x^i
	\end{align}
	is Cram\'er's series whose first two coefficients are
	\begin{align}
	    a_0 &= \frac{\ske}{6} \label{eq:a0}\\
	    a_1 &= \frac{\kappa_4 \kappa_2 - 3 \kappa_3^2}{24 \kappa_2^3}. \label{eq:a1}
	\end{align}
	\end{theorem}

    The condition in \eqref{eq:Cramercond} is called Cram\'er's condition. 
    Petrov presents \eqref{eq:Cramercond} for complex functions as ``in the circle $\{z \in \mathbb{C} \colon |z| \leq t_0$\}, $\phi^{(X_i)}(z)$ is analytic and $|\phi^{(X_i)}(z)| \leq H_i$.'' However, this is equivalent to \eqref{eq:Cramercond} (see e.g., \cite[Th.~1.7.1]{CF1999}). Note that Cram\'er's condition also implies that all moments of $X_i$ are finite.

    Let $X_1, \dots, X_n$ be supported on a common finite alphabet with $\Var{X_i} > \sigma^2 > 0$ for all $i 
    \in [n]$. Then, there exists an $H > 0$ such that \eqref{eq:Cramercond} is satisfied with $H_i \leq H$ for all $i \in [n]$; and $\kappa_2 > \sigma^2 > 0$. Therefore, the conditions of \thmref{thm:moderatePetrov} are satisfied for this class of random variables, which is the case in the application of \thmref{thm:moderatePetrov}.
 
	The $O(\cdot)$ terms in \eqref{eq:petrov1}--\eqref{eq:petrov2} constitute a bottleneck in deriving the $O(1)$ terms in \eqref{eq:achievability} and \eqref{eq:converse}; that is, one needs to compute the leading term of the $O(\cdot)$ terms in \eqref{eq:petrov1}--\eqref{eq:petrov2} in order to compute the $O(1)$ terms in our achievability and converse bounds. In the CLT regime, i.e., $x = O(1)$, \thmref{thm:moderatePetrov} reduces to the Berry-Esseen theorem for the sum of independent random variables without explicitly giving the constant, that is, the $\left(1 + O\left(\frac{1}{\sqrt{n}}\right)\right)$ term in \eqref{eq:petrov1}--\eqref{eq:petrov2} dominates the $\exp\{\cdot\}$ term. 
	
	Inverting \thmref{thm:moderatePetrov} (that is, obtaining an expansion for $x$ in terms $y$ where $F_n(-x) = Q(y)$) is advantageous in many applications. For $Q(y) = \epsilon_n$, where $\{\epsilon_n\}_{n = 1}^{\infty}$ is an SMD sequence of probabilities \eqref{eq:range}, \lemref{lem:cornish}, below, gives the corresponding sequence of quantiles. In the CLT regime, in which $F_n(-x) \in (0, 1)$ is equal to a value independent of $n$, that expansion is known as the Corner-Fisher theorem \cite[Sec.~8]{cornish}, which inverts the Edgeworth expansion. Note that \cite[Sec.~8]{cornish} applies under the assumption that the elements in the sum are i.i.d. and strongly non-lattice random variables; these assumptions need not hold for our application.  
	\begin{lemma}\label{lem:cornish}
	Let $X_1, \dots, X_n$ satisfy the conditions in \thmref{thm:moderatePetrov}. Let $y \triangleq Q^{-1}(\epsilon_n) = o(\sqrt{n})$. Suppose that $F_n(-x) = Q(y) = \epsilon_n$, then
	\begin{align}
	    x = y - \frac{b_0 y^2}{\sqrt{n}} + \frac{b_1 y^3}{n} + \bigo{\frac{y^4}{n^{3/2}}} + \bigo{\frac{1}{\sqrt{n}}}, \label{eq:xy2}
	\end{align} \label{eq:eqx}
	where
	\begin{align}
	    b_0 &\triangleq \frac{\ske}{6} \\
	    b_1 &\triangleq \frac{3 \kappa_4 \kappa_2 - 4 \kappa_3^2}{72 \kappa_2^3}.
	\end{align}
	\end{lemma}
	\begin{IEEEproof}
	See \appref{app:cornish}.
	\end{IEEEproof}
	A weaker version of \lemref{lem:cornish}, with only the first two terms in \eqref{eq:xy2} and with $\epsilon_n$ decaying polynomially with $n$, is proved in \cite[Lemma 7]{sakai2021Third}. We use \thmref{thm:moderatePetrov} and \lemref{lem:cornish} to bound the probability $\Prob{\imath(\mb{X}; \mb{Y}) \leq \tau}$, where $\tau$ is a threshold satisfying the condition in \thmref{thm:moderatePetrov}, and the resulting probability is an SMD sequence \eqref{eq:range}. Although the MD approximation to the CDF of the normalized sum in \thmref{thm:moderatePetrov} is seemingly different than the CLT approximation to the same CDF (the Edgeworth expansion), their inverted theorems, i.e.,  \lemref{lem:cornish} and the Cornish-Fisher theorem \cite[Sec.~8]{cornish}, respectively, have similar forms; for continuous random variables, the Cornish-Fisher theorem admits the formula in \eqref{eq:xy2}, where $\bigo{\frac{1}{\sqrt{n}}}$ is replaced by $\frac{b_0}{\sqrt{n}} + \bigo{\frac{1}{n}}$. This is the main reason why the channel skewness bounds computed in the CLT regime extend to the MD regime without change. 
    
    \subsection{Strong LD Asymptotics}	\label{sec:large}
    For the results in this section, we consider a sequence of $d$-dimensional random vectors $\mb{S}_n = (S_{n, 1}, \dots, S_{n, d})$, $n = 1, 2, \dots$. Let $\phi_n(\cdot)$ denote the MGF of $\mb{S}_n$, and let $\kappa_n(\cdot)$ be the normalized CGF of $\mb{S}_n$ denoted by
    \begin{align}
        \phi_n(\mb{z}) &\triangleq \phi^{(\mb{S}_n)}(\mb{z}) \\
        \kappa_n(\mb{z}) &\triangleq \frac{1}{n} \log \phi_n(\mb{z}). \label{eq:kappan}
    \end{align}
    The Fenchel--Legendre transform of $\kappa_n(\cdot)$ is given by
	\begin{align}
	    \Lambda_n(\mb{x}) \triangleq \sup_{\mb{t} \in \mathbb{R}^d} \left\{\langle \mb{t}, \mb{x} \rangle - \kappa_n(\mb{t}) \right\}, \label{eq:lambdax}
	\end{align}
	where $\mb{x} \in \mathbb{R}^d$. The quantity \eqref{eq:lambdax} is commonly known as the \emph{rate function} in the LD literature \cite[Ch. 2.2]{dembobook}.

	\thmref{thm:chaganty}, below, is a strong LD result for an arbitrary sequence of random vectors $\mb{S}_n$ in $\mathbb{R}^d$; here, \emph{strong} refers to the fact that \thmref{thm:chaganty} characterizes the exact prefactor in front of the LD exponent. 
	\begin{theorem}[{Chaganty and Sethuraman \cite[Th. 3.4] {chaganty}}]\label{thm:chaganty}
	Let $\{\mb{a}_n\}_{n = 1}^{\infty}$ be a bounded sequence of $d$-dimensional vectors. Assume that the following conditions hold. \\
	\textbf{Smoothness (S):} $\kappa_n(\mb{z})$ is bounded below and above, and is analytic in $\mc{D}^d$, where $\mc{D} \triangleq \{\mb{z} \in \mathbb{C} \colon |\mb{z}| < c\}$ and $c$ is a finite constant. \\
    \textbf{Non-Degenerate (ND):} There exist a real sequence $\{\mb{s}_n\}_{n = 1}^{\infty}$ and constants $c_0$ and $c_1$ that satisfy
	    \begin{align}
	        &\nabla \kappa_n(\mb{s}_n) = \mb{a}_n \\
	        &0 < c_0 < s_{n,j} < c_1 < c \text{ for all } j \in [d] \text{ and } n \geq 1, \label{eq:sbound}
	    \end{align}
	 where $c$ is the constant given in condition (S), and the Hessian matrix $\nabla^2 \kappa_n(\mb{s}_n)$, which is a covariance matrix of a tilted distribution obtained from $\mb{S}_n$, is positive definite with a minimum eigenvalue bounded away from zero for all $n$. \\
	 \textbf{Non-Lattice (NL):} There exists $\delta_0 > 0$ such that for any given $\delta_1$ and $\delta_2$ such that $0 < \delta_1 < \delta_0 < \delta_2$
	 \begin{align}
	     \sup_{\mb{t} \colon \delta_1 < \| \mb{t} \|_{\infty} \leq \delta_2} \left \lvert \frac{\phi_n(\mb{s}_n + \mathrm{i} \mb{t})}{\phi_n(\mb{s}_n)} \right \rvert = o\left(n^{-d/2}\right), \label{eq:absphi}
	 \end{align}
	 where $\mathrm{i} = \sqrt{-1}$ is the imaginary unit.
	 Then,
	 \begin{align}
	     \Prob{\mb{S}_n \geq n \mb{a}_n} = \frac{E_{\mr{NL}}}{n^{d/2}} \exp\{-n \Lambda_n(\mb{a}_n)\} (1 + o(1)), \label{eq:expNL}
	 \end{align}
	 where 
	 \begin{align}
	     E_{\mr{NL}} \triangleq  \frac{1}{(2 \pi)^{d/2} \left(\prod\limits_{j = 1}^d s_{n,j} \right) \sqrt{\det( \nabla^2\kappa_n(\mb{s}_n))}}.
	 \end{align}
	
	\end{theorem}
	
	Condition (S) of \thmref{thm:chaganty} is a \emph{smoothness} assumption for the CGF $\kappa_n$, which is a generalization of Cram\'er's condition that appears in the LD theorem for the sum of i.i.d. random vectors \cite[Th. 2.2.30]{dembobook}. Condition (S) implies that all moments of the tilted distribution obtained from $\mb{S}_n$ are finite. Condition (ND) is used to ensure that $\mb{S}_n$ is a \emph{non-degenerate} random vector, meaning that it does not converge in distribution to a random vector with $\ell < d$ dimensions, and that the rate function $\Lambda_n(\mb{a}_n)$ is bounded and does not decay to zero. The latter follows from the boundedness condition in \eqref{eq:sbound}, and implies that the probability of interest is in the LD regime. The ratio $\frac{\phi_n(\mb{s}_n + \mathrm{i} \mb{t})}{\phi_n(\mb{s}_n)}$ in  \eqref{eq:absphi} is equal to the characteristic function of a random vector that is obtained by tilting $\mb{S}_n$ by $\mb{s}_n$ \cite{chaganty}. A random variable is non-lattice if and only if its characteristic function satisfies $|\phi(\ms{i} t)| < 1$ for all real $t \neq 0$ \cite[Ch. XV, Sec. 1, Lemma 4]{feller1971introduction}. Therefore, since tilting does not affect the support of a distribution, condition (NL) requires $\mb{S}_n$ to be a \emph{non-lattice} random vector.
	Condition (NL) is used to guarantee that the absolute value of that characteristic function decays to zero quickly enough outside a neighborhood of the origin, which makes the random vector $\mb{S}_n$ behave like a sum of $n$ non-lattice random vectors. 
	
	Let $\overline{\mb{X}}$ be a random codeword that is independent of both the transmitted codeword $\mb{X}$ and the channel output $\mb{Y}$. If $\imath(\mb{X}; \mb{Y})$ and $\imath(\overline{\mb{X}}; \mb{Y})$ are non-lattice, we apply \thmref{thm:chaganty} to the sequence of 2-dimensional non-lattice random vectors $(\imath(\mb{X}; \mb{Y}), \imath(\overline{\mb{X}}; \mb{Y}) - \imath(\mb{X}; \mb{Y}))$ to bound the probability $\Prob{\imath(\overline{\mb{X}}; \mb{Y})) \geq \imath(\mb{X}; \mb{Y}) \geq \tau_n}$ for some sequence~$\tau_n$. 
	
	When applied to the sum of $n$ i.i.d. random variables $\mb{S}_n = \sum_{i = 1}^n \mb{A}_i$, $\kappa_n$ in \eqref{eq:kappan} reduces to the CGF of $\mb{A}_1$ as
	\begin{align}
	    \kappa(\mb{z}) = \log \E{\exp\{\langle \mb{z}, \mb{A}_1 \rangle\}}. \label{eq:kappaz}
	\end{align}
	In our application, since $\mb{A}_1 = (\imath(X_1; Y_1), \imath(\overline{X}_1; Y_1) - \imath(X_1; Y_1))$ has a finite support, the expectation in \eqref{eq:kappaz} is bounded, and all moments of $\mb{A}_1$ are finite;
	therefore, condition (S) of \thmref{thm:chaganty} is satisfied. Further, the characteristic function of the sum of $n$ i.i.d. random vectors is equal to $n$-th power of the characteristic function of one of the summands. Therefore, the left-hand side of \eqref{eq:absphi} decays to zero exponentially quickly for the sum of i.i.d. non-lattice random vectors. This means that in our application, condition (NL) of \thmref{thm:chaganty} is satisfied with room to spare.
	
	We use the following strong LD result to bound the probability $\Prob{\imath(\overline{\mb{X}}; \mb{Y})) \geq \imath(\mb{X}; \mb{Y}) \geq \tau_n}$ with lattice $\imath(\mb{X}; \mb{Y})$ and $\imath(\overline{\mb{X}}; \mb{Y})$. 
	\begin{theorem}\label{thm:lattice}
	Suppose that $\mb{S}_n = (S_{n, 1}, \dots, S_{n, d})$, and $S_{n, j}$ is a lattice random variable with span $h_{n, j}$, i.e., $\Prob{S_{n, j} \in \{b_{n, j} + k h_{n, j} \colon k \in \mathbb{Z}\}} = 1$ for some $b_{n, j}$, such that there exist positive constants $\underline{h}_{j}$ and $\overline{h}_j$ satisfying $\underline{h}_{j} < h_{n, j} < \overline{h}_j$ for all $j \in [d]$, $n \geq 1$. Assume that conditions (S) and (ND) in \thmref{thm:chaganty} hold, and replace condition (NL) by the following. \\
    \textbf{Lattice (L):} There exists $\bs{\lambda} > \bs{0}$ such that, for any given $\bs{\delta}$ satisfying $\bs{0} < \bs{\delta} < \bs{
    \lambda}$, 
    \begin{align}
        \sup_{\mb{t} \colon \delta_{j} < |t_j| \leq \frac{\pi}{h_{n, j}} \text{ for } j \in [d] } \left \lvert \frac{\phi_n(\mb{s}_n + \mathrm{i} \mb{t})}{\phi_n(\mb{s}_n)} \right \rvert = o\left(n^{-d/2}\right). \label{eq:momentlattice}
    \end{align}
    If $n \mb{a}_n$ is in the range of the random vector $\mb{S}_n$, then
	 \begin{align}
	     \Prob{\mb{S}_n \geq n \mb{a}_n} = \frac{E_{\mr{L}}}{n^{d/2}} \exp\{-n \Lambda_n(\mb{a}_n)\} (1 + o(1)), \label{eq:chagantyLat}
	 \end{align}
	 where 
	 \begin{align}
	     E_{\mr{L}} \triangleq   \frac{1}{(2 \pi)^{d/2}  \sqrt{\det( \nabla^2\kappa_n(\mb{s}_n))}} \left(\prod\limits_{j = 1}^d \frac{h_{n, j}}{1-\exp\{-s_{n,j} h_{n, j}\}} \right).
	 \end{align}
	\end{theorem}
	\begin{IEEEproof}
	The one-dimensional lattice case, i.e., $d = 1$, is proved in \cite[Th. 3.5]{chagantyOneDimensional}. The proof of the $d$-dimensional lattice case follows by inspecting the proofs for the $d$-dimensional non-lattice random vectors in \cite[Th. 3.4]{chaganty} and the one-dimensional lattice random variables in \cite[Th. 3.5]{chagantyOneDimensional}. Specifically, in the proof of \cite[Th. 3.4]{chaganty}, we replace \cite[Th. 2.4]{chaganty} by extending the lattice result in \cite[Th.~2.10]{chagantyOneDimensional} to $d$-dimensional random vectors.  The modification in the proof yields \thmref{thm:lattice}. The full proof of \thmref{thm:lattice} appears in \appref{app:lattice}.
	\end{IEEEproof}
	
	If $\mb{S}_n = (S_{n, 1}, \dots, S_{n, d})$ is a sum of $n$ i.i.d. random vectors, where
	\begin{align}
	    S_{n, j} = \sum_{i = 1}^n A_{i, j}, \quad j \in [d],
	\end{align}
	and $A_{1, j}$ is a lattice random variable with span $h_j$ for $j \in [d]$, then it holds that
	\begin{align}
	     \sup_{\delta_{j} < |t_j| \leq \frac{\pi}{h_{j}}} \left \lvert \frac{\phi^{(X_{1, j})}(s_j + \mathrm{i} t_j)}{\phi^{(X_{1, j})}(s_j)} \right \rvert < 1, \quad j \in [d]. \label{eq:less1lat}
	\end{align}
	The bound \eqref{eq:less1lat} follows from \cite[Ch. 15, Sec. 1, Lemma 4]{feller1971introduction} since $\frac{\phi^{(A_{1, j})}(s_j + \mathrm{i} t_j)}{\phi^{(A_{1, j})}(s_j)}$ is a characteristic function of a lattice random variable with span $h_j$. The condition in \eqref{eq:momentlattice} modifies the condition in \eqref{eq:absphi} for lattice random vectors by considering a single period of that characteristic function. If $\mb{S}_n$ is an i.i.d.\ sum, then the left-hand side of \eqref{eq:momentlattice} decays exponentially with $n$, and condition (L) is satisfied. Note that if $h_{n, j} \to 0$ for all $(n, j)$ pairs, then $\mb{S}_n$ converges to a non-lattice random vector, and the prefactor $E_{\mr{L}}$ converges to the prefactor for the non-lattice random vectors,~$E_{\mr{NL}}$. 

   Although the strong LD theorems (Theorems~\ref{thm:chaganty} and \ref{thm:lattice}) are used only for the sum of independent random variables in this paper, their applications extend to Gaussian-like distributions that do not necessarily arise from a sum, and to the sum of weakly dependent random vectors. An example in the first distributional family is the multidimensional noncentral $t$-distribution, which appears in the analysis of the Gaussian channel. In \cite{chaganty}, the multidimensional $F$-distribution is also given as an example. Examples in the second family include the information density $\imath(\mb{X}; \mb{Y})$ for constant-composition codes. One drawback of the use of Theorems~\ref{thm:chaganty} and \ref{thm:lattice} is that they require knowledge of an expansion of the MGF of the random vector, which might not be available in some cases.
	
In \cite[Lemma 3]{altug2014refinement}, Altu\u{g} and Wagner derive a non-asymptotic upper bound on the probability $\Prob{\mb{S}_n \geq n \mb{a}_n}$ as
    \begin{align}
        \Prob{\mb{S}_n \geq n \mb{a}_n} \leq \frac{E_{\mr{AW}}}{n} \exp\{-n \Lambda_n(\mb{a}_n)\}, \label{eq:AWbound}
    \end{align}
    where
    $\mb{S}_n = \sum_{i = 1}^n \mb{A}_{i}$ is a sum of $n$ i.i.d. 2-dimensional random vectors; the random variables $A_{i, 1}$ and $A_{i, 2}$ satisfy $\max\{|A_{i, 1}|, |A_{i, 2}|\} \leq B$ almost surely for some $B < \infty$ and can be either lattice or non-lattice. Comparing \eqref{eq:AWbound} to \eqref{eq:expNL} and \eqref{eq:chagantyLat}, we see that \eqref{eq:AWbound} is asymptotically tight in the order of the prefactor. However, the constant $E_{\mr{AW}}$ depends on an unspecified universal constant.   
    Since our achievability proof in \secref{sec:proofach} relies only on the fact that the prefactor in the LD bound is a bounded constant (see \eqref{eq:strongbound}, below) and that $\imath(X; Y)$ is bounded for DMCs, \eqref{eq:AWbound} is also applicable in our achievability proof. If one seeks to derive the $O(1)$ term in \eqref{eq:achievability}, the tightness of the prefactor used in the probability bound is important. 
	
	\subsection{Proof of \thmref{thm:mainAch}} \label{sec:proofach}
	The proof consists of two parts and follows steps similar to the achievability proof in \cite{moulin2017log}. First, we derive a refined asymptotic achievability bound for an arbitrary input distribution $\PX \in \mc{P}$. Then, we optimize that achievability bound over all $\PX \in \mc{P}$. 
	
	\begin{lemma}\label{lem:achP}
	Suppose that $\epsilon_n$ is an SMD sequence \eqref{eq:range}. Fix some $\PX \in \mc{P}$ such that 
    $(\PX, \W)$ is a nonsingular pair and $V_{\mr{u}}(\PX) > 0$ for all $n$.  It holds that
\begin{IEEEeqnarray}{rCl}
    \IEEEeqnarraymulticol{3}{l}{\log M^*(n, \epsilon_n)} \notag \\
    &\geq& n I(\PX) - \sqrt{n V_{\mathrm{u}}(\PX)} Q^{-1}(\epsilon_n) +  \frac{1}{2} \log n \notag \\
    &&+ Q^{-1}(\epsilon_n)^2 
    \bigg( \frac{\ske_{\mathrm{u}}(\PX) \sqrt{V_{\mr{u}}(\PX)}}{6} + \frac{1 - \eta(\PX)}{2(1+\eta(\PX))}  \bigg) \notag \\
    &&+\bigo{\frac{Q^{-1}(\epsilon_n)^3}{\sqrt{n}}} + O(1).\label{eq:achievP}
\end{IEEEeqnarray}
	\end{lemma}
	We require  $V_{\mr{u}}(\PX) > 0$ in order to apply Theorems \ref{thm:moderatePetrov}--\ref{thm:lattice}. 
\begin{IEEEproof}[Proof of \lemref{lem:achP}]
We generate $M$ i.i.d. codeword according to the input distribution $\PX^n$ and employ a maximum likelihood decoder. Let $W$ be the transmitted message that is equiprobably distributed on $[M]$, and let $\hat{W}$ be the decoder output.
Define the random variables 
\begin{align}
    Z &\triangleq  \imath(\mb{X}; \mb{Y}) \\
    \overline{Z} &\triangleq \imath(\overline{\mb{X}}; \mb{Y}),
\end{align}
where $(\mb{X}, \overline{\mb{X}}, \mb{Y})$ is distributed according to
$P_{\mb{X}, \overline{\mb{X}}, \mb{Y}}(\mb{x}, \overline{\mb{x}}, \mb{y}) = \PX^n(\mb{x}^n) \PX^n(\overline{\mb{x}}) \W^n(\mb{y}|\mb{x})$. 
The random variable $\overline{Z}$ represents the information density obtained from a random codeword that is independent of both transmitted codeword $\mb{X}$ and the received channel output $\mb{Y}$.

\subsubsection{Error analysis}
Fix a threshold value $\tau_n$ 
\begin{align}
    \tau_n \triangleq n I(\PX) - \sqrt{n V_{\mr{u}}(\PX)} t_n, \label{eq:taun}
\end{align}
with $t_n$ to be specified in \eqref{eq:tildeeps}, below. Define the event
\begin{align}
    \mathcal{D} \triangleq \{Z < \tau_n\}. \label{eq:tau}
\end{align}

We weaken the RCU bound from \cite[Th.~16]{polyanskiy2010Channel}  and bound the average error probability as
\begin{IEEEeqnarray}{rCl}
   \IEEEeqnarraymulticol{3}{l}{\Prob{\hat{W} \neq W}}\notag \\
   &\leq& \E{\min \left\{1, M-1 \, \Prob{\overline{Z} \geq Z | \mb{X}, \mb{Y}} \right\}} \label{eq:RCUpol}\\
    &\leq& \Prob{\mc{D}} + (M-1) \Prob{\overline{Z} \geq Z \geq \tau_n}. \label{eq:RCU}
\end{IEEEeqnarray}
Define the function
\begin{align}
    h(x) \triangleq \frac{1}{\sqrt{2 \pi}} \exp \cB{ -\frac{Q^{-1}(x)^2}{2}}
\end{align}
and the sequences
\begin{align}
    h_n &\triangleq  \frac{1}{\sqrt{n V_{\mr{u}}(\PX)}} h(\epsilon_n) \label{eq:hn} \\
    \tilde{\epsilon}_n &\triangleq \epsilon_n - h_n. \label{eq:tildeepsn}
\end{align}

Below, we show that the first and second terms in \eqref{eq:RCU} are bounded by $\tilde{\epsilon}_n$ and $h_n$, respectively.
Here, $h_n$ is chosen so that $\log M$ is maximized up to the $O(Q^{-1}(\epsilon_n)^2)$ term given that the right-hand side of \eqref{eq:RCU} is equal to $\epsilon_n$. 

We set $t_n$ in \eqref{eq:taun} as
\begin{align}
    \Prob{\mc{D}} &= \Prob{\frac{Z - n I(\PX)}{\sqrt{n V_u(\PX)}} \leq - t_n} = \tilde{\epsilon}_n. \label{eq:tildeeps}
\end{align}
Since the channel is a DMC, the random variables $\imath(X_i; Y_i)$ are supported on a finite alphabet, thereby satisfying Cram\'er's condition in \eqref{eq:Cramercond}. Further, since, by assumption, $V_{\mr{u}}(P_X) > 0$, \thmref{thm:moderatePetrov} and \lemref{lem:cornish} are applicable.
Applying the MD result in \lemref{lem:cornish} to \eqref{eq:tildeeps}, we get
\begin{align}
    t_n &= Q^{-1}(\tilde{\epsilon}_n) - \frac{\ske_{\mr{u}}(\PX) Q^{-1}(\tilde{\epsilon}_n)^2}{6 \sqrt{n}}  \notag \\
    &\quad + \bigo{\frac{Q^{-1}(\tilde{\epsilon}_n)^3}{n}} +  \bigo{\frac{1}{\sqrt{n}}}. \label{eq:tneq1}
\end{align}
We compute the first two derivatives of the $Q^{-1}(x)$ function~as
    \begin{align}
        (Q^{-1})'(x) &=  \frac{1}{Q'(Q^{-1}(x))} = \frac{-1}{h(x)} \label{eq:Qderiv1}\\
        (Q^{-1})''(x) &= - \frac{Q^{-1}(x)}{h(x)^2}. \label{eq:Qderiv2}
    \end{align}
By taking the Taylor series expansion of $Q^{-1}(\cdot)$ around $\epsilon_n$ and using \eqref{eq:tneq1}--\eqref{eq:Qderiv2}, we get
\begin{align}
    t_n &= Q^{-1}(\epsilon_n) - \frac{\ske_{\mr{u}}(\PX) Q^{-1}({\epsilon}_n)^2}{6 \sqrt{n}}  \notag \\
    &\quad + \bigo{\frac{Q^{-1}(\epsilon_n)^3}{n}} + \bigo{\frac{1}{\sqrt{n}}}. \label{eq:tn}
\end{align}

Next, we bound the probability $\Prob{\overline{Z} \geq Z \geq \tau_n}$. Define the random vector $\mb{U} \triangleq (U_1, U_2) = (Z, \overline{Z} - Z)$. 
and the sequence 
\begin{align}
    \mb{a}_n = (a_{n, 1}, a_{n, 2}) = \left(\frac{\tau_n}{n}, 0\right). \label{eq:an}
\end{align}
Applying Theorems \ref{thm:chaganty} or \ref{thm:lattice} (depending on whether $\imath(X; Y)$ is non-lattice or lattice), we get
\begin{align}
    \Prob{\overline{Z} \geq Z \geq \tau_n} &= \Prob{\mb{U} \geq n \mb{a}_n} \\
    &\leq \frac{E}{n} \exp\{-n \Lambda(\mb{a}_n)\} (1 + o(1)), \label{eq:strongbound}
\end{align}
where 
\begin{align}
    E &= \begin{cases} E_{\mr{NL}} &\text{if } \imath(X; Y) \text{ is non-lattice} \\ E_{\mr{L}} &\text{if }\imath(X; Y) \text{ is lattice.}
    \end{cases} \\
    \Lambda(\mb{a}_n) &= \sup_{\mb{s}_n \in \mathbb{R}^2} \{\langle \mb{a}_n, \mb{s}_n 
    \rangle - \kappa(\mb{s}_n) \} \\
    \kappa(\mb{s}_n) &= \frac{1}{n} \log \E{\exp\{\langle \mb{s}_n, \mb{U} \rangle\}}.
\end{align}
Note that the functions $\kappa(\cdot)$ and $\Lambda(\cdot)$ do not depend on $n$ since $\mb{U}$ is an i.i.d.\ sum.
The rate function $\Lambda(\mb{a}_n)$ has the Taylor series expansion
\begin{align}
    \Lambda(\mb{a}_n) &= I(\PX) + (a_{n, 1} - I(\PX)) + \frac{(a_{n, 1} - I(\PX))^2}{(1 + \eta(\PX)) V_{\mr{u}}(\PX)}   \notag \\
    &\quad + O(|a_{n, 1} - I(\PX) |^3)\\
    &= a_{n, 1} + \frac{1}{n} \frac{Q^{-1}(\epsilon_n)^2}{1 + \eta(\PX)} + \bigo{\frac{Q^{-1}(\epsilon_n)^3}{n^{3/2}}} + \bigo{\frac{1}{n}}. \label{eq:lambdaexp}
\end{align}

In the application of Theorems \ref{thm:chaganty} and \ref{thm:lattice}, conditions (S), (NL), and (L) are already satisfied since $U_1$ and $U_2$ have finite supports. The verification of condition (ND) and the derivation of \eqref{eq:lambdaexp} appear in \appref{app:NDproof}.

We set
\begin{align}
    \log M &= n I(\PX) - \sqrt{n V_{\mathrm{u}}(\PX)} Q^{-1}(\epsilon_n) +  \frac{1}{2} \log n \notag \\
    &+ Q^{-1}(\epsilon_n)^2 
    \bigg( \frac{\ske_{\mathrm{u}}(\PX) \sqrt{V_{\mr{u}}(\PX)}}{6}  +\frac{1 - \eta(\PX)}{2(1+\eta(\PX))}  \bigg) \notag \\
    &+ \bigo{\frac{Q^{-1}(\epsilon_n)^3}{\sqrt{n}}} + O(1).
\end{align}
We put \eqref{eq:taun} into \eqref{eq:tn}, and then \eqref{eq:an} into \eqref{eq:strongbound} to bound the probability $\Prob{\overline{Z} \geq Z \geq \tau_n}$. Then, from the expansion \eqref{eq:lambdaexp}, we get
\begin{align}
    M \Prob{\overline{Z} \geq Z \geq \tau_n} \leq h_n, \label{eq:Mprob}
\end{align}
where $h_n$ is defined in \eqref{eq:hn}. Combining \eqref{eq:RCU}, \eqref{eq:tildeeps}, and \eqref{eq:Mprob} completes the proof of \lemref{lem:achP}.
\end{IEEEproof}

To complete the proof of \thmref{thm:mainAch}, it only remains to maximize the right-hand side of \eqref{eq:achievP} over $\PX \in \mc{P}$. The following arguments extend the proof of \cite[Lemma 9]{moulin2017log} to the MD regime.
Define 
\begin{align}
    G(\PX) \triangleq -\sqrt{V_{\mr{u}}(\PX)} Q^{-1}(\epsilon_n).
\end{align}
Let $\mb{h}$ be a vector whose components approach zero with a rate $\bigo{\frac{Q^{-1}(\epsilon_n)}{\sqrt{n}}}$ satisfying $\mb{h}^\top \bs{1} = 0$. Let $f(\mb{h})$ be the right-hand side of  \eqref{eq:achievP} evaluated at $\PX = \PXs + \mb{h} \in \mc{P}$ for some $\PXs \in \mc{P}^*$. We expand $f(\mb{h})$ by using the second-, first-, and zeroth-order Taylor series expansions of $I(P_X)$, $G(P_X)$, and $\ske_{\mr{u}}(P_X)$ and $\eta(P_X)$, respectively
\begin{align}
    f(\mb{h}) &\triangleq n I(\PXs) + n \mb{h}^\top \nabla I(\PXs) + \frac{n}{2} \mb{h}^\top \nabla^2 I(\PXs) \mb{h} \notag \\
    &\quad + O(n \| \mb{h} \|_{\infty}^3) 
    + \sqrt{n} G(\PXs) + \sqrt{n} \mb{h}^\top \nabla G(\PXs) \notag \\
    &\quad + \sqrt{n} O(\| \mb{h} \|_{\infty}^2)+ \frac{1}{2} \log n \notag \\
    &\quad + Q^{-1}(\epsilon_n)^2 
    \bigg( \frac{\ske_{\mathrm{u}}(\PXs) \sqrt{V_{\mr{u}}(\PXs)}}{6} + \frac{1 - \eta(\PXs)}{2(1+\eta(\PXs))}  \bigg) \notag \\
    &\quad + \bigo{\frac{Q^{-1}(\epsilon_n)^3}{\sqrt{n}}} + O(1) \label{eq:fexp} \\
    &=  n \mb{h}^\top \nabla I(\PXs)  + \frac{n}{2} \mb{h}^\top \nabla^2 I(\PXs) \mb{h} \notag \\
    &\quad + \sqrt{n}\mb{h}^\top \nabla G(\PXs) + b, \label{eq:163}
\end{align}
where $b$ is the right-hand side of \eqref{eq:achievP} evaluated at $P_X = \PXs$. Here, the terms involving the first derivatives of $\ske_{\mr{u}}(P_X)$ and $\eta(P_X)$ are absorbed in the $O(\cdot)$ terms in \eqref{eq:fexp}.

For every $\mb{h}$ such that $\PXs + \mb{h}$ is a valid probability distribution, $\mb{h}^\top \nabla I(\PXs) \leq 0$ by \eqref{eq:derivI} and \cite[Th.~4.5.1]{gallager1968book}; equality holds if and only if $\mb{h}$ is supported on $\mc{X}^\dagger$. Therefore, for any valid $\mb{h}$ and $n$ large enough, 
\begin{align}
    &f(\mb{h}) \notag \\
    &\leq \sup_{\substack{\mb{h'} \colon \mb{h}'^\top \boldsymbol{1} = 0 \\ \mr{supp}(\mb{h}') \subseteq \mc{X}^\dagger\\ \mb{h}'_{\mc{X}^\dagger \setminus \mc{X}^*} \geq \bs{0}}} \bigg\{ - \frac{n}{2} \mb{h'}^\top \ms{J}_{\mc{X}^\dagger} \mb{h'} + \sqrt{n}\mb{h'}^\top \nabla G(\PXs) + b \bigg\}. \label{eq:fg}
\end{align}
Instead of maximizing over all valid $\mb{h}$ as in \eqref{eq:fg}, we further restrict $\mr{supp}(\mb{h})$ to $\mc{X}^*$, which yields the optimization problem
\begin{align}
    \sup_{\substack{\mb{h'} \colon \mb{h}'^\top \boldsymbol{1} = 0 \\ \mr{supp}(\mb{h'}) \subseteq \mc{X}^*} } \bigg\{ - \frac{n}{2} \mb{h'}^\top \ms{J}_{\mc{X}^*} \mb{h'} + \sqrt{n}\mb{h'}^\top \nabla G(\PXs) + b \bigg\}. \label{eq:fg2}
\end{align}
The following arguments follow from the proof of \cite[Lemma~9]{moulin2017log}. For any $\mb{h'}$ in the kernel of $\ms{J}_{\mc{X}^*}$, the first two terms in \eqref{eq:fg} are zero. Therefore, the optimal $\mb{h}^*$ must lie in the row space of $\ms{J}_{\mc{X}^*}$. From \cite[eq.~(2.48)]{moulin2017log}, $\mb{h'}^\top \nabla V_{\mr{u}}(\PXs) = \mb{h'}^\top \nabla V(\PXs)$ for any feasible $\mb{h'}$. Thus, the problem captured by \eqref{eq:fg2} reduces to \eqref{eq:opt}, whose solution is given in \eqref{eq:hstar}.
See \appref{app:lagrange} for details. Notice that if $\mc{X}^\dagger = \mc{X}^*$ holds, then the right-hand sides of \eqref{eq:fg} and \eqref{eq:fg2} are equal, meaning that for DMCs with $\mc{X}^\dagger = \mc{X}^*$, \eqref{eq:hstar} yields the optimal direction (up to the skewness term) with respect to maximizing the RCU bound. 

Combining the value of $b$ and the value of \eqref{eq:opt} gives the maximum of \eqref{eq:achievP} over all input distributions $\PX \in \mc{P}$ and completes the proof of \thmref{thm:mainAch}. 


	\subsection{Proof of \thmref{thm:mainConv}} \label{sec:proofconv}
 
	The proof analyzes Tomamichel and Tan's non-asymptotic converse bound from \cite[Prop. 6]{tomamichel2013converse} using techniques from \cite{moulin2017log}. 
	
	The main difference between our proof and Moulin's proof in \cite{moulin2017log} is that while Moulin analyzes the meta-converse bound \cite[Th.~27]{polyanskiy2010Channel}, we analyze a relaxation of the meta-converse given in \lemref{lem:tom}, below. In general, the analysis of the meta-converse is more involved since it requires splitting the code into subcodes according to the types of the codewords and then carefully combining the bounds for each subcode. The advantage of \lemref{lem:tom} over the meta-converse bound is that the optimization problem in \lemref{lem:tom} can be converted into a simpler single-letter minimax problem as we show in \lemref{lem:Dsasymp}, and the type-splitting step is avoided. 
		A similar simplification to a single-letter problem using the meta-converse is possible (i) under the average error probability criterion for channels that satisfy certain symmetry conditions  \cite[Th.~28]{polyanskiy2010Channel} (e.g., Cover--Thomas-symmetric channels satisfy these symmetry conditions) and (ii) under the maximal error probability criterion for arbitrary DMCs \cite[Th.~31]{polyanskiy2010Channel}.  
		While both approaches yield the same upper bound $\overline{S}$ on the skewness (in the CLT regime in Moulin's work and in the MD regime in our work), we note that \lemref{lem:tom} is not tight enough to obtain the tightest $O(1)$ term in the converse \eqref{eq:converse}, which we do not focus on here.

	
	
	We define the divergence spectrum  \cite[Ch. 4]{hanbook}, \cite{tomamichel2013converse}, which gives a lower bound on the minimum type-II error probability of the binary hypothesis test, $\beta_{1-\epsilon}(P, Q)$, 
	\begin{align}
	    D_s^{\epsilon}(P \| Q) \triangleq \sup \left\{ \gamma \in \mathbb{R} \colon \Prob{\log \frac{P(X)}{Q(X)} \leq \gamma} \leq \epsilon \right\},
	\end{align}
	where $\epsilon \in (0, 1)$, $P, Q \in \mc{P}$, and $X \sim P$. 

 We define the function $\xi^{\epsilon_n} \colon \mc{P} \times \mc{P} \to \mathbb{R}$ in \eqref{eq:xi} below, where $P_X^{(\mr{o})} \to \W \to P_Y^{(\mr{o})}$. The function $\xi^{\epsilon_n}$ is related to the asymptotic expansion of the divergence spectrum. In particular, for $P_X \in \mc{P}$ and $P_X^* \in \mc{P}^*$, it evaluates to
 \begin{align}
 &\xi^{\epsilon_n}(P_X, P_X) = n I(P_X) - \sqrt{n V(P_X)} \Qinv \notag \\
        &\quad + \frac{\ske(P_X) \sqrt{V(P_X)}}{6} \Qinv^2 \\
&\xi^{\epsilon_n}(\PXs, \PXs) = \notag \\
        &nC - \sqrt{n V_{\epsilon_n}} Q^{-1}(\epsilon_n) +  
    \frac{\ske_{\mathrm{u}}(\PXs) \sqrt{V_{\epsilon_n}}}{6} Q^{-1}(\epsilon_n)^2.
 \end{align}
 Note that by \cite[Lemma~2]{moulin2017log}, $\ske_{\mr{u}}(\PXs) = \ske(\PXs)$.
 
    \newcounter{MYtempeqncnt}
    \begin{figure*}[!t]
\normalsize
\setcounter{MYtempeqncnt}{\value{equation} + 1}
\begin{align}
        \xi^{\epsilon_n}(\PXi, \PXo) &\triangleq  n D(\W \| P_Y^{(\mr{o})} | \PXi) - \sqrt{n V(\W \| P_Y^{(\mr{o})} | \PXi)} Q^{-1}(\epsilon_n) \notag \\
        &\quad + \frac{\ske(\W \| P_Y^{(\mr{o})} | \PXi)\sqrt{V(\W \| P_Y^{(\mr{o})} | \PXi)}}{6} Q^{-1}(\epsilon_n)^2 \label{eq:xi}
\end{align}
\setcounter{equation}{\value{MYtempeqncnt}}
\hrulefill
\vspace*{4pt}
\end{figure*}
	
	The main tools to prove \thmref{thm:mainConv}, presented below, are \lemref{lem:Dsasymp}, which gives an asymptotic expansion of the divergence spectrum in the MD regime and \lemref{lem:tom}, which gives a channel coding converse based on the divergence spectrum. 
	

    \begin{lemma}{\label{lem:Dsasymp}}
    Let $a > 0$. Define $\mc{Q}(a) \triangleq \{Q_Y \in \mc{Q} \colon Q_Y(y) \geq a \, \, \forall y \in \mc{Y} \text{ and } \exists \, Q_X \in \mc{P} \text{ such that } Q_X \to \W \to Q_Y\}$. Assume that $\{\epsilon_n\}_{n = 1}^{\infty}$ is an SMD sequence \eqref{eq:range}. 
    
    (i)\,\, Then, for $n$ large enough, there exist constants $K_1$ and $K_2$ that depend only on $P_{Y|X}$ and $a$ such that
    \begin{align}
        &\max_{\substack{\mb{x} \in \mc{X}^n \\ Q_Y \in \mc{Q}(a)}} \left\lvert D_s^{\epsilon_n}(\Wxn \| \QY^n) - \xi^{\epsilon_n}(\Phat, Q_X) \right \rvert \notag \\
        & \leq K_1 \frac{|Q^{-1}(\epsilon_n)|^3}{\sqrt{n}} + K_2. \label{eq:Dsasy}
    \end{align}
    
    (ii)\,\, Let $\Phat \to \W \to \hat{Q}_{\mb{x}}$ for all $\mb{x} \in \mc{X}^n$. For $n$ large enough, there exist constants $K_3$ and $K_4$ that depend only on $P_{Y|X}$ such that
    \begin{align}
        &\max_{\mb{x} \in \mc{X}^n } \big\lvert D_s^{\epsilon_n}(\Wxn \| \hat{Q}_{\mb{x}}^n) - \xi^{\epsilon_n}(\Phat, \Phat) \big \rvert \notag \\
        & \leq K_3 \frac{|Q^{-1}(\epsilon_n)|^3}{\sqrt{n}} + K_4. \label{eq:Dsasy2}
    \end{align}
    \end{lemma}
    \begin{IEEEproof}
    See \appref{app:Dsasymp}.
    \end{IEEEproof}
    Note that the argument of the absolute value on the left side of \eqref{eq:Dsasy} depends on $\mb{x}$ only through its empirical distribution $P_{\mb{x}} \in \mc{P}_n$. By \lemref{lem:Dsasymp} (i), for $P_Y^{(\mr{o})}$ such that $P_Y^{(\mr{o})}(y) > 0$ for all $y \in \mc{Y}$, $\xi^{\epsilon_n}(\Phat, P_X^{(\mr{o})}) = D_s^{\epsilon_n}(\Wxn \| (P_Y^{(\mr{o})})^n) + \bigo{\frac{\Qinv^3}{\sqrt{n}}} + O(1)$.
	
	\begin{lemma}[{\cite[Prop. 6]{tomamichel2013converse}}] \label{lem:tom}
	Let $\epsilon_n$ be any sequence in $(0, 1)$ and $\W$ be a DMC. Then, for any $\delta_n \in (0, 1-\epsilon_n)$, we have
	\begin{align}
	    &\log M^*(n, \epsilon_n) \notag \\
	    &\leq \min_{\QY^{(n)} \in \mc{Q}^n} \max_{\mb{x} \in \mc{X}^n} D_s^{\epsilon_n + \delta_n}(\Wxn \| \QY^{(n)}) - \log \delta_n, \label{eq:tomeq}
	\end{align}
	where $\Wxn = \prod_{i = 1}^n P_{Y|X = x_i}$.
	\end{lemma}
 
    Define
    \begin{align}
        \rho_n \triangleq  \max \left\{\frac{c_0 |Q^{-1}(\epsilon_n)|}{\sqrt{n}}, \frac{\log^2(n)}{\sqrt{n} |Q^{-1}(\epsilon_n)|} \right\}, \label{eq:rhon}
    \end{align}
     where
   $c_0 > 0$ is a sufficiently large constant that will be determined later. 
 Define the set of input distributions
    \begin{align}
       \mc{P}^*(\nu) &\triangleq \left \{ \PX \in \mc{P} \colon \norm{ \PX - \PXs }_{\infty} \leq \nu  \text{ for some } \PXs \in \mc{P}^* \right\}.
    \end{align}
    The auxiliary output distribution
    \begin{align}
        {Q_Y^{(n)}}^* = \frac{1}{2} (\QYs + {\tilde{\mb{h}}})^n + \frac{1}{2} \sum_{\hat{P}_{\mb{x}} \in \mc{P}_n} \frac{1}{|\mc{P}_n|} \hat{Q}_{\mb{x}}^n, \label{eq:convexQY} 
    \end{align}
    which is a convex combination of product distributions, is inspired by Hayashi \cite[Th.~2]{hayashi} and Tomamichel and Tan \cite[eq.~6]{tomamichel2013converse}.
    Here, $\QYs$ is the unique capacity-achieving output distribution, and $\Phat \to \W \to \hat{Q}_{\mb{x}}$. The vector $\tilde{\mb{h}}$, supported on $\mc{Y}$, is intended to be optimized under the constraints that its entries sum to 0 and $\tilde{\mb{h}} \to \bs{0}$. 
    The first term in \eqref{eq:convexQY} targets the sequences $\mb{x} \in \mc{X}^n$ in the maximization in \eqref{eq:tomeq} such that the corresponding empirical distribution is close to $\PXs$ in the sense that $\hat{P}_{\mb{x}} \in \mc{P}^*(\rho_n)$, where $\rho_n$ is given in \eqref{eq:rhon}. The second term in \eqref{eq:convexQY} targets the sequences $\mb{x} \in \mc{X}^n$ that are far away from $\PXs$, i.e., $\hat{P}_{\mb{x}} \notin {\mc{P}^*(\rho_n)}$.

    We upper bound the right-hand side of \eqref{eq:tomeq} by setting the auxiliary output distribution $Q_Y^{(n)}$ to ${Q_Y^{(n)}}^*$ and get
    \begin{align}
        \log M^*(n, \epsilon_n) \leq \max_{\mb{x} \in \mc{X}^n} D_s^{\epsilon_n + \delta_n}(\Wxn \| {\QY^{(n)}}^*) - \log \delta_n. \label{eq:tomeqfixed}        
    \end{align}

    To bound the right-hand side of \eqref{eq:tomeqfixed}, we present three auxiliary lemmas.

       The following lemma by Tomamichel and Tan bounds $D_s^{\epsilon_n}(P \| Q)$ where $Q$ is a convex combination of distribitions.
    \begin{lemma}[{\cite[Lemma~3]{tomamichel2013converse}}] \label{lem:tom3}
        Let $\epsilon_n$ be any sequence in $(0, 1)$. Let $\mc{I}$ be a countable index set. Let $q$ be a distribution on $\mc{I}$. Let $P, \{Q^i\}_{i \in \mc{I}}$ be distributions on a common alphabet. Let $Q = \sum_{i \in \mc{I}} q(i) Q^i$. Then,
        \begin{align}
            D_s^{\epsilon_n}(P \| Q) \leq \inf_{i \in \mc{I}} \{ D_s^{\epsilon_n}(P \| Q^i) - \log q(i)\}.
        \end{align}
    \end{lemma}

      Define
    \begin{align}
        \Delta_x &\triangleq D(\Wx \| \QYs) - C  \quad \forall \, x \in \mc{X}\\
        \hi &\triangleq \PXi - \PXs \\
        \ho &\triangleq \PXo - \PXs \\
        \tilde{\mb{v}}_x &\triangleq V(\Wx \| \QYs) \quad \forall \, x \in \mc{X} \label{eq:vtilde}\\
        \overline{\mb{g}} &\triangleq -\frac{Q^{-1}(\epsilon_n)}{2\sqrt{V_{\epsilon_n}}} 
        \overline{\mb{v}}(\PXs) \\
        \tilde{\mb{g}}  &\triangleq -\frac{Q^{-1}(\epsilon_n)}{2\sqrt{V_{\epsilon_n}}} \tilde{\mb{v}} \\
        \Gamma(\hi, \ho) &\triangleq \frac{1}{2} {\mb{h}^{(\mr{o})}}^\top \ms{J}_{\mc{X}^*} {\mb{h}^{(\mr{o})}} +  \overline{\mb{g}}^\top \mb{h}^{(\mr{o})} \notag \\
        &\quad -  {\mb{h}^{(\mr{i})}}^\top (\ms{J}_{\mc{X}^\dagger}  \mb{h}^{(\mr{o})} - \tilde{\mb{g}}), 
    \end{align}
    where $\PXs$ is a dispersion-achieving input distribution, and $\overline{\mb{v}}(\PXs)$ is defined in \eqref{eq:barvP}. Recall the definitions of $\ms{J}$ and $\tilde{\ms{J}}$ from \eqref{eq:J} and \eqref{eq:tildeJ}. The quadratic form $\Gamma(\hi, \ho)$ arises after taking the second-order Taylor series expansion of the function $\xi^{\epsilon_n}(\PXi, \PXo)$ around the point $(\PXs, \PXs)$. Lemma~\ref{lem:moulinsaddle}, below, gives the saddlepoint solution to the minimax of $\Gamma(\hi, \ho)$, where the maximization is over $\hi$ and the minimization is over $\ho$.

    \begin{lemma}[{\cite[Lemma~14]{moulin2017log}}] \label{lem:moulinsaddle}
    Consider the minimax problem  
    \begin{align}
         \min_{\substack{\ho \colon \mr{supp}(\ho) \subseteq \mc{X}^* \\ {\ho}^\top \bs{1} = 0 \\ \ho \in \mr{row}(\ms{J}_{\mc{X}^*})}} \max_{\substack{\hi \colon \mr{supp}(\hi) \subseteq \mc{X}^\dagger \\ {\hi}^\top \bs{1} = 0
        \\
        \hi_{\mc{X}^\dagger \setminus \mc{X}^*} \geq \bs{0}}}  \Gamma(\hi, \ho). \label{eq:minmaxGamma}
    \end{align}
    The point $({\hi}^*, {\ho}^*)$, where $\hi = \tilde{\ms{J}} \mb{g}$ and $\ho = \tilde{\ms{J}} \tilde{\mb{g}}$, and $\mb{g} = \overline{\mb{g}} + \tilde{\mb{g}}$, admits the saddlepoint property 
    \begin{align}
        \Gamma(\hi, {\ho}^*) = \Gamma({\hi}^*, {\ho}^*) \leq \Gamma({\hi}^*, \ho) 
    \end{align}
    for all feasible $\hi$ and $\ho$. The value of the saddlepoint is $\frac{1}{2} \mb{g}^\top \tilde{\ms{J}} \mb{g} - \frac{1}{2} \overline{\mb{g}}^\top \tilde{\ms{J}} \overline{\mb{g}}$. 
    \end{lemma}

    In \eqref{eq:minmaxGamma}, the constraints ${\hi}^\top \bs{1} = 0$ and ${\ho}^\top \bs{1} = 0$ are due to $\PXi, \PXo$, and $\PXs$ being distributions. The constraints $\mr{supp}(\hi) \subseteq \mc{X}^\dagger$ and $\hi_{\mc{X}^\dagger \setminus \mc{X}^*} \geq \bs{0}$ are due to the optimality of $\PXs$. The constraints $\mr{supp}(\ho) \subseteq \mc{X}^*$ and $\ho \in \mr{row}(\ms{J}_{\mc{X}^*})$ are by our choice.
 
     Define the function
     \begin{align}
     {\psi^{\epsilon_n}}^*(\PXs) \triangleq \xi^{\epsilon_n}(\PXs, \PXs) + Q^{-1}(\epsilon_n)^2  (A_0(\PXs) - A_1(\PXs)). \label{eq:saddlevalue}
    \end{align}
    
     The following lemma bounds $\xi^{\epsilon_n}(\PX, \PX)$ for $\PX \in \mc{P}$ that are sufficiently far away from the set of dispersion-achieving input distributions.
    \begin{lemma}\label{lem:xibound}
        Let $\epsilon_n$ be an SMD sequence. There exist constants $c_0 > 0$ and $c_1 > 0$ such that for all $\rho_n \geq \frac{c_0 |\Qinv|}{\sqrt{n}}$ and for all $\PX \notin {\mc{P}^*\left(\rho_n\right)}$,
        \begin{align}
            &\xi^{\epsilon_n}(\PX, \PX) \notag \\
            &\leq \max_{\PXs \in \mc{P}^*} {\psi^{\epsilon_n}}^*(\PXs)   - c_1 \sqrt{n} \rho_n |\Qinv| (1 + o(1)). \label{eq:xic1}
        \end{align}
    \end{lemma}
    \begin{IEEEproof}
        The proof extends the result in \cite[Lemma~9 (iii)]{moulin2017log} to SMD sequences and uses the quadratic decay property of mutual information, which is formalized in \cite[Th.~1]{cao2023}. See \appref{app:xibound} for details.
    \end{IEEEproof}

  We bound the right-hand side of \eqref{eq:tomeqfixed} in two steps. 
\begin{enumerate}
\item We optimize the value of the perturbation $\tilde{\mb{h}}$ in the auxiliary distribution given in \eqref{eq:convexQY}. To do this, we take the Taylor series expansion of $\xi^{\epsilon_n}(\PXi, \PXo)$ around the point $(\PXs, \PXs)$ and then use \lemref{lem:moulinsaddle}.
\item We bound the right-hand side of \eqref{eq:tomeqfixed} separately depending on whether $\Phat \in \mc{P}^*(\rho_n)$ or $\Phat \notin \mc{P}^*(\rho_n)$. For the case $\Phat \in \mc{P}^*(\rho_n)$, we apply Lemmas~\ref{lem:tom3}, \ref{lem:Dsasymp}, and \ref{lem:moulinsaddle} in order. For the case $\Phat \notin \mc{P}^*(\rho_n)$, we apply Lemmas~\ref{lem:tom3}, \ref{lem:Dsasymp}, and \ref{lem:xibound} in order.
\end{enumerate}
In the following, we detail these two proof steps.

\subsubsection{Optimization of the value of $\tilde{\mb{h}}$}    
    The minimax of the first term $n D(\W \| P_Y^{(\mr{o})} | \PXi)$
    in \eqref{eq:xi} satisfies the saddlepoint property (e.g., \cite[Cor. 4.2]{polyanskiyLectureNotes})
    \begin{align}
        D(\W \| \QYs | \PX) \leq D(\W \| \QYs | \PXdag) \leq D(\W \| \QY | \PXdag)
    \end{align}
    for all $\PX \in \mc{P}, \QY \in \mc{Q}$, where $\PXdag \in \mc{P}^\dagger$ is a capacity-achieving input distribution, and $\QYs$ is the capacity-achieving output distribution; the minimax solution for the first term in \eqref{eq:xi} is $\PXi = \PXo = \PXdag$, and the saddlepoint value is $D(\W \| \QYs | P^{\dagger}) = C$. 
    
    Let $\PXs \in \mc{P}^*$ be a dispersion-achieving input distribution. To set the perturbation $\tilde{\mb{h}}$ in \eqref{eq:tomeq}, we consider the problem
    \begin{align}
        \min_{\substack{\PXo \in \mc{P} \colon \\ \norm{ \PXo - \PXs }_{\infty} \leq \rho_n}} \,\, \max_{\substack{\PXi \in \mc{P} \colon \\ \norm{ \PXi - \PXs }_{\infty} \leq \rho_n}} \xi^{\epsilon_n}(\PXi,  \PXo) \label{eq:saddleN}.
    \end{align}
   To be able to apply \lemref{lem:moulinsaddle}, we further restrict the perturbation $\PXo - \PXs \in \mr{row}(\ms{J}_{\mc{X}^*})$, which yields the minimax problem
    \begin{align}
        \min_{\substack{\PXo \in \mc{P}\colon \norm{ \PXo - \PXs }_{\infty} \leq \rho_n \\ \mr{supp}(\PXo - \PXs) \subseteq \mc{X}^* \\
        \PXo - \PXs \in \mr{row}(\ms{J}_{\mc{X}^*})}} \,\, \max_{\substack{\PXi \in \mc{P} \colon \\ \norm{ \PXi - \PXs }_{\infty} \leq \rho_n}} \xi^{\epsilon_n}(\PXi,  \PXo). \label{eq:saddleNres}
    \end{align}
    Note that \eqref{eq:saddleNres} gives an upper bound on \eqref{eq:saddleN}.
    
    Assume that $\norm{\hi}_{\infty} \leq \rho_n$, $\norm{\ho}_{\infty} \leq \rho_n$ and $\mr{supp}(\ho) \subseteq \mc{X}^*$. 
    In \cite[Lemma~12]{moulin2017log}, Moulin derives the Taylor series expansion of $\xi^{\epsilon_n}(\PXi, \PXo)$ around the point $(\PXs, \PXs)$ and obtains
    \begin{align}
        \xi^{\epsilon_n}(\PXi, \PXo) &= \xi^{\epsilon_n}(\PXs, \PXs) + n {\mb{h}^{(\mr{i})}}^\top \Delta + \frac{n}{2} {\mb{h}^{(\mr{o})}}^\top \ms{J}_{\mc{X}^*} {\mb{h}^{(\mr{o})}} \notag \\
        &+ \sqrt{n} \overline{\mb{g}}^\top \mb{h}^{(\mr{o})} - \sqrt{n} {\mb{h}^{(\mr{i})}}^\top (\ms{J}_{\mc{X}} \sqrt{n} \mb{h}^{(\mr{o})} - \tilde{\mb{g}}) \notag \\
        &+ O(\rho_n^2 \sqrt{n} Q^{-1}(\epsilon_n))\label{eq:taylorxi}. 
    \end{align}
    The term $O(\rho_n^2 \sqrt{n} Q^{-1}(\epsilon_n))$ in \eqref{eq:taylorxi} is bounded by $\bigo{\frac{\Qinv^3}{\sqrt{n}}} + O(1)$. 
    
    The following arguments follow the steps in the proof of \cite[Prop.~30]{moulin2017log}.
    We decompose $\hi$ as $\hi = \hi_{\mc{X}^\dagger} + \hi_{\mc{X} \setminus \mc{X}^\dagger}$. Then, the right-hand side of \eqref{eq:taylorxi} becomes
    \begin{align}
        &\xi^{\epsilon_n}(\PXi, \PXo) = \xi^{\epsilon_n}(\PXs, \PXs) + n {\hi_{\mc{X} \setminus \mc{X}^\dagger}}^\top \Delta \notag \\
        &\quad + \frac{n}{2} {\mb{h}^{(\mr{o})}}^\top \ms{J}_{\mc{X}^*} {\mb{h}^{(\mr{o})}} + \sqrt{n} \overline{\mb{g}}^\top \mb{h}^{(\mr{o})} \notag \\
        &\quad - \sqrt{n} (\hi_{\mc{X}^\dagger} + \hi_{\mc{X} \setminus \mc{X}^\dagger})^\top (\ms{J}_{\mc{X}^\dagger} \sqrt{n} \mb{h}^{(\mr{o})} - \tilde{\mb{g}})  \notag \\
        &\quad + \bigo{\frac{Q^{-1}(\epsilon_n)^3}{\sqrt{n}}} + O(1). \label{eq:xidecompose}
    \end{align}
    Consider the maximization of the right-hand side of \eqref{eq:xidecompose} over $\hi$. Note that $\max_{x \in \mc{X} \setminus \mc{X}^\dagger} \Delta_x < 0$ and $\hi_{\mc{X} \setminus \mc{X}^\dagger} \geq \bs{0}$. The term $n {\hi_{\mc{X} \setminus \mc{X}^\dagger}}^\top \Delta$ is negative for any nonzero ${\hi_{\mc{X} \setminus \mc{X}^\dagger}}$. Since $\rho_n = o(1)$, it dominates the term $- \sqrt{n} (\hi_{\mc{X}^\dagger} + \hi_{\mc{X} \setminus \mc{X}^\dagger})^\top (\ms{J}_{\mc{X}^\dagger} \sqrt{n} \mb{h}^{(\mr{o})} - \tilde{\mb{g}})$ for $n$ large enough. This means that, for $n$ large enough, the maximizer $\hi$ must satisfy $\hi_{\mc{X} \setminus \mc{X}^\dagger} = \bs{0}$. Therefore, we have
    \begin{align}
        \xi^{\epsilon_n}(\PXi, \PXo) &\leq \xi^{\epsilon_n}(\PXs, \PXs) + \Gamma(\sqrt{n} \hi, \sqrt{n} \ho) \notag \\
        &\quad + \bigo{\frac{Q^{-1}(\epsilon_n)^3}{\sqrt{n}}} + O(1).  \label{eq:taylorlast}
    \end{align}
    Lastly, we apply \lemref{lem:moulinsaddle} to $\Gamma(\sqrt{n} \hi, \sqrt{n} \ho)$. A saddlepoint solution to the right-hand side of \eqref{eq:taylorlast} ignoring the $O(\cdot)$ terms is given by
    \begin{align}
        {\hi}^* &= -\frac{Q^{-1}(\epsilon_n)}{2 \sqrt{n V_{\epsilon_n}}} \tilde{\ms{J}} \mb{v}(\PXs) \\
        {\ho}^* &= -\frac{Q^{-1}(\epsilon_n)}{2 \sqrt{n V_{\epsilon_n}}}
       \tilde{\ms{J}} \tilde{\mb{v}},
    \end{align}
    where $\mb{v}(\PXs)$ and  $\tilde{\mb{v}}$ are defined in \eqref{eq:vP} and \eqref{eq:vtilde}, respectively. Notice that ${\ho}^*$ is uniquely defined even if the dispersion-achieving $\PXs$ is not unique in general.
    The value of the saddlepoint without the $O(\cdot)$ terms is ${\psi^{\epsilon_n}}^*(\PXs)$, which is defined in \eqref{eq:saddlevalue}.
    We set the perturbation vector $\tilde{\mb{h}}$ supported on $\mc{Y}$ such that ${\ho}^* \to \W \to \tilde{\mb{h}}$.

    \subsubsection{Bounding the right-hand side of \eqref{eq:tomeqfixed}} We bound the right-hand side of \eqref{eq:tomeqfixed} separately for $\mb{x} \in \mc{X}^n$ whose empirical distribution is close to some $\PXs$ and far away from all $\PXs$.

    \textbf{Case 1:} $\hat{P}_{\mb{x}} \in \mc{P}^*(\rho_n)$. For this case, we apply \lemref{lem:tom3} to the function $D_s^{\epsilon_n}(\Wxn \| {Q_Y^{(n)}}^*)$ with $Q^i = (\QYs + \tilde{\mb{h}})^{n}$ and $q(i) = \frac{1}{2}$ and get
    \begin{align}
        &D_s^{\epsilon_n}(\Wxn \| {Q_Y^{(n)}}^*) \notag \\ 
        &\leq D_s^{\epsilon_n}(\Wxn \| (\QYs + \tilde{\mb{h}})^{n}) + \log 2 \\
        &\leq \xi^{\epsilon_n}(\hat{P}_{\mb{x}}, \PXs + {\ho}^*) + \bigo{\frac{Q^{-1}(\epsilon_n)^3}{\sqrt{n}}} + \bigo{1} \label{eq:xistep} \\
        &\leq \max_{\PXs \in \mc{P}^*}{\psi^{\epsilon_n}}^*(\PXs) + \bigo{\frac{Q^{-1}(\epsilon_n)^3}{\sqrt{n}}} + \bigo{1}, \label{eq:psistep}
    \end{align}
    where \eqref{eq:xistep} follows from \lemref{lem:Dsasymp} (i) and the fact that $\QYs(y) > 0$ for all $y \in \mc{Y}$, and \eqref{eq:psistep} follows since \eqref{eq:saddlevalue} is the saddlepoint value for $\hat{P}_{\mb{x}} \in \mc{P}^*(\rho_n)$. Note that in \eqref{eq:psistep}, the maximization over dispersion-achieving input distributions is needed in case there are multiple dispersion-achieving input distributions.

    \textbf{Case 2:} $\hat{P}_{\mb{x}} \notin {\mc{P}^*(\rho_n)}$. For this case, we apply \lemref{lem:tom3} to the function $D_s^{\epsilon_n}(\Wxn \| {Q_Y^{(n)}}^*)$ with $Q^i = \hat{Q}_{\mb{x}}^n$ and $q(i) = \frac{1}{2 |\mc{P}_n |}$ and get
    \begin{align}
        &D_s^{\epsilon_n}(\Wxn \| {Q_Y^{(n)}}^*) \notag \\ 
        &\leq D_s^{\epsilon_n}(\Wxn \| \hat{Q}_{\mb{x}}^n) + (|\mc{X}|-1) \log (n + 1) + \log 2 \label{eq:methodoftypes}\\
        &\leq \xi^{\epsilon_n}(\hat{P}_{\mb{x}}, \hat{P}_{\mb{x}}) + \bigo{\frac{Q^{-1}(\epsilon_n)^3}{\sqrt{n}}} + \bigo{1} + O(\log n), \label{eq:lognterm}
    \end{align}
    where in \eqref{eq:methodoftypes}, we use the well-known bound on the number of types $|\mc{P}_n| \leq (n + 1)^{|\mc{X}| - 1}$. Inequality \eqref{eq:lognterm} follows from \lemref{lem:Dsasymp} (ii). 

    We set the constant $c_0 > 0$ in \eqref{eq:rhon} as dictated by \lemref{lem:xibound}. Then, for all
    $\Phat \notin {\mc{P}^*(\rho_n)}$, 
    \begin{align}
        \xi^{\epsilon_n}(\hat{P}_{\mb{x}}, \hat{P}_{\mb{x}}) &\leq  \max_{\PXs \in \mc{P}^*}{\psi^{\epsilon_n}}^*(\PXs) \notag \\
        &\quad - c_1 \sqrt{n} \rho_n |\Qinv| (1 + o(1)). \label{eq:xiexp}
    \end{align}
    By \eqref{eq:rhon}, $\sqrt{n} \rho_n |\Qinv| \geq \log^2 n$. Hence, the $O(\log n)$ term in \eqref{eq:lognterm} is dominated by the $- c_1 \sqrt{n} \rho_n |\Qinv| (1 + o(1))$ term in \eqref{eq:xiexp}. This property, together with \eqref{eq:lognterm}, ensures that for all $\Phat \notin {\mc{P}^*(\rho_n)}$, 
   \begin{align}
        D_s^{\epsilon_n}(\Wxn \| {Q_Y^{(n)}}^*) &\leq \max_{\PXs \in \mc{P}^*}{\psi^{\epsilon_n}}^*(\PXs) + \bigo{\frac{Q^{-1}(\epsilon_n)^3}{\sqrt{n}}} \notag \\
        &\quad + \bigo{1}. \label{eq:case2}
   \end{align}
   From \eqref{eq:psistep} and \eqref{eq:case2}, we conclude that 
   \begin{align}
       \max_{\mb{x} \in \mc{X}^n} D_s^{\epsilon_n}(\Wxn \| {Q_Y^{(n)}}^*) &\leq \max_{\PXs \in \mc{P}^*}{\psi^{\epsilon_n}}^*(\PXs) \notag \\
       &\quad + \bigo{\frac{Q^{-1}(\epsilon_n)^3}{\sqrt{n}}}+ \bigo{1}.\label{eq:Dsmax}
   \end{align}
    
    Finally, we set the parameter $\delta_n$ in \eqref{eq:tomeqfixed} so that 
    \begin{align}
        \log \delta_n = - \frac{Q^{-1}(\epsilon_n)^2}{2} - \frac{1}{2} \log n. \label{eq:deltan}
    \end{align}
    We replace $\epsilon_n$ in \eqref{eq:Dsmax} with $\epsilon_n + \delta_n$. Expanding the Taylor series of $Q^{-1}(\cdot)$ around $\epsilon_n$ completes the proof of \thmref{thm:mainConv}.

    	\section{Proof of \thmref{thm:NP}}\label{sec:NPproof}
	Assume that the random variable $\sum_{i = 1}^n Z_i$ is lattice with span $h > 0$. Let $\underline{\gamma}$ and $\overline{\gamma}$ satisfy
	\begin{align}
	    \Prob{\sum_{i = 1}^n Z_i \geq \underline{\gamma}} &= 1 - \underline{\epsilon}_n \geq 1-\epsilon_n \label{eq:epslow} \\
	    \Prob{\sum_{i = 1}^n Z_i \geq \overline{\gamma}} &= 1 - \overline{\epsilon}_n \leq 1-\epsilon_n, \label{eq:epshigh}
	\end{align}
	where $\underline{\gamma}$ and $\overline{\gamma}$ are in the range of $\sum_{i = 1}^n Z_i$,  $\overline{\gamma} - \underline{\gamma} = h$, and $\underline{\epsilon}_n \leq \epsilon_n \leq \overline{\epsilon}_n$. Let $\lambda \in [0, 1]$ satisfy
	\begin{align}
	    \Prob{\sum_{i = 1}^n Z_i \geq \underline{\gamma}} \lambda + \Prob{\sum_{i = 1}^n Z_i \geq \overline{\gamma}} (1-\lambda) = 1 - \epsilon_n.
	\end{align}
	By the Neyman-Pearson Lemma (see \cite[eq. (101)]{polyanskiy2010Channel}), 
	\begin{IEEEeqnarray}{rCl}
	    \IEEEeqnarraymulticol{3}{l}{\beta_{1-\epsilon_n}(P^{(n)}, Q^{(n)})} \notag\\
	    &=& \Prob{\sum_{i = 1}^n \overline{Z}_i \geq \underline{\gamma}} \lambda 
	    + \Prob{\sum_{i = 1}^n \overline{Z}_i \geq \overline{\gamma}} (1-\lambda). \label{eq:balambda}
	\end{IEEEeqnarray}
	Define the asymptotic expansion
	\begin{align}
	    \chi(\epsilon) &\triangleq D - \sqrt{\frac{V}{n}} Q^{-1}(\epsilon) + \frac{\ske \sqrt{V}}{6n}  Q^{-1}(\epsilon)^2 \notag \\
	    &\,\,-  \frac{3 (\mu_4 - 3 V^2) V - 4 \mu_3^2}{72 V^{5/2}} \frac{Q^{-1}(\epsilon)^3}{n^{3/2}} \notag \\
        &\,\,+ \bigo{\frac{Q^{-1}(\epsilon)^4}{n^2}} + \bigo{\frac{1}{n}}. 
	\end{align}
	By conditions (A) and (B) of \thmref{thm:NP}, the conditions of \thmref{thm:moderatePetrov} are satisfied for the sum $\sum_{i = 1}^n Z_i$. We apply \lemref{lem:cornish} to \eqref{eq:epslow}--\eqref{eq:epshigh}, and get the asymptotic expansions
	\begin{align}
	    \underline{\gamma} &= n \chi(\underline{\epsilon}_n) \label{eq:chi1}\\
         \overline{\gamma} &= n \chi(\overline{\epsilon}_n). \label{eq:chi2}
	\end{align}
	From the Taylor series expansion of $\chi(\cdot)$ around $\epsilon_n$, \eqref{eq:chi1}--\eqref{eq:chi2}, and $\overline{\gamma} - \underline{\gamma} = O(1)$, it holds that
	\begin{align}
	    \underline{\gamma} &= n \chi(\epsilon_n) + O(1) \label{eq:ugamma} \\
	    \overline{\gamma} &= n \chi(\epsilon_n) + O(1). \label{eq:ogamma}
	\end{align}
    The arguments above also hold in the non-lattice case (i.e., $h = 0$) with $\underline{\gamma} = \overline{\gamma}$.
    
    Next, we evaluate the probability $\Prob{\sum_{i = 1}^n \overline{Z}_i \geq \underline{\gamma}}$ in \eqref{eq:balambda} separately in the lattice and non-lattice cases.
    \subsubsection{Lattice case} 
    We here apply \thmref{thm:lattice} to evaluate the probability of interest.
    By \cite[Appendix D]{moulin2017log}, 
    \begin{align}
        \kappa(1) &= 0 \\
        \kappa'(1) &= D \label{eq:kappa1}\\
        \kappa''(1) &= V \\
        \kappa'''(1) &= \mu_3.
    \end{align}
    From \eqref{eq:ugamma}, we have
    $\frac{1}{n} \underline{\gamma} = D + o(1)$. Therefore, by \eqref{eq:kappa1}, condition (ND) of \thmref{thm:chaganty} is satisfied with $s = 1 + o(1)$. Condition (S) of \thmref{thm:chaganty} is satisfied by condition (C) of \thmref{thm:NP}. Therefore, it only remains to verify condition (L) of \thmref{thm:lattice} in the one-dimensional case. Since $\sum_{i = 1}^n \overline{Z}_i$ is lattice with span $h$, each $\overline{Z}_i$ is also lattice with a span that is a multiple of $h$. By \cite[p. 1687]{chagantyOneDimensional}, we have
    \begin{align}
	     \sup_{\delta < |t| \leq \frac{\pi}{h}} \left \lvert \frac{\phi_i(s + \mathrm{i} t)}{\phi_i(s)} \right \rvert \leq c_1 < 1, \quad i \in [n]
    \end{align}
    for every $0 < \delta \leq \frac{\pi}{h}$,
    where $\phi_i(\cdot)$ is the MGF of $\overline{Z}_i$. Since $\overline{Z}_1, \dots, \overline{Z}_n$ are i.i.d., the MGF $\phi(\cdot)$ of $\sum_{i = 1}^n \overline{Z}_i$ satisfies
    \begin{align}
        \sup_{\delta < |t| \leq \frac{\pi}{h}} \left \lvert \frac{\phi(s + \mathrm{i} t)}{\phi(s)} \right \rvert &=  \sup_{\delta < |t| \leq \frac{\pi}{h}} \left \lvert \prod_{i = 1}^n \frac{\phi_i(s + \mathrm{i} t)}{\phi_i(s)} \right \rvert \\
        &\leq c_1^n = o(n^{-1/2}).
    \end{align}
    Therefore, condition (L) of \thmref{thm:lattice} is satisfied. Applying \thmref{thm:lattice} to $\Prob{\sum_{i = 1}^n \overline{Z}_i \geq \underline{\gamma}}$, we have
    \begin{align}
        \Prob{\sum_{i = 1}^n \overline{Z}_i \geq \underline{\gamma}} = \exp\left \{-n \Lambda(a_n) - \frac{1}{2} \log n + O(1) \right\}, \label{eq:taylorL}
    \end{align}
    where
    \begin{align}
        \Lambda(a_n) &= \sup_{t \in \mathbb{R}} \{t a_n - \kappa(t)\} \\
        a_n &= \chi(\epsilon_n) + \bigo{\frac{1}{n}}.
    \end{align}
    We expand the Taylor series of $\Lambda(\cdot)$ around $D$ as
    \begin{align}
        \Lambda(a_n) &= \Lambda(D) + (a_n - D) \Lambda'(D) + \frac{(a_n - D)^2}{2} \Lambda''(D) \notag \\
        &\quad + \frac{(a_n - D)^3}{6} \Lambda'''(D) + O(|a_n - D|^4).
    \end{align}
    By \cite[Appendix D]{moulin2017log},
    \begin{align}
        \Lambda(D) &= D \label{eq:LamD}\\
        \Lambda'(D) &= 1 \\
        \Lambda''(D) &= \frac{1}{V} \\
        \Lambda'''(D) &= - \frac{\mu_3}{V^3}. \label{eq:Lam3D}
    \end{align}
    Combining \eqref{eq:taylorL} and \eqref{eq:LamD}--\eqref{eq:Lam3D}, we get
    \begin{align}
        \Lambda(a_n) = a_n + \frac{Q^{-1}(\epsilon_n)^2}{2n} + \bigo{\frac{Q^{-1}(\epsilon_n)^4}{n^2}} + \bigo{\frac{1}{n}}. \label{eq:asyL}
    \end{align}
    By \eqref{eq:ugamma}--\eqref{eq:ogamma}, the asymptotic expansion on the right-hand side of \eqref{eq:taylorL} also holds for the probability $ \Prob{\sum_{i = 1}^n \overline{Z}_i \geq \overline{\gamma}}$. Combining \eqref{eq:balambda}, \eqref{eq:taylorL}, and \eqref{eq:asyL} completes the proof for the lattice case. 
    
    \subsubsection{Non-lattice case} 
    The proof for the non-lattice case is identical to the proof for the lattice case except for the verification of condition (NL) in \thmref{thm:chaganty}. Define
    \begin{align}
        \tilde{S}_j \triangleq \sum_{i \in \mc{I}_j} Z_i, \quad j \in [w_n],
    \end{align}
    where $\tilde{S}_j$ are non-lattice by condition (D) of \thmref{thm:NP}. By \cite[p. 1687]{chagantyOneDimensional},
    \begin{align}
        \sup_{j \in [w_n]} \sup_{\delta < |t| \leq \lambda} \left \lvert \frac{\tilde{\phi}_j(s + \mathrm{i} t)}{\tilde{\phi}_j(s)} \right \rvert \leq c_2 < 1 \label{eq:c2small}
    \end{align}
    for every $0 < \delta < \lambda$,
    where $\tilde{\phi}_j$ denotes the MGF of $\tilde{S}_j$. Since $\overline{Z}_1, \dots, \overline{Z}_n$ are i.i.d., we have
    \begin{align}
        \sup_{\delta < |t| \leq \lambda} \left \lvert \frac{\phi(s + \mathrm{i} t)}{\phi(s)} \right \rvert &=  \sup_{\delta < |t| \leq \lambda} \left \lvert \prod_{j = 1}^{w_n} \frac{\tilde{\phi}_j(s + \mathrm{i} t)}{\tilde{\phi}_j(s)} \right \rvert \cdot 1 \label{eq:c2step1} \\
        &\leq c_2^{w_n} \label{eq:c2step2}\\
        &= o(n^{-1/2}), \label{eq:c1cond}
    \end{align}
    where \eqref{eq:c2step1} follows since $\frac{\tilde{\phi}_j(s + \mathrm{i} t)}{\tilde{\phi}_j(s)}$ is a characteristic function of a non-lattice random variable \cite{chagantyOneDimensional}, \eqref{eq:c2step2} follows from \eqref{eq:c2small}, and \eqref{eq:c1cond} follows from condition (D) and $c_2 < 1$. This verifies condition (NL) of \thmref{thm:chaganty}. Applying \thmref{thm:chaganty} in a way that is similar to \eqref{eq:taylorL} completes the proof.

        \section{Proof of \thmref{thm:Gaussian}}\label{app:Gaussian}
    We begin by presenting the preliminary definitions of the subsets of an $n$-dimensional sphere. 
    A centered, unit sphere embedded in $\mathbb{R}^n$ (the manifold dimension is $n-1$) is defined as
    \begin{align}
        \mathbb{S}^{n-1} \triangleq \{\mb{x} \in \mathbb{R}^n \colon \norm{\mb{x}} = 1\}.
    \end{align}
    A centered, unit-radius spherical cap embedded in $\mathbb{R}^n$ is defined as
    \begin{align}
        \mathrm{cap}(\mb{x}, a) \triangleq \{\mb{y} \in \mathbb{R}^n \colon \langle \mb{x}, \mb{y} \rangle \geq a, \norm{\mb{y}}_2 = 1\},
    \end{align}
    where $\mb{x} \in \mathbb{S}^{n-1}$ is the center point of the cap, and $a \in [-1, 1]$ defines the size of the cap, which is equal to the cosine of the half-angle of the cap. For example, $\mathrm{cap}(\mb{x}, -1) = \mathbb{S}^{n-1}$ and $\mathrm{cap}(\mb{x}, 0)$ is a half-sphere. We use $\mathrm{Area}(\cdot)$ to denote the surface area of an $(n-1)$-dimensional manifold embedded in $\mathbb{R}^n$. For example, the surface area of a unit sphere is
    \begin{align}
        \mathrm{Area}(\mathbb{S}^{n-1}) = \frac{2 \pi^{\frac{n}{2}}}{\Gamma(\frac{n}{2})},
    \end{align}
    where $\Gamma(\cdot)$ denotes the Gamma function. Below, we use $\hat{\mb{X}} \triangleq \frac{\mb{X}}{\norm{\mb{X}}_2}$ to denote the projection of $\mb{X}$ onto $\mathbb{S}^{n-1}$.
    
    \subsection{Shannon's Random Coding Bound}
    Shannon's random coding bound from \cite{shannon1959Probability} can viewed as a relaxation of the RCU bound \eqref{eq:RCUpol}, but the relaxation is different than the one in \eqref{eq:RCU}. We generate $M$ independent codewords uniformly distributed on the power sphere $\sqrt{nP} \mathbb{S}^{n-1}$. Since all codewords lie on the power sphere and since the maximum likelihood decoding rule is equal to the minimum-distance decoder for the Gaussian channel, \eqref{eq:RCUpol} is equivalent to
    \begin{align}
        \epsilon \leq \Prob{ \bigcup_{m = 2}^M \{\langle \hat{\mb{X}}(m), \hat{\mb{Y}} \rangle \geq \langle \hat{\mb{X}}(1), \hat{\mb{Y}} \rangle\} | W = 1}. \label{eq:Gaussbound1}
    \end{align}
    We bound the right-hand side of \eqref{eq:Gaussbound1} by
    \begin{align}
        \Prob{\langle \hat{\mb{X}}, \hat{\mb{Y}} \rangle < a} + M \Prob{\langle \hat{\overline{\mb{X}}}, \hat{\mb{Y}} \rangle \geq \langle \hat{\mb{X}}, \hat{\mb{Y}} \rangle \geq a} \label{eq:GShannonach}
    \end{align}
    for some $a \in [-1, 1]$ to be determined later. Here, $\mb{X}$ is uniformly distributed on $\sqrt{nP} \mathbb{S}^{n-1}$, $\mb{Y} = \mb{X} + \mb{Z}$, $\mb{Z} \sim \mathcal{N}(\mb{0}, \ms{I}_n)$ and is independent of $\mb{X}$, and $\overline{\mb{X}}$ is distributed identically to $\mb{X}$ and is independent of $\mb{X}$ and $\mb{Y}$. The bound in \eqref{eq:GShannonach} is exactly equal to \cite[eq.~(19)]{shannon1959Probability} and \cite[eq.~(61)]{erseghe2016}. Both of \cite{shannon1959Probability} and \cite{erseghe2016} set the threshold $a$ to satisfy
    \begin{align}
        \Prob{\langle \hat{\overline{\mb{X}}}, \hat{\mb{Y}} \rangle \geq a} = \frac{1}{M}
    \end{align}
    to analyze the bound in the LD regime. We here set $a$ slightly differently for the CLT regime, namely, as
    \begin{align}
        \Prob{\langle \hat{\mb{X}}, \hat{\mb{Y}} \rangle < a} = \tilde{\epsilon} = \epsilon - \frac{1}{\sqrt{2 \pi n V(P)}} \exp\cB{-\frac{Q^{-1}(\epsilon)^2}{2}}, \label{eq:Gaussfirst}
    \end{align}
    which is the same choice that we make in \eqref{eq:tildeepsn}. 
    
    Using the same steps as \cite[eq.~(16)-(17)]{shannon1959Probability} and \cite[Appendix G]{erseghe2016}, we express the probability \eqref{eq:Gaussfirst} in terms of a CDF of a noncentral $t$-distribution with noncentrality parameter $\sqrt{nP}$ and $n-1$ degrees of freedom as\footnote{To see this, set $\mb{X}$ to $(\sqrt{nP}, 0, \dots, 0)$ and use spherical symmetry.}
    \begin{align}
        \Prob{\langle \hat{\mb{X}}, \hat{\mb{Y}} \rangle < a} = \Prob{\rho < \sqrt{n-1} \frac{a}{\sqrt{1 - a^2}}}, \label{eq:rhoa}
    \end{align}
    where $\rho \sim \mathrm{noncentral-}t(n-1, \sqrt{nP})$, which is defined as
    \begin{align}
        \frac{A_1 + \sqrt{nP}}{\sqrt{\frac{1}{n-1}\sum_{i = 2}^n A_i^2}},
    \end{align}
    where $A_1, \dots, A_n$ are i.i.d. $\mathcal{N}(0, 1)$. 
    
    Due to spherical symmetry, $\langle \hat{\overline{\mb{X}}}, \hat{\mb{Y}} \rangle$ is independent of $\langle \hat{\mb{X}}, \hat{\mb{Y}} \rangle$, and from \cite[Sec.~IV]{shannon1959Probability},
    \begin{align}
        \Prob{\langle \hat{\overline{\mb{X}}}, \hat{\mb{Y}} \rangle \geq b} = \frac{\mathrm{Area}(\mathrm{cap}(\mb{x}_0, b))}{\mathrm{Area}(\mathbb{S}^{n-1})} \quad \text{for } b \in (-1, 1), \label{eq:Area}
    \end{align}
    where $\mb{x}_0$ is any point on the unit-sphere. Shannon proves the following asymptotic expansion of \eqref{eq:Area}
    \begin{align}
        v_n(b) &\triangleq \frac{1}{n} \log \frac{\mathrm{Area}(\mathrm{cap}(\mb{x}_0, b))}{\mathrm{Area}(\mathbb{S}^{n-1})} \\
        &= \frac{1}{2} \log (1 - b^2) \notag \\
        &\quad - \frac{1}{2n} \log n  - \frac{1}{2n} \log (2 \pi b^2 (1-b^2)) + \bigo{\frac{1}{n^2}}. \label{eq:vasymp}
    \end{align}
    
    To find the value of $a$ in \eqref{eq:Gaussfirst} as a function of $\epsilon$, we first derive a Cornish-Fisher expansion of the random variable $\rho$. Fisher and Cornish \cite{fisher1960} extend the Cornish-Fisher expansion of the random variables with known cumulants that do not need to be sums of independent random variables; they give the expansions for $t$ and chi-squared distributions as examples. Van Eeden \cite{eeden1961} uses the same technique for the noncentral $t$-distribution, where the noncentrality parameter is fixed and the number of degrees of freedom approaches infinity. In our application, $\rho$ has a noncentrality parameter $\sqrt{nP}$ growing to infinity.
    
    To extend \cite{eeden1961} to the case where the noncentrality parameter also grows with $n$, we realize that a sufficient condition for the expansion in \cite{fisher1960} to be valid is that the random variable is continuous and its first $s+1$ cumulants satisfy $\kappa_j = \bigo{\frac{1}{n^{\frac{j}{2}-1}}}$, $j \leq s+1$, which holds in our application (see \eqref{eq:kappa1G}--\eqref{eq:kappa3G}). Therefore, the expansion in \cite{eeden1961} extends to the case with a noncentrality parameter $\sqrt{nP}$ with the change of corresponding cumulant expansions. 

    The version that we are interested in is studied in a recent paper \cite[Th.~6.2]{gil2023new}, where an asymptotic expansion of the tail probability of the noncentral $t$-distribution with noncentrality parameter $\sqrt{nP}$ is provided. Inverting that result using the Lagrange inversion theorem also verifies the following result.

    From \cite{cornish, fisher1960, eeden1961}, the quantile $t$ of $\rho$ at the value $\tilde{\epsilon}$ admits the expansion
    \begin{align}
       \tilde{\epsilon} &=  \Prob{\rho < t} \label{eq:rhot} \\
        t &= \kappa_1 - \sqrt{\kappa_2} \left(Q^{-1}(\tilde{\epsilon}) - \frac{\ske}{6}(Q^{-1}(\tilde{\epsilon})^2 - 1)\right) + \bigo{\frac{1}{n}}, \label{eq:expt}
    \end{align}
    where $\kappa_1 = \E{\rho}$, $\kappa_2 = \Var{\rho}$, and $\ske = \frac{\E{(\rho - \kappa_1)^3}}{\kappa_2^{3/2}}$ is the skewness. 
    
    From the moments of the noncentral $t$-distribution \cite{hogben1961moments} and Taylor series expansions, we calculate the asymptotic expansions for $\kappa_1, \kappa_2$, and $\ske$ as
    \begin{align}
        \kappa_1 &= \sqrt{nP} + \frac{3}{4} \sqrt{\frac{P}{n}} + \bigo{n^{-3/2}} \label{eq:kappa1G}\\
        \kappa_2 &= \left(1 + \frac{P}{2}\right) + \frac{2 + \frac{19 P}{8}}{n} + \bigo{n^{-3/2}} \label{eq:kappa2G} \\
        \ske &= \frac{12 \sqrt{P} + 5 P^{3/2}}{ \sqrt{2n} \, (2 + P)^{3/2}} + \bigo{n^{-3/2}} \label{eq:kappa3G}
    \end{align}
    and check that the fourth cumulant satisfies $\kappa_4 = O(n^{-1})$.
    Applying the Taylor series expansion to $Q^{-1}(\tilde{\epsilon})$, we get
    \begin{align}
        Q^{-1}(\tilde{\epsilon}) = Q^{-1}(\epsilon) + \frac{1}{\sqrt{n V(P)}} + \bigo{\frac{1}{n}}. \label{eq:taylorQtilde}
    \end{align}
    Juxtaposing \eqref{eq:rhoa} and \eqref{eq:rhot}, we note
    \begin{align}
        a = \frac{\frac{t}{\sqrt{n-1}}}{\sqrt{1 + \frac{t^2}{n-1}}}. \label{eq:Gaussa}
    \end{align}
    Substituting \eqref{eq:kappa1G}--\eqref{eq:taylorQtilde} into \eqref{eq:expt}, and the latter into \eqref{eq:Gaussa}, we get
    \begin{align}
        a =& \frac{\sqrt{P}}{\sqrt{1+P}} - \frac{1}{\sqrt{n}}\frac{\sqrt{2 + P} Q^{-1}(\epsilon)}{
 \sqrt{2} (1 + P)^{3/2}} \notag \\ &+\frac{1}{n}\frac{
  18 \sqrt{P} + 28 P^{3/2} + 10 P^{5/2}}{12 (1 + P)^{5/2} (2 + P)} \notag \\
  &- \frac{Q^{-1}(\epsilon)^2}{n}\frac{24 \sqrt{P} +
   19 P^{3/2} + 4 P^{5/2}}{12 (1 + P)^{5/2} (2 + P)} \notag \\
  &- \frac{1}{n}\frac{\sqrt{
  2 + P}}{\sqrt{2} (1 + P)^{3/2} \sqrt{V(P)}}. \label{eq:Aexp}
  \end{align}
  
  It only remains to find the asymptotic expansion of the probability $\Prob{\langle \hat{\overline{\mb{X}}}, \hat{\mb{Y}} \rangle \geq \langle \hat{\mb{X}}, \hat{\mb{Y}} \rangle \geq a}$. Note that this probability is in the LD regime. Using the analysis in \cite[Sec.~V-B]{erseghe2016}, we find the density of $\langle \hat{\mb{X}}, \hat{\mb{Y}} \rangle$ as
  \begin{align}
      f_{\langle \hat{\mb{X}}, \hat{\mb{Y}} \rangle}(a) &= \exp\{n u_n(a)\} \\
      u_n(a) &= u_0(a) + \frac{\log n}{2n} - \frac{u_1(a)}{2n} + O(n^{-2}) \\
      u_0(a) &= \frac{1}{2} \log(1 - a^2) - 2 \alpha^2 + (\alpha a)^2 + \alpha a \sqrt{1 + (\alpha a)^2} \notag \\
      &\quad + \log(\alpha a + \sqrt{1 + (\alpha a)^2}) \\
      u_1(a) &= \log(1 + (\alpha a)^2 + \alpha a \sqrt{1 + (\alpha a)^2}) \notag \\
      &\quad + 3 \log (1- a^2) + \log(2 \pi), \label{eq:u1}
  \end{align}
where $\alpha \triangleq \sqrt{\frac{P}{4}}$. 

In \cite{erseghe2016}, the asymptotic expansion to the probability $\Prob{\langle \hat{\overline{\mb{X}}}, \hat{\mb{Y}} \rangle \geq \langle \hat{\mb{X}}, \hat{\mb{Y}} \rangle \geq a}$ is derived using the Laplace integration method as
\begin{align}
    &\Prob{\langle \hat{\overline{\mb{X}}}, \hat{\mb{Y}} \rangle \geq \langle \hat{\mb{X}}, \hat{\mb{Y}} \rangle \geq a} \notag \\
    &\quad = \int_a^1 f_{\langle \hat{\mb{X}}, \hat{\mb{Y}} \rangle}(b) \frac{\mathrm{Area}(\mathrm{cap}(\mb{x}_0, b))}{\mathrm{Area}(\mathbb{S}^{n-1})} \mathrm{d}b \\
    &\quad = \int_{a}^1 \exp\{n g_n(b)\} \mathrm{d}b \\
    &\quad =  \exp\{n g_n(a)\} \left(\frac{1}{-n g_n'(a)} + O(n^{-2}) \right),
\end{align}
where $g_n(b) = u_n(b) + v_n(b)$ and $g_n'(a)$ is the derivative of $g_n(\cdot)$ evaluated at $a$. 

Finally, equating the second term in \eqref{eq:GShannonach} to $\frac{1}{\sqrt{2 \pi n V(P)}} \exp\cB{-\frac{Q^{-1}(\epsilon)^2}{2}}$, giving
\begin{align}
    M \exp\{n g_n(a)\} \frac{1}{-n g_n'(a)} = \frac{1}{\sqrt{2 \pi n V(P)}} \exp\cB{-\frac{Q^{-1}(\epsilon)^2}{2}},
\end{align}
and using \eqref{eq:Aexp}--\eqref{eq:u1} along with several Taylor series expansions, we complete the proof for the lower bound \eqref{eq:Gausslower}.

    \subsection{Vazquez-Vilar's Converse}
    We here analyze the meta-converse bound in \eqref{eq:metaG} in the CLT regime. The arguments of the $\beta_{\alpha}$ function in \eqref{eq:metaG} satisfy the conditions of \cite[Th.~18]{moulin2017log}. Let $P_Y = \mathcal{N}(\sqrt{P}, 1)$, $Q_Y^{(\delta_n)} = \mathcal{N}(0, 1 + P + \delta_n)$, and $Z = \log \frac{\mathrm{d} P_Y}{\mathrm{d} Q_Y^{(\delta_n)}}(Y)$, where $Y \sim P_Y$. 

    We compute
    \begin{align}
        D(P, \delta_n) &\triangleq \E{Z} = C(P + \delta_n) - \frac{\delta_n}{2(1 + P + \delta_n)} \label{eq:Ddel}\\
        V(P, \delta_n) &\triangleq \Var{Z} = \frac{P^2 + 2 P \delta_n + 2 P + \delta_n^2}{2 (1 + P + \delta_n)^2} \\
        \mu_3(P, \delta_n) &\triangleq \E{(Z - \E{Z})^3} = - \frac{P^2 (P + 3)}{(1 + P)^3} + O(\delta_n) \\
        S(P, \delta_n) &\triangleq \frac{\mu_3(P, \delta_n)}{V(P, \delta_n)^{3/2}}.\label{eq:Sdel}
    \end{align}
    Note that $D(P, 0) = C(P)$ and $V(P, 0) = V(P)$. Putting \eqref{eq:Ddel}--\eqref{eq:Sdel} in \cite[Th.~18]{moulin2017log} gives
    \begin{align}
        F(\delta_n) &\triangleq - \log \beta_{1-\epsilon}(P_Y^{\otimes n}, (Q_Y^{(\delta_n)})^{\otimes n}) \\
        &= n D(P, \delta_n) - \sqrt{V(P, \delta_n)} Q^{-1}(\epsilon) + \frac{1}{2} \log n \notag \\
        &\quad + \left(\frac{S(P, \delta_n) \sqrt{V(P, \delta_n)}}{6} \right) \left(Q^{-1}(\epsilon)^2 - 1\right) \notag \\
        &\quad + \frac{1}{2} Q^{-1}(\epsilon)^2 + \frac{1}{2} \log(2 \pi V(P, \delta_n)) + o(1).  \label{eq:logbetaG}
    \end{align}
    Next, we take the Taylor series expansion of \eqref{eq:logbetaG} around $\delta_n = 0$ and get
    \begin{align}
        &F(\delta_n) =    n C(P) - \sqrt{n V(P)} Q^{-1}(\epsilon) + \frac{1}{2} \log n \notag \\
        &\quad + \frac{n}{2} \delta_n^2 \frac{1}{2 (1 + P)^2} + \sqrt{n} \delta_n \frac{\sqrt{P} Q^{-1}(\epsilon)}{(1 + P)^2 \sqrt{2 (P + 2)}} \notag \\
        &\quad + \left(- \frac{P (P + 3)}{3 (1 + P)(2 + P)} \right) \left(Q^{-1}(\epsilon)^2 - 1\right) \notag \\
        &\quad + \frac{1}{2} Q^{-1}(\epsilon)^2 + \frac{1}{2} \log( 2 \pi V(P)) + o(1) + O(\delta_n).
    \end{align}
    Setting $\frac{\partial F}{\partial \delta_n} = 0$ shows that the $\delta_n^*$ that minimizes $F(\delta_n)$ is given by
    \begin{align}
        \delta_n^* = - \frac{Q^{-1}(\epsilon)}{\sqrt{n}} \sqrt{\frac{2 P}{P + 2}}.
    \end{align}
    Evaluating $F(\delta_n^*)$ completes the proof of \eqref{eq:Gaussupper}.

    \subsection{Shannon's Cone-Packing Converse}
    In \cite[eq.~(15)]{shannon1959Probability}, Shannon derives a converse bound for the Gaussian channel with an equal-power constraint using a cone-packing idea, which is equal to Polyanskiy's later minimax bound \cite[Th.~28]{polyanskiy2010Channel}. Shannon's bound is still the tightest known bound for any error probability. We here analyze \cite[eq.~(15)]{shannon1959Probability} in the CLT regime.
    
    Shannon's cone-packing converse for the equal-power case is given by
    \begin{align} 
        \epsilon \geq \Prob{\langle \hat{\mb{X}}, \hat{\mb{Y}} \rangle < a^*}, \label{eq:Shannonconv}
    \end{align}
    where $a^*$ satisfies
    \begin{align}
        \frac{1}{M} = \frac{\mathrm{Area}(\mathrm{cap}(\mb{x}_0, a^*))}{\mathrm{Area}(\mathbb{S}^{n-1})}. \label{eq:ShannonMArea}
    \end{align}
    To evaluate \eqref{eq:Shannonconv}, we express $a^*$ in terms of $\epsilon$ using the Cornish-Fisher expansion in \eqref{eq:expt}. Then, we plug the value of $a^*$ into \eqref{eq:ShannonMArea} and write the right-hand side of \eqref{eq:ShannonMArea} using the asymptotic expansion in \eqref{eq:vasymp}. 
    After several Taylor series expansions, combining \eqref{eq:Shannonconv} and \eqref{eq:ShannonMArea} yields the  
    bound on the right-hand side of  \eqref{eq:Gaussupper} for the equal-power constraint with $S(P)$ unchanged and $\overline{B}(P)$ replaced with $\underline{B}(P) + 1$. Combining this converse with \cite[eq.~(23)]{vazquez2021}, which is a refinement of \eqref{eq:shannonmaxeq}, we get a converse bound for the maximal-power constraint with $S(P)$ in \eqref{eq:Gaussupper} unchanged and $\overline{B}(P)$ replaced with $\underline{B}(P) + 1 + C(P) - \frac{P}{2 (1 + P)}$.

    \section{Conclusion} \label{sec:conclusion}
    This paper investigates the third-order characteristic of nonsingular DMCs, the Gaussian channel with a maximal-power constraint, and binary hypothesis tests, defining a new term, the \emph{channel skewness} for this purpose. Since the channel skewness is multiplied by $Q^{-1}(\epsilon)^2$ in the asymptotic expansion of the logarithm of the maximum achievable message set size, including the channel skewness term in the approximation is particularly important to accurately approximate non-asymptotic bounds in the small-$\epsilon$ regime. In most of the paper (except the Gaussian channel extension), we derive tight bounds on the non-Gaussianity \eqref{eq:CLT} in the MD regime. We show in Theorems~\ref{thm:mainAch}--\ref{thm:mainConv} that Moulin's CLT approximations in \eqref{eq:achrho}--\eqref{eq:convrho} up to the skewness terms remain valid when the constant $\epsilon$ is replaced by an SMD sequence $\epsilon_n$ \eqref{eq:range}. 
    For a BSC(0.11) and most pairs $(n, \epsilon)$ pairs satisfying $\epsilon \in [10^{-10}, 10^{-1}]$ and $n \in [100, 500]$, we observe that our skewness approximation in Theorems~\ref{thm:mainAch}-\ref{thm:mainConv} is more accurate than the CLT approximation from \cite{polyanskiy2010Channel} and the state-of-the-art LD approximations from \cite{honda2018, segura2018}. While the prefactor in those LD approximations requires solution of a different optimization problem for each $(n, \epsilon)$ pair, our skewness approximations are easily computable, and the skewness term informs us about the accuracy of the CLT approximation for a particular channel.
    For Cover--Thomas-symmetric channels,  our bounds determine the channel skewness $S$ exactly; in \thmref{thm:refinedAchConv}, we refine Theorems~\ref{thm:mainAch}--\ref{thm:mainConv} by computing the term that is one order higher than the channel skewness. 
    
    By analyzing Shannon's random coding bound in \cite{shannon1959Probability} and Vazquez-Vilar's meta-converse bound in \cite{vazquez2021} in the CLT regime, we exactly compute the channel skewness for the Gaussian channel with a maximal-power constraint. \thmref{thm:NP} implies that the converse bound generalizes to SMD sequences. We leave to future work the MD analysis of the achievability bound for the Gaussian channel, which calls for new tools for approximating the probabilities of sums of dependent random variables.
    
    Our techniques also apply to BHT in the MD regime, where the third- and fourth-order terms in the type-II error probability exponent have forms similar to the third- and fourth-order terms in the expansion of the logarithm of the maximum achievable message set size for Cover--Thomas-symmetric channels. For example, the skewness of the log-likelihood ratio in BHT plays the role of information skewness in channel coding. Using our new MD approximations to BHT, several information-theoretic results that rely on BHT asymptotics such as \cite{kostina2012lossyJ, kostina2013joint, chen2020lossless, yavas2020Random} can be extended to the MD regime.

    The asymptotic expansions in \eqref{eq:expNL} and \eqref{eq:chagantyLat} for the tail probability of the $d$-dimensional Gaussian-like random vectors (Theorems~\ref{thm:chaganty} and \ref{thm:lattice}, respectively)  are quite useful and may find many applications within and beyond information theory. For example, together with \thmref{thm:moderatePetrov}, 
    one can prove an MD version of the asymptotics of the rate-distortion function in \cite{kostina2012lossyJ}. A lossless source coding asymptotics result is already proven in \cite{sakai2021Third} by applying \thmref{thm:moderatePetrov}, which was one of the motivations for the analysis in the current paper. 
    Theorems~\ref{thm:chaganty}--\ref{thm:lattice} can also be used to refine the asymptotic expansions for several network information theory problems, including the characterization of the performance of the multiple access channel; for these problems in particular, the challenging task for the extension to the MD regime is to prove the $d$-dimensional version of \thmref{thm:moderatePetrov}. 
    As Theorems~\ref{thm:chaganty}--\ref{thm:lattice} apply to Gaussian-like distributions that are not necessarily the sum of independent random vectors, they could allow the refinements of the performance of constant-composition codes and universal codes (e.g., the maximum empirical mutual information decoder) in channel coding.


    \appendices
    
    
    \section{Proof of \lemref{lem:cornish}}
    \label{app:cornish}
 For $y = O(1)$, \lemref{lem:cornish} states that $F_n(-x) = Q(x) \left(1 + O\left( \frac{1}{\sqrt{n}} \right) \right)$. By the Taylor series expansion of $Q(y)$, we solve
 \begin{align}
     x = y + \bigo{\frac{1}{\sqrt{n}}},
 \end{align}
 which confirms the statement of the lemma for finite $y$.

 Next, we focus on the cases $y \to \infty$ or $y \to -\infty$ with $y \in o(\sqrt{n})$. We here prove the case where $y \to \infty$. The case $y \to -\infty$ follows similarly using \eqref{eq:petrov1}.  From \eqref{eq:petrov2}, we have
	\begin{align}
	    F_n(-x) &= Q(x) \exp\bigg\{-a_0 \frac{x^3}{\sqrt{n}} + a_1 \frac{x^4}{n} \notag \\
	    &\quad - O\left(\frac{x^5}{n^{3/2}}\right) + O\left(\frac{x}{\sqrt{n}}\right) \bigg\}. \label{eq:Fnx}
	\end{align}
	
	Let $x = y + \delta$ where $\delta / y \to 0$. 
	Substituting $F_n(-x) = Q(y)$ into \eqref{eq:Fnx}, we get
	\begin{IEEEeqnarray}{rCl}
	    \IEEEeqnarraymulticol{3}{l}{\frac{Q(y + \delta)}{Q(y)}} \notag \\
	    &=& \exp\left \{a_0 \frac{x^3}{\sqrt{n}} - a_1 \frac{x^4}{n} + O\left(\frac{y^5}{n^{3/2}}\right)+ O\left(\frac{y}{\sqrt{n}}\right) \right\}. \IEEEeqnarraynumspace \label{eq:Qfrac}
	\end{IEEEeqnarray}
	As $y \to \infty$, we have the asymptotic expansion \cite[eq. 26.2.12]{abramowitz1972handbook}
	\begin{align}
	    Q(y) = \frac{1}{\sqrt{2 \pi}} \exp\left\{-\frac{y^2}{2}\right\} \frac{1}{y} \left( 1 - \bigo{\frac{1}{y^2}} \right). \label{eq:Qyexp}
	\end{align}
 
	Substituting \eqref{eq:Qyexp} into the left-hand side of \eqref{eq:Qfrac} and
	taking the logarithm of both sides of \eqref{eq:Qfrac}, we get
	\begin{IEEEeqnarray}{rCl}
	    \IEEEeqnarraymulticol{3}{l}{-\delta y - \frac{\delta^2}{2} - \frac{\delta}{y} + \bigo{\frac{\delta^2}{y^2}}} \notag \\
	    &=& a_0 \frac{y^3}{\sqrt{n}} + a_0 \frac{3 y^2 \delta}{\sqrt{n}} + a_0 \frac{3 y \delta^2}{\sqrt{n}} + a_0 \frac{ \delta^3}{\sqrt{n}} - a_1 \frac{y^4}{n} \notag\\
	    &&+ \bigo{\frac{y^5}{n^{3/2}}} + \bigo{\frac{y^3 \delta }{n}} + \bigo{\frac{y}{\sqrt{n}}}. \label{eq:longeq}
	\end{IEEEeqnarray}
	Equating the coefficients of $\frac{y^3}{\sqrt{n}}$ and $\frac{y^4}{n}$ of both sides of \eqref{eq:longeq}, we get
	\begin{align}
	    b_0 &= a_0 \\
	    b_1 &= \frac{5}{2} a_0^2 + a_1,
	\end{align}
	which completes the proof.

 \section{Proof of \thmref{thm:lattice}} \label{app:lattice}
First, for lattice random vectors, we prove two auxiliary results that are similar to those in \cite{chaganty, chagantyOneDimensional}.

\begin{lemma}\label{lem:lat1}
Let $\mb{Y}_n \in \mathbb{R}^d$ be a lattice random vector taking values in the $d$-dimensional lattice $\Pi_{j = 1}^d \{k h_{n, j} \colon k \in \mathbb{Z}\}$, where $h_{n, j} > 0$ is the span in coordinate $j$. Assume that $h_{n, j} \to 0$ for all $j \in [d]$ as $n \to \infty$. Let $b_n$ be a sequence satisfying $0 < \liminf_{n \to \infty} b_n h_{n, j} < \infty$ for all $j \in [d]$. Suppose that $\mb{Y}_n$ converges in distribution to a random vector $\mb{Y}$ with a well-defined probability distribution function $f_{\mb{Y}}$ satisfying
\begin{align}
    \sup_{n \geq 1, \mb{y} \in \mathbb{R}^d} \frac{1}{\prod_{j = 1}^d h_{n, j}} \Prob{\mb{Y}_n = \mb{y}} \leq M \label{eq:condlat1}
\end{align}
for some $M < \infty$, and if $\mb{y}_n$ is in the range of $\mb{Y}_n$ and $\mb{y}_n$ converges to $\mb{y}$, then
\begin{align}
    \frac{1}{\prod_{j = 1}^d h_{n, j}} \Prob{\mb{Y}_n = \mb{y}_n} \to f_{\mb{Y}}(\mb{y}). \label{eq:condlat2}
\end{align}
Then, as $n \to \infty$
\begin{align}
    &\prod_{j = 1}^d \frac{1 - \exp\{-b_n h_{n, j}\}}{h_{n, j}} \E{\exp\left\{-b_n \sum_{j = 1}^d Y_{n, j} \right\} 1\{\mb{Y}_n \geq 0\}} \notag \\
    &\quad = f_\mb{Y}(\mb{0}) (1 + o(1)).
\end{align}

\end{lemma}
\begin{IEEEproof}[Proof of \lemref{lem:lat1}]
The proof extends \cite[Th.~2.10]{chagantyOneDimensional} to the multidimensional scenario. We use arguments similar to those in \cite[Th.~2.10]{chagantyOneDimensional}. Let
\begin{align}
    I_n &\triangleq \E{\exp\left \{ - b_n \sum_{j = 1}^d Y_{n, j} \right\} 1 \{\mb{Y}_n \geq 0\}} \\
    &= \sum_{k_1 = 0}^{\infty} \dots \sum_{k_d = 0}^{\infty} \exp\left\{- b_n \sum_{j = 1}^d k_j h_{n, j}\right\} \notag \\
    &\quad \Prob{\mb{Y}_n = (k_1 h_{n, 1}, \dots, k_d h_{n, d})}.
\end{align}
Fix some integer $N > 0$. We bound $I_n$ as 
\begin{align}
    I_n &\geq \sum_{k_1 = 0}^{N-1} \dots \sum_{k_d = 0}^{N-1} \exp\left\{- b_n \sum_{j = 1}^d k_j h_{n, j}\right\} \notag \\
    &\quad \Prob{\mb{Y}_n = (k_1 h_{n, 1}, \dots, k_d h_{n, d})}.
\end{align}
Using the assumption in \eqref{eq:condlat2}, we get
\begin{align}
    &\liminf_n \left( \prod_{j = 1}^d \frac{1 - \exp\{- b_n h_{n, j}\}}{h_{n, j}} \right) I_n \notag \\
    &\geq f_{\mb{Y}}(\boldsymbol{0}) \liminf_{n} \prod_{j = 1}^d (1- \exp\{- N b_n h_{n, j}\}).
\end{align}
Similarly, using assumption \eqref{eq:condlat1} from the same statement of the lemma, we bound $I_n$ as
\begin{align}
    I_n&\leq \sum_{k_1 = 0}^{N-1} \dots \sum_{k_d = 0}^{N-1} \exp\left\{- b_n \sum_{j = 1}^d k_j h_{n, j}\right\} \notag \\
    &\quad \Prob{\mb{Y}_n = (k_1 h_{n, 1}, \dots, k_d h_{n, d})} \\
    &\quad + M \prod{j = 1}^d h_{n, j} \sum_{k = N}^{\infty} \exp\{-k b_n h_{n, j}\},
\end{align}
giving
\begin{align}
     &\limsup_n \left( \prod_{j = 1}^d \frac{1 - \exp\{- b_n h_{n, j}\}}{h_{n, j}} \right) I_n \notag \\
     &\leq f_{\mb{Y}}(\boldsymbol{0}) +M \limsup_n M \prod_{j = 1}^d \exp \{- N b_n h_{n, j}\}.
\end{align}
Letting $N \to \infty$ and using the fact that $0 < \liminf_n b_n h_{n, j} < \infty$ for all $j$, we conclude that
\begin{align}
    \lim_{n \to \infty} \left( \prod_{j = 1}^d \frac{1 - \exp\{- b_n h_{n, j}\}}{h_{n, j}} \right) I_n = f_{\mb{Y}}(\boldsymbol{0}).
\end{align}
\end{IEEEproof}

\begin{lemma}\label{lem:lat2}
    Let $\mb{Y}_n$ be a lattice random vector as defined in \lemref{lem:lat1}. Let $\mb{Y}_n$ converge in distribution to $\mb{Y}$. Let $\phi_n(\cdot)$ and $\phi(\cdot)$ be the MGF of $\mb{Y}_n$ and $\mb{Y}$, respectively. Assume that there exists an integrable function $f^*$ such that
    \begin{align}
        \sup_n |\phi_n(\mathsf{i} \mb{t})| 1 \{\norm{t}_{\infty} \leq \beta_n \} \leq f^*(\mb{t}) \label{eq:condlat21}
    \end{align}
    for each $\mb{t} \in \mathbb{R}^d$, and
    \begin{align}
        \sup_{\beta_{n, j} < t_j \leq \pi / h_{n, j} \colon j \in d} |\phi_n(\mathsf{i} \mb{t})| = o\left(\prod_{j = 1}^d h_{n, j}\right) \label{eq:condlat22}
    \end{align}
    for some $\beta_{n, j} \to \infty$ for all $j \in [d]$.
    Then, the conditions in \eqref{eq:condlat1}--\eqref{eq:condlat2} hold for $\mb{Y}_n$ and $\mb{Y}$.
\end{lemma}

\begin{IEEEproof}[Proof of \lemref{lem:lat2}]
The proof follows steps identical to the proof of \cite[Th.~2.9]{chagantyOneDimensional}, where the inversion formula for the one-dimensional case is replaced by the one for the $d$-dimensional case. 
\end{IEEEproof}

We are now equipped to complete the proof of \thmref{thm:lattice}. Using the identity in \cite[eq.~(3.6)]{chaganty}, we get
\begin{align}
    & \Prob{\mb{S}_n \geq n \mb{a}_n} = \exp\{-n \Lambda_n(\mb{a}_n)\} \notag \\
    &\quad  \E{\exp\left\{ - \sqrt{n} \sum_{j = 1}^d Y_{n, j} 1\{\mb{Y}_n \geq 0\} \right \}},
\end{align}
where $\mb{Y}_n = (Y_{n, 1}, \dots, Y_{n, d})$, $\mb{Y}_n = \ms{D}_n \frac{\mb{S}_n^* - n \mb{a}_n}{\sqrt{n}}$, $\ms{D}_n$ is the diagonal matrix with the diagonal entries $s_{n, 1}, \dots, s_{n, d}$, and $\mb{S}_n^*$ is the tilted version of $\mb{S}_n$, defined as 
\begin{align}
    dF_{\mb{S}_n^*}(\mb{y}) = \frac{\exp\{\mb{y}^\top \mb{s}_n\}}{\phi_n(\mb{s}_n)} d F_{\mb{S}_n}(\mb{y}) \quad \forall \, \mb{y} \in \mathbb{R}^d.
\end{align}
It follows that $\mb{Y}_n$ is also lattice with a span vector $\frac{1}{\sqrt{n}} (s_{n, 1} h_{n, 1}, \dots, s_{n, d} h_{n, d})$. From \cite[Lemma 3.1]{chaganty} and Conditions (S) and (ND) in \thmref{thm:chaganty}, it follows that $\mb{Y}_n$ converges to $\mb{Y} \sim \mc{N}(\mb{0}, \ms{D}_n \nabla^2 \kappa_n (\mb{s}_n) \ms{D}_n)$ in distribution. Therefore, the density of $\mb{Y}$ satisfies 
\begin{align}
    f_{\mb{Y}}(\mb{0}) = \frac{1}{(2 \pi)^{d/2}} \frac{1}{\prod_{j = 1}^d {s_{n, j}} \sqrt{\det(\nabla^2 \kappa_n (\mb{s}_n))}}. \label{eq:fY}
\end{align}
Applying \lemref{lem:lat1} with $f_{\mb{Y}}(\mb{0})$ from \eqref{eq:fY}, $b_n = \sqrt{n}$, and $h_{n, j}$ replaced with $\frac{1}{\sqrt{n}} s_{n, j} h_{n, j}$, we obtain \eqref{eq:chagantyLat}.

To complete the proof, it only remains to verify the conditions of \lemref{lem:lat1}. By \lemref{lem:lat2}, \eqref{eq:condlat21}--\eqref{eq:condlat22} are sufficient. 

The condition in \eqref{eq:condlat21} holds with $\beta_n = \sqrt{n} \delta$ for some $\delta > 0$ by \cite[Lemma 3.1]{chaganty} and Conditions (S) and (ND). 

From \cite[eq. (3.8) and (3.21)]{chaganty}, we have
\begin{align}
    |\phi^{(\mb{Y}_n)}(\ms{i} \mb{t})| \leq  \left \lvert \frac{\phi_n(\mb{s}_n + \mathrm{i} \ms{D}_n \mb{t} / \sqrt{n})}{\phi_n(\mb{s}_n)} \right \rvert, \label{eq:rightphi}
\end{align}
where $\phi_n$ denotes the MGF of $\mb{S}_n$. Rewriting Condition (L) implies that \eqref{eq:rightphi} is bounded by $o(\frac{1}{n^{d/2}})$ for $\mb{t}$ such that $\sqrt{n} \delta_j < |t_j| \leq \frac{\pi \sqrt{n} }{s_{n, j} h_{n, j}}$. Since the span of $\mb{Y}_n$ scales as $O\left(\frac{1}{\sqrt{n}}\right)$, the $o\left(\frac{1}{n^{d/2}}\right)$ bound verifies \eqref{eq:condlat22}, which completes the proof.

    \section{Proof of \eqref{eq:lambdaexp}} \label{app:NDproof}
    From \eqref{eq:taun} and \eqref{eq:an}, we get $\mb{a}_n \to (I(\PX), 0)$ as $n \to \infty$. 
To evaluate the gradient and the Hessian of $\Lambda(\mb{a}_n)$, we start from the equation in condition (ND)
\begin{align}
    \nabla \kappa(\mb{s}_n) = \mb{a}_n. \label{eq:gradeq}
\end{align}
Viewing $\mb{a}_n$ as a vector-valued function of $\mb{s}_n$ and differentiating both sides of  \eqref{eq:gradeq} with respect to $\mb{s}_n$, we get
\begin{align}
    J_{\mb{s}_n}(\mb{a}_n) = \nabla^2 \kappa(\mb{s}_n), \label{eq:Jsa}
\end{align}
where $J_{\mb{s}_n}(\mb{a}_n) \triangleq \begin{bmatrix} \frac{\partial a_{n, 1}}{\partial s_{n, 1}} &  \frac{\partial a_{n, 1}}{\partial s_{n, 2}} \\ \frac{\partial a_{n, 2}}{\partial s_{n, 1}} &  \frac{\partial a_{n, 2}}{\partial s_{n, 2}} \end{bmatrix}$ is the Jacobian of $\mb{a}_n$ with respect to $\mb{s}_n$. 

Differentiating the equation $\Lambda(\mb{a}_n) = \langle \mb{s}_n, \nabla \kappa(\mb{s}_n) \rangle - \kappa(\mb{s}_n)$ with respect to $\mb{s}_n$, we get a 2-dimensional row vector
\begin{align}
    J_{\mb{s}_n}(\Lambda(\mb{a}_n)) = \mb{s}_n^\top \nabla^2 \kappa(\mb{s}_n).
\end{align}
Applying the function inversion theorem and using \eqref{eq:Jsa}, we reach
\begin{align}
    J_{\mb{a}_n}(\Lambda(\mb{a}_n)) &= J_{\mb{s}_n}(\Lambda(\mb{a}_n)) J_{\mb{a}_n}(\mb{s}_n) \\
    &= \mb{s}_n^\top \nabla^2 \kappa(\mb{s}_n) (\nabla^2 \kappa)^{-1}(\mb{s}_n) \\
    &= \mb{s}_n^\top, \label{eq:Jaa}
\end{align}
equivalently
\begin{align}
    \nabla \Lambda(\mb{a}_n) = \mb{s}_n. \label{eq:La}
\end{align}
Differentiating \eqref{eq:La} with respect to $\mb{a}_n$, we get
\begin{align}
    \nabla^2 \Lambda(\mb{a}_n) &= \nabla (\nabla \Lambda(\mb{a}_n)) \\
    &= J_{\mb{a}_n}(\mb{s}_n) \\
    &= (\nabla^2 \kappa)^{-1}(\mb{s}_n). \label{eq:G2La}
\end{align}

We would like to obtain the Taylor series expansion of $\Lambda(\cdot)$ around $\mb{a} = (I(\PX), 0)$. By direct computation, we get
\begin{align}
   \Lambda(\mb{a}) &= I(\PX) \label{eq:LIP} \\
   \nabla \Lambda(\mb{a}) &= (1, 1) \label{eq:Ld}\\
   \nabla \kappa((1, 1)) &= \mb{a},
\end{align}
giving $\mb{s}_n \to \mb{s} \triangleq (1, 1)$, which verifies condition (ND). Define 
\begin{align}
     \mb{T} &\triangleq (T_{1}, T_{2})\\
     T_{1} &\triangleq \log \frac{\W(Y| X)}{\PY(Y)} \\
     T_{2} &\triangleq \log \frac{\W({Y} | \overline{{X}})}{\W(Y | X)},
\end{align}
where $P_{X, \overline{X}, Y}(x, \overline{x}, y) = \PX(x) \PX(\overline{x}) \W(y|x)$.
We have 
\begin{align}
    \nabla^2 \kappa (\mb{s}) = \mathrm{Cov}(\tilde{\mb{T}})^{-1}, \label{eq:kappa2s}
\end{align}
where $\tilde{\mb{T}}$ is distributed according to the tilted distribution
\begin{align}
    P_{\mb{\tilde{T}}} = \exp\{\langle \mb{s}, \mb{T} \rangle\} P_{\mb{T}} =  \frac{\W(Y|\overline{X})}{\PY(Y)} P_{\mb{T}},
\end{align}
and $P_{\mb{T}}$ denotes the distribution of $\mb{T}$.
We compute the inverse of the covariance matrix of $\tilde{\mb{T}}$ as 
\begin{align}
    \mathrm{Cov}(\tilde{\mb{T}})^{-1} = \begin{bmatrix} \frac{2}{1 + \eta(\PX)} & \frac{1}{1+\eta(\PX)} \\ \frac{1}{1 + \eta(\PX)} & \frac{1}{1-\eta(\PX)^2}\end{bmatrix} \frac{1}{V_{\mathrm{u}}(\PX)}. \label{eq:Ldd}
\end{align}
From \eqref{eq:LIP}, \eqref{eq:Ld}, and \eqref{eq:Ldd}, we get
\begin{align}
    \Lambda(\mb{a}_n) &= I(\PX) + (a_{n, 1} - I(\PX)) \notag \\
    &+ \frac{1}{2} (a_{n, 1} - I(\PX))^2 \mathrm{Cov}(\tilde{\mb{T}})^{-1}_{1, 1}  + O(|a_{n, 1} - I(\PX) |^3)\\
    &= a_{n, 1} + \frac{1}{n} \frac{Q^{-1}(\epsilon_n)^2}{1 + \eta(\PX)} + \bigo{\frac{Q^{-1}(\epsilon_n)^3}{n^{3/2}}} + \bigo{\frac{1}{n}}. 
\end{align}
	
	\section{Solution of \eqref{eq:opt}} \label{app:lagrange}
We solve the convex optimization problem in \eqref{eq:opt} by writing the Lagrangian
\begin{align}
    L(\mb{h}, \lambda) = \mb{h}^\top \mb{g} - \frac{1}{2} \mb{h}^\top \ms{J}_{\mc{X}^*} \mb{h} - \lambda \mb{h}^\top \bs{1}. \label{eq:lagran}
\end{align}
The Karush--Kuhn--Tucker condition $\nabla L(\mb{h}, \lambda) = 0$ gives
\begin{align}
\ms{J}_{\mc{X}^*} \mb{h} &= \mb{g} - \lambda \bs{1} \label{eq:Jeq}\\
\mb{h}^\top \bs{1} &= 0, \label{eq:gt}
\end{align}
where $\ms{J}$ is given in \eqref{eq:J}.
Since $\mb{h}$ belongs to $\mr{row}(\ms{J}_{\mc{X}^*})$ by assumption, the Lagrangian in \eqref{eq:lagran} depends on $\mb{g}$ only through its projection onto $\mr{row}(\ms{J}_{\mc{X}^*})$. Therefore, without loss of generality, assume that $\mb{g} \in \mr{row}(\ms{J}_{\mc{X}^*})$.

The equation \eqref{eq:Jeq} has a solution since both $\mb{g}$ and $\bs{1}$ are in the row space of $\ms{J}_{\mc{X}^*}$.  Solving the system of equations in \eqref{eq:Jeq} and \eqref{eq:gt}, we get the dual variable
\begin{align}
    \lambda^* = \frac{\bs{1}^\top \ms{J}_{\mc{X}^*}^+ \mb{g}}{\bs{1}^\top \ms{J}_{\mc{X}^*}^+ \bs{1}}. \label{eq:lambdasol}
\end{align}
Plugging \eqref{eq:lambdasol} in \eqref{eq:gt}, we get
\begin{align}
    \mb{h}^* &= \tilde{\ms{J}} \mb{g}\\
    &= -  \frac{Q^{-1}(\epsilon_n)}{2 \sqrt{n V_{\epsilon_n}}} \tilde{\ms{J}} \mb{v}(\PXs), \label{eq:gg}
\end{align}
where $\tilde{\ms{J}}$ and $\mb{v}$ are given in \eqref{eq:tildeJ} and \eqref{eq:vP}.
An equivalent characterization of \eqref{eq:gg} in terms of the eigenvalue decomposition of $\ms{J}_{\mc{X}^*}$ is given in \cite[Lemma 1 (v)]{moulin2017log}.
The value of the supremum in \eqref{eq:opt} is $\frac{1}{2} {\mb{g}}^{\top} \tilde{\ms{J}} \mb{g} = A_0(\PXs) Q^{-1}(\epsilon_n)^2$, where $A_0(\cdot)$ is given in \eqref{eq:A0}.

	\section{Proof of \lemref{lem:Dsasymp}}\label{app:Dsasymp}

    Define 
    \begin{align}
        D_{\mb{x},Q_Y} &\triangleq D(\W \| \QY | \hat{P}_{\mb{x}}) \\
        V_{\mb{x},Q_Y} &\triangleq V(\W \| \QY | \hat{P}_{\mb{x}}) \\
        T_{\mb{x},Q_Y} &\triangleq T(\W \| \QY | \hat{P}_{\mb{x}}).
    \end{align}
 
    (i)\,\,First consider the case $V_{\mb{x}, Q_Y} = 0$. In this case, the random variable $\log \frac{\Wxn(\mb{Y})}{Q_Y^n(\mb{Y})} = \sum_{i = 1}^n \log \frac{\W(Y_i | x_i)}{\QY(Y_i)}$, where $\mb{Y} \sim \Wxn$, is almost surely equal to $n D_{\mb{x}, Q_Y}$, and the inequality in \eqref{eq:Dsasy} trivially holds for any SMD sequence $\epsilon_n$.

    Next, consider the case $V_{\mb{x}, Q_Y} > 0$. The random variable $\sum_{i = 1}^n \log \frac{\W(Y_i | x_i)}{\QY(Y_i)}$ is a sum of $n$ independent, but not necessarily identically distributed random variables. Since $Q_Y(y) \geq a > 0$ for all $y \in \mc{Y}$, it follows for each $x \in \mc{X}$ that $\log \frac{P_{Y|X}(Y|x)}{Q_Y(Y)}$ is a bounded random variable, hence all of its cumulants are finite.  

    Notice that the distribution of the random variable $\frac{1}{n}\sum_{i = 1}^n \log \frac{\W(Y_i | x_i)}{\QY(Y_i)}$ depends on $\mb{x}$ only through its type $\Phat$. Let $\kappa_k(\Phat, Q_Y)$ be $k$-th cumulant of $\frac{1}{n} \sum_{i = 1}^n \log \frac{\W(Y_i | x_i)}{\QY(Y_i)}$, and let $\kappa_k(x, Q_Y)$ be the $k$-th cumulant of $\log \frac{P_{Y|X}(Y|x)}{Q_Y(Y)}$. Then, we have
    \begin{align}
        \kappa_k(\Phat, Q_Y) = \sum_{x \in \mc{X}} \Phat(x) \kappa_k(x, Q_Y) 
    \end{align}
    Hence, there exist constants $c_k$ such that 
    \begin{align}
        \max_{\substack{\Phat \in \mc{P}_n \\ Q_Y \in \mc{Q}(a)}} \left \lvert \kappa_k(\Phat, Q_Y) \right \rvert \leq c_k, \quad k \geq 1. \label{eq:ci}
    \end{align}
    In particular,
    \begin{align}
        \kappa_1(\Phat, Q_Y) &= \frac{1}{n} \E{\sum_{i = 1}^n \log \frac{\W(Y_i | x_i)}{\QY(Y_i)}} = D_{\mb{x}, Q_Y} \\
         \kappa_2(\Phat, Q_Y) &= \frac{1}{n} \Var{\sum_{i = 1}^n \log \frac{\W(Y_i | x_i)}{\QY(Y_i)}} =  V_{\mb{x}, Q_Y} \\ 
         \kappa_3(\Phat, Q_Y) &= \E{\left(\frac{1}{n} \sum_{i = 1}^n \log \frac{\W(Y_i | x_i)}{\QY(Y_i)} - D_{\mb{x}}\right)^3} \notag \\
         &=  T_{\mb{x}, Q_Y}.
    \end{align}
   
    Note that Cram\'er's condition \eqref{eq:Cramercond} in \thmref{thm:moderatePetrov} is satisfied since $\log \frac{\W(Y|x)}{\QY(Y)}$ is a bounded random variable for all $x \in \mc{X}$, and \eqref{eq:kappacond} is satisfied since $V_{\mb{x}, Q_Y} > 0$. Applying \lemref{lem:cornish} by setting $X_i$ to $\log \frac{\W(Y_i|x_i)}{\QY(Y_i)}$ gives \eqref{eq:Dsasy}. The universality of the constants in \eqref{eq:Dsasy} follows from the uniform bound on the cumulants in \eqref{eq:ci}.

    (ii)\,\,Similar to part (i), the case where $V(\Phat) = 0$ trivially follows. Suppose that $V(\Phat) > 0$. Let $\kappa_i(\Phat)$ be the $i$-th cumulant of $\frac{1}{n} \sum_{i = 1}^n \log \frac{P_{Y|X}(Y_i|x_i)}{\hat{Q}_{\mb{x}}(Y_i)}$, which is a sum of $n$ independent random variables. We compute
    \begin{align}
        \kappa_1(\Phat) &= D(\W \| \hat{Q}_{\mb{x}} | \Phat) = I(\Phat) \\
        \kappa_2(\Phat) &= V(\W \| \hat{Q}_{\mb{x}} | \Phat) = V(\Phat) \\
        \kappa_3(\Phat) &= T(\W \| \hat{Q}_{\mb{x}} | \Phat) = T(\Phat).
    \end{align}
    
    Next, following the steps in the proof of \cite[Lemma~46]{polyanskiy2010Channel} and using the notation $\left \lVert Z \right \rVert_k = \E{|Z|^k}^{1/k}$, we get for every $P_X \in \mc{P}$ 
    \begin{align}
        &\left \lVert \log \frac{P_{Y|X}(Y|X)}{P_Y(Y)} - D(P_{Y|X} \| P_Y | P_X) \right \rVert_k \\
        &\leq \left \lVert \log \frac{1}{P_{Y|X}(Y|X)} \right \rVert_k +  \left \lVert \log \frac{1}{P_{Y}(Y)} \right \rVert_k + I(P_X) \\
        &\leq 2 |\mc{Y}| \left( \frac{k}{e} \right) + \log |\mc{Y}| < \infty. \label{eq:Zkbound} 
    \end{align}
    Since the $k$-th cumulant is a polynomial function of the first $k$ central moments, \eqref{eq:Zkbound} implies that there exist constants $d_k$ such that
    \begin{align}
        \max_{\Phat \in \mc{P}_n} |\kappa_k(\Phat)| \leq d_k, \quad k \geq 1. \label{eq:di}
    \end{align}
    Applying \lemref{lem:cornish} by setting $X_i$ to $\log \frac{\W(Y_i|x_i)}{\hat{Q}_{\mb{x}}(Y_i)}$ gives \eqref{eq:Dsasy2}. The universality of the constants in \eqref{eq:Dsasy2} follows from the uniform bound on the cumulants in \eqref{eq:di}.

\section{Proof of \lemref{lem:xibound}} \label{app:xibound}

The proof follows the proof of \cite[Lemma~9 (iii)]{moulin2017log} closely. The difference is that we consider SMD sequences $\epsilon_n$, which means that $\Qinv$ is not necessarily $O(1)$.
Fix any sequence of distributions $P_{X,n} \notin {\mc{P}^*(\rho_n)}$ such that $P_{X,n} \to P_X' \in \mc{P}$. Define
\begin{align}
    P_{0,n} &\triangleq \argmin_{\tilde{P}_X \in \mc{P}^\dagger} \norm{P_{X,n} - \tilde{P}_X}_2 \\
    P_{1,n} &\triangleq \argmin_{\tilde{P}_X \in \mc{P}^*} \norm{P_{X,n} - \tilde{P}_X}_2 \\
    P_{2,n} &\triangleq \argmin_{\tilde{P}_X \in \mc{P}^*} \norm{P_{0,n} - \tilde{P}_X}_2 \\
    \delta_n &\triangleq \norm{P_{X,n} - P_{0,n}}_{\infty} \\
    \nu_n &\triangleq \norm{P_{X,n} - P_{1,n}}_{\infty}.
\end{align}
and
\begin{align}
    G(P_{X,n}) &\triangleq \begin{cases}
        - \sqrt{V(P_{X,n})} &\text{ if } \epsilon_n \leq \frac{1}{2} \\
        \sqrt{V(P_{X,n})} &\text{ otherwise}
        \end{cases} \\
    L(P_{X,n}) &\triangleq \frac{S(P_{X,n}) \sqrt{V(P_{X,n})}}{6}.
\end{align}
Note that 
\begin{align}
    \nu_n &\geq \rho_n \\
    \xi^{\epsilon_n}(P_{X,n}) &\triangleq \xi^{\epsilon_n}(P_{X,n}, P_{X,n}) \notag \\
    &= n I(P_{X,n}) + \sqrt{n} G(P_{X,n}) |\Qinv| \notag \\
    &\quad + L(P_{X,n}) \Qinv^2.
\end{align}

We analyze $\xi^{\epsilon_n}(P_{X,n})$ according to cases based on the value of $\nu_n$.

\textbf{Case 1: } $\nu_n \geq a$ for some constant $a > 0$, equivalently, $P_X' \notin \mc{P}^*$. There are two sub-cases.

\textbf{Case 1.a: } $\delta_n \geq b$ for some constant $b > 0$, equivalently, $P_X' \notin \mc{P}^\dagger$. Then, by the continuity of $I(\cdot)$, $I(P_X) \leq C'$ for some $C' < C$. Hence,
\begin{align}
    \xi^{\epsilon_n}(P_X) \leq n C' + o(n),
\end{align}
and the claim in \eqref{eq:xic1} holds.

\textbf{Case 1.b: } $\delta_n \to 0$ and $\nu_n \geq a$, equivalently, $P_X' \in \mc{P}^\dagger$ but $P_X' \notin \mc{P}^*$. In this case, by the continuity of $V(\cdot)$, there exists some $V'$ such that $V' > V_{\epsilon_n}$ if $\epsilon_n < \frac{1}{2}$, and $V' < V_{\epsilon_n}$ if $\epsilon_n > \frac{1}{2}$.\footnote{The case $\epsilon_n = \frac{1}{2}$ belongs to Case 1.a since $\mc{P}^* = \mc{P}^\dagger$ if $\epsilon_n = \frac{1}{2}$.} Hence,
\begin{align}
    \xi^{\epsilon_n}(P_X) \leq nC - \sqrt{n V'} \Qinv + o(\sqrt{n} \Qinv),
\end{align}
and the claim holds. 

\textbf{Case 2: } $\nu_n \to 0$, equivalently $P_X' \in \mc{P}^*$. There are two sub-cases depending on whether $P_{0,n} \in \mc{P}^*$. 

\textbf{Case 2.a: } $P_{0,n} \in \mc{P}^*$, which implies $\delta_n = \nu_n$. By the quadratic decay property of mutual information from \cite[Th.~1]{cao2023} and the fact that $\norm{x}_2 \geq \norm{x}_{\infty}$, there exists a constant $\alpha > 0$ such that
\begin{align}
    I(P_{X,n}) \leq C - \alpha \nu_n^2. \label{eq:quaddecay}
\end{align}
The property in \eqref{eq:quaddecay} is claimed in the proof of \cite[Th.~48]{polyanskiy2010Channel}. In \cite[Th.~1]{cao2023}, Cao and Tomamichel close a gap in the proof of \eqref{eq:quaddecay}.
From the Taylor series expansion of $G(P_{X,n})$ and $L(P_{X,n})$ around $P_{0,n}$, we get
\begin{align}
    G(P_{X,n}) &\leq G(P_{0,n}) + \norm{\nabla G(P_{0,n})}_{1} \nu_n + o(\nu_n) \\
    L(P_{X,n}) &= L(P_{0,n}) + o(\nu_n).
\end{align}
Then,
\begin{align}
    \xi^{\epsilon_n}(P_{X,n}) &\leq nC - \sqrt{n V_{\epsilon_n}} \Qinv \\
    &\quad - n \alpha \nu_n^2 + \sqrt{n} \norm{\nabla G(P_{0,n})}_{1} \nu_n |\Qinv| \notag \\
    &\quad + L(P_{0,n}) \Qinv^2 \label{eq:quadr} \\
    &\quad + o(\sqrt{n} \nu_n \Qinv).
\end{align}
The terms in \eqref{eq:quadr} form a second-order polynomial of $\nu_n$ with a strictly negative leading coefficient. We fix a constant $c_1 > 0$ to be determined later. Then, there exists a constant $c_{01} > 0$ such that for all $\nu_n \geq \frac{c_{01} |\Qinv|}{\sqrt{n}}$, 
    \begin{align}
        &- n \alpha \nu_n^2 + \sqrt{n} \norm{\nabla G(P_{0,n})}_{1} \nu_n |\Qinv| \notag \\
        &\quad + L(P_{0,n}) \Qinv^2 \notag \\
        &\leq \left(\max_{\PXs \in \mc{P}^*} (L(\PXs) + A_0(\PXs) - A_1(\PXs))\right) \Qinv^2 \notag \\
        &\quad - c_1 \sqrt{n} \nu_n |\Qinv|,
    \end{align}
    and the claim holds for $c_0 \geq c_{01}$.

    \textbf{Case 2.b: } $P_{0,n} \notin {\mc{P}^*}$. The analysis of this sub-case is from \cite[eq. (B.9)--(B.14)]{moulin2017log}. This sub-case implies that $\mc{P}^* \neq \mc{P}^\dagger$. Since $V(P_X) = P_X^\top \tilde{\mb{v}}$ for all $P_X \in \mc{P}^\dagger$, where $\tilde{\mb{v}}$ is defined in \eqref{eq:vtilde}, the projection of $\nabla G(P_X^*)$ onto $\mathrm{ker}(\ms{J}_{\mc{X}^\dagger})$ is a vector $\mb{g}_0 \neq 0$ independent of $\PXs \in \mc{P}^*$. By the extremal property of $\mc{P}^*$, there exists a constant $c' > 0$ such that
    \begin{align}
        (P_{0,n} - P_{2,n})^\top \mb{g}_0 \leq -c' \norm{\mb{g}_0}_2 \norm{P_{0,n} - P_{2,n}}_2. \label{eq:extremal}
    \end{align}
    By the triangle inequality,
    \begin{align}
        \nu_n &\leq \norm{P_{X,n} - P_{2,n}}_{\infty} \\
        &\leq \norm{P_{X,n} - P_{0,n}}_{\infty} + \norm{P_{0,n} - P_{2,n}}_{\infty}. 
    \end{align}
    Let
    \begin{align}
        \lambda \triangleq \frac{1}{2} \frac{c' \norm{\mb{g}_0}_2}{c' \norm{\mb{g}_0}_2 + \sqrt{|\mc{X}|} \max_{\PXs \in \mc{P}^*} \norm{\nabla G(\PXs) }_2} \in \left( 0, \frac{1}{2} \right). \label{eq:lambdadef}
    \end{align}
    Then, one of the following two statements is true.
    \begin{enumerate}
    \item $\norm{P_{X,n} - P_{0,n}} \geq \lambda \nu_n$: In this case, $P_X$ is sufficiently far away from $\mc{P}^\dagger$.
    \item $\norm{P_{0,n} - P_{2,n}} \geq (1-\lambda) \nu_n \geq \lambda \nu_n > \norm{P_{X,n} - P_{0,n}}_{\infty}$: In this case, $P_{X,n}$ may be arbitrarily close to $\mc{P}^\dagger$, but it is sufficiently far away from $\mc{P}^*$. 
    \end{enumerate}
    In case 1), by \cite[Th.~1]{cao2023},
    \begin{align}
        I(P_{X,n}) &\leq C - \alpha \lambda^2 \nu_n^2, \label{eq:lambdaI}
    \end{align}
    and from the Taylor series expansion
    \begin{align}
        G(P_{X,n}) &\leq G(P_{1,n}) + \norm{\nabla G(P_{1,n})}_1 \nu_n + o(\nu_n) \\
        L(P_{X,n}) &= L(P_{1,n}) + o(1). \label{eq:lambdaLPX}
    \end{align}
    Applying the arguments in Case 2.a with \eqref{eq:lambdaI}--\eqref{eq:lambdaLPX}, we conclude that there exists a constant $c_{02} > 0$ such that for $\nu_n \geq \frac{c_{02} |\Qinv|}{\sqrt{n}}$, the claim holds for $c_0 \geq c_{02}$.

    In case 2), we expand $G(P_{X,n})$ as
    \begin{align}
        G(P_{X,n}) &= G(P_{2,n}) + (P_{X,n} - P_{0,n} + P_{0,n} \notag \\
        &\quad - P_{2,n})^\top \nabla G(P_{2,n}) + o(\nu_n) \\
        &\leq G(P_{2,n}) + (P_{0,n} - P_{2,n})^\top \mb{g}_0 \notag \\
        &\quad + \norm{\nabla G(P_{2,n})}_2 \norm{P_{X,n} - P_{0,n}}_2 + o(\nu_n) \\
        &\leq G(P_{2,n}) - (c' \norm{\mb{g}_0}_2 \norm{P_{0,n} - P_{2,n}}_2 \notag \\
        &\quad -  \norm{\nabla G(P_{2,n})}_2 \norm{P_{X,n} - P_{0,n}}_2) + o(\nu_n) \label{eq:stepextremal}\\
        &\leq G(P_{2,n}) - \nu_n \big( c' (1-\lambda) \norm{\mb{g}_0}_2 \notag \\
        &\quad - \lambda \sqrt{|\mc{X}|} \max_{\PXs \in \mc{P}^*} \norm{\nabla G(\PXs) }_2 \big) + o(\nu_n) \label{eq:lambdacond} \\
        &= G(P_{2,n}) - \frac{c'}{2} \norm{\mb{g}_0}_2 \nu_n + o(\nu_n). \label{eq:lambdadefused}
    \end{align}
    Here, \eqref{eq:stepextremal} follows from \eqref{eq:extremal}. \eqref{eq:lambdacond} follows since $\norm{P_{0,n} - P_{2,n}}_2 \geq \norm{P_{0,n} - P_{2,n}}_\infty \geq (1-\lambda) \nu_n$ and $\norm{P_{X,n} - P_{0,n}}_2 \leq \sqrt{|\mc{X}|} \norm{P_{X,n} - P_{0,n}}_\infty$, and \eqref{eq:lambdadefused} follows from \eqref{eq:lambdadef}.

   From $I(P_{X,n}) \leq C$, $L(P_{X,n}) = L(P_{1,n}) + o(1)$, and \eqref{eq:lambdadefused}, we get
    \begin{align}
        \xi^{\epsilon_n}(P_{X,n}) &\leq nC - \sqrt{nV_{\epsilon_n}} \Qinv + L(P_{1,n}) \Qinv^2 \notag \\
        & - \sqrt{n} \nu_n \frac{c'}{2} \norm{\mb{g}_0}_2  |\Qinv| + o(\sqrt{n} \nu_n \Qinv). \label{eq:xilinear}
    \end{align}
    Since the right-hand side of \eqref{eq:xilinear} decays linearly with $\nu_n$ with the scaling $\sqrt{n} \Qinv$, there exists a constant $c_{03} > 0$ such that for $\nu_n \geq \frac{c_{03} |\Qinv|}{\sqrt{n}}$, the claim in \eqref{eq:xic1} holds for $c_0 \geq c_{03}$ and some constant $c_1$ satisfying $0 < c_1 < \frac{c'}{2} \norm{\mb{g}_0}_2$. 

    By setting $c_0 = \max\{c_{01}, c_{02}, c_{03}\}$ and $\rho_n \geq \frac{c_0 |\Qinv|}{\sqrt{n}}$, we conclude that \eqref{eq:xic1} holds for all $P_{X.n} \notin \mc{P}^*(\rho_n)$.
    
    \section{Proof of Theorem \ref{thm:refinedAchConv}} \label{app:refined}
        \begin{IEEEproof}[Proof of the achievability]
    To prove the achievability, we derive the coefficient of $\bigo{\frac{Q^{-1}(\epsilon_n)^3}{\sqrt{n}}}$ in \lemref{lem:achP}, and invoke the refined \lemref{lem:achP} with $\PX = \PXs$. For this purpose, we need to modify the proof of \lemref{lem:achP} at two steps. First, using \lemref{lem:cornish}, the expansion for $t_n$ in \eqref{eq:tn} is refined as
    \begin{align}
        t_n &= Q^{-1}(\epsilon_n) - \frac{\ske_{\mr{u}} Q^{-1}({\epsilon}_n)^2}{6 \sqrt{n}} \notag \\
        &\quad + \frac{3 (\mu_4 - 3 V^2) V - 4 \mu_3^2}{72 V^3}\frac{ Q^{-1}({\epsilon}_n)^3}{n}  \notag \\
        &\quad +\bigo{\frac{ Q^{-1}({\epsilon}_n)^4}{n^{3/2}}} +  \bigo{\frac{1}{\sqrt{n}}}. \label{eq:tnref}
    \end{align}
    Second, we refine the expansion in \eqref{eq:lambdaexp} by computing the third-order gradient $\nabla^3 \Lambda(\mb{a}_n)$. Taking the gradient of \eqref{eq:G2La}, we get
    \begin{IEEEeqnarray}{rCl}
        &&\nabla^3 \Lambda(\mb{a}_n)_{i, j, k} = - \sum_{(a, b, c) \in [2]^3} \nabla^3 \kappa(\mb{s}_n)_{a, b, c} (\nabla^2 \kappa)^{-1}(\mb{s}_n)_{a, i}  \notag \\
        && \cdot (\nabla^2 \kappa)^{-1}(\mb{s}_n)_{b, j}  (\nabla^2 \kappa)^{-1}(\mb{s}_n)_{c, k}, \quad (i, j, k) \in [2]^3. \label{eq:kappa3mat}
    \end{IEEEeqnarray}
    In the case $\eta(\PXs) = 0$, the inverse of the Hessian $(\nabla^2 \kappa)^{-1}(\mb{s})$ in \eqref{eq:kappa2s} becomes
    \begin{align}
        (\nabla^2 \kappa)^{-1}(\mb{s}) = \begin{bmatrix} 2 & 1 \\ 1 & 1 \end{bmatrix} \frac{1}{V}, \label{eq:mat2}
    \end{align}
    and we compute
    \begin{align}
        \nabla^3 \kappa(\mb{s})_{1, 1, 1} &= \mu_3 \label{eq:kappa31}\\
        \nabla^3 \kappa(\mb{s})_{1, 1, 2} &= - \mu_3 \\
        \nabla^3 \kappa(\mb{s})_{1, 2, 2} &= \mu_3 \\
        \nabla^3 \kappa(\mb{s})_{2, 2, 2} &= 0. \label{eq:kappa34}
    \end{align}
    Note that \eqref{eq:kappa31}--\eqref{eq:kappa34} is sufficient to determine $\nabla^3 \kappa(\mb{s})$ since it is a symmetric order-3 tensor. From \eqref{eq:kappa3mat}--\eqref{eq:kappa34}, we compute
    \begin{align}
        \nabla^3 \Lambda(\mb{a})_{1, 1, 1} = - \frac{2 \mu_3}{V^3}. \label{eq:kappa3111}
    \end{align}
    Using \eqref{eq:mat2} and \eqref{eq:kappa3111}, we refine \eqref{eq:lambdaexp} as
    \begin{align}
         \Lambda(\mb{a}_n) &= a_{n, 1} + \frac{(a_{n, 1} - I(\PXs))^2}{V} - \frac{1}{6} (a_{n, 1} - I(\PXs))^3 \frac{2 \mu_3}{V^3}  \notag \\
         &\quad + O(|a_{n, 1} - I(\PXs)|^4) \\
         &= a_n + \frac{Q^{-1}(\epsilon_n)^2}{n} + \bigo{\frac{Q^{-1}(\epsilon_n)^4}{n^2}} + \bigo{\frac{1}{n}}. \label{eq:Lambdaref}
    \end{align}
     Following the steps in the proof \lemref{lem:achP} and using \eqref{eq:tnref} and \eqref{eq:Lambdaref} completes the proof.
    \end{IEEEproof}
    
    \begin{IEEEproof}[Proof of the converse]
    Set $\QY^{(n)} = (\QYs)^n$, where $\QYs$ is the equiprobable capacity-achieving output distribution. Since Cover--Thomas-symmetric channels have rows that are permutation of each other, we have that
       $ \beta_{1-\epsilon_n}(\Wxn, \QY^{(n)}) $
    is independent of $\mb{x} \in \mc{X}^n$. By \cite[Th. 28]{polyanskiy2010Channel}, we have 
    \begin{align}
        \log M^*(n, \epsilon_n) \leq - \log \beta_{1-\epsilon_n}(\Wxn, \QY^{(n)}), \label{eq:28}
    \end{align}
    where $\mb{x} = (x_0, \dots, x_0)$ for some $x_0 \in \mc{X}$. Applying \thmref{thm:NP} to the right-hand side of \eqref{eq:28} completes the proof.
    \end{IEEEproof}

\section*{Acknowledgment}
    We are grateful to Dr. Vincent Y. F. Tan and the anonymous reviewers for their helpful comments and suggestions, in particular for pointing out the papers \cite{chubb2017quantum, vazquez2021, cao2023}.
    
\bibliographystyle{IEEEtran}
\bibliography{mac} 

\begin{IEEEbiographynophoto}{Recep Can Yavas}
(S'18--M'22) received the B.S. degree (Hons.) in electrical engineering from Bilkent University, Ankara, Turkey, in 2016. He received the M.S. and Ph.D. degrees in electrical engineering from the California Institute of Technology (Caltech) in 2017 and 2023, respectively. He is currently a research fellow at CNRS at CREATE, Singapore. His research interests include information theory, probability theory, and multi-armed bandits.
\end{IEEEbiographynophoto}

\begin{IEEEbiographynophoto}{Victoria Kostina}
    (S'12--M'14--SM'22)
    is a professor of electrical engineering and of computing and mathematical sciences at Caltech. She received the bachelor's degree from Moscow Institute of Physics and Technology (MIPT) in 2004, the master's degree from University of Ottawa in 2006, and the Ph.D. degree from Princeton University in 2013.  During her studies at MIPT, she was affiliated with the Institute for Information Transmission Problems of the Russian Academy of Sciences. 
 
 Her research interests lie in information theory, coding, communications, learning, and control.
She has served as an Associate Editor for IEEE Transactions of Information Theory, and as a Guest Editor for the IEEE Journal on Selected Areas in Information Theory.  She received the Natural Sciences and Engineering Research Council of Canada postgraduate scholarship during 2009--2012, Princeton Electrical Engineering Best Dissertation Award in 2013, Simons-Berkeley research fellowship in 2015 and the NSF CAREER award in 2017.  
 \end{IEEEbiographynophoto}


\begin{IEEEbiographynophoto}{Michelle Effros}
(S'93--M'95--SM'03--F'09) is the George Van Osdol
Professor of Electrical Engineering and Vice Provost at the California
Institute of Technology.  She was a co-founder of Code On Technologies, a
technology licensing firm, which was sold in 2016.  Dr. Effros is a fellow
of the IEEE and has received a number of awards including Stanford’s
Frederick Emmons Terman Engineering Scholastic Award (for excellence in
engineering), the Hughes Masters Full-Study Fellowship, the National
Science Foundation Graduate Fellowship, the AT\&T Ph.D. Scholarship, the
NSF CAREER Award, the Charles Lee Powell Foundation Award, the Richard
Feynman-Hughes Fellowship, an Okawa Research Grant, and the Communications
Society and Information Theory Society Joint Paper Award.  She was cited
by Technology Review as one of the world’s top 100 young innovators in
2002, became a fellow of the IEEE in 2009, and is a member of Tau Beta Pi,
Phi Beta Kappa, and Sigma Xi.  She received the B.S.~(with distinction),
M.S., and Ph.D.~degrees in electrical engineering from Stanford
University.  Her research interests include information theory (with a
focus on source, channel, and network coding for multi-node networks)
and theoretical neuroscience (with a focus on neurostability and memory).

Dr. Effros served as the Editor of the IEEE Information Theory Society
Newsletter from 1995 to 1998 and as a Member of the Board of Governors of
the IEEE Information Theory Society from 1998 to 2003 and from 2008 to
2017. She served as President of the IEEE Information Theory Society in
2015 and as Executive Director for the film ``The Bit Player,'' a movie
about Claude Shannon, which came out in 2018.  She was a member of the
Advisory Committee and the Committee of Visitors for the Computer and
Information Science and Engineering (CISE) Directorate at the National
Science Foundation from 2009 to 2012 and in 2014, respectively. She served
on the IEEE Signal Processing Society Image and Multi-Dimensional Signal
Processing (IMDSP) Technical Committee from 2001 to 2007 and on ISAT from
2006 to 2009. She served as Associate Editor for the joint special issue
on Networking and Information Theory in the IEEE Transactions on
Information Theory and the IEEE/ACM Transactions on Networking, as
Associate Editor for the special issue honoring the scientific legacy of
Ralf Koetter in the IEEE Transactions on Information Theory and, from 2004
to 2007 served as Associate Editor for Source Coding for the IEEE
Transactions on Information Theory.  She has served on numerous technical
program committees and review boards, including serving as general
co-chair for the 2009 Network Coding Workshop and technical program
committee co-chair for the 2012 IEEE International Symposium on
Information Theory and the 2023 IEEE Information Theory Workshop.
\end{IEEEbiographynophoto}

\end{document}